\documentclass[usenatbib]{mn2e}
\usepackage{times}
\usepackage{graphicx}
\usepackage{url}
\usepackage[caption=false]{subfig}
\usepackage{amsmath}
\usepackage{amssymb}
\usepackage{tabularx}
\usepackage{ifpdf}
\usepackage{rotating}

\usepackage{fixltx2e} 

\ifpdf\else\usepackage{graphics}\fi

\newcommand{\degree}{\ensuremath{^\circ}}

\title[Post-correlation filtering techniques]{Post-correlation filtering techniques for off-axis source and RFI removal}
\author[A.R. Offringa et al.]{
A.R.~Offringa$^1$,
A.G.~de~Bruyn$^{1,2}$,
S.~Zaroubi$^1$
\\
\small{$^1$Kapteyn Astronomical Institute, University of Groningen, PO Box 800, 9700 AV Groningen, The Netherlands} \\
\small{e-mail: \texttt{offringa@astro.rug.nl} } \\
\small{$^2$ASTRON, PO Box 2, 7990 AA Dwingeloo, The Netherlands} \\
}
\pagerange{\pageref{firstpage}--\pageref{lastpage}}
\date{Accepted 25 January 2012. Received in original form 25 November 2011. The definitive version is available at www.blackwell-synergy.com.}
\pubyear{2012}

\begin{document}
\label{firstpage}
\maketitle
\begin{abstract}
Techniques to improve the data quality of interferometric radio observations are considered. Fundaments of fringe frequencies in the $uv$-plane are discussed and filters are used to attenuate radio-frequency interference (RFI) and off-axis sources. Several new applications of filters are introduced and tested. A low-pass filter in time and frequency direction on single baseline data is successfully used to lower the noise in the area of interest and to remove sidelobes coming from unmodelled off-axis sources and RFI. Related side effects of data integration, averaging and gridding are analysed, and shown to be able to cause ghosts and an increase in noise, especially when using long baselines or interferometric elements that have a large field of view. A novel projected fringe low-pass filter is shown to be potentially useful for first order source separation. Initial tests show that the filters can be several factors faster compared to common source separation techniques such as peeling and a variant of peeling that is currently being tested on LOFAR observations called ``demixed peeling''. Further testing is required to support the performance of the filters.
\end{abstract}


\section{Introduction}
For several decades, it has been a challenge to increase the dynamic range of images produced by interferometric radio telescopes. The raw sensitivity improvements and advanced understanding of calibration errors have pushed the limits on the dynamic range of modern telescopes to unprecedented levels \citep{revisiting-me-i}. The final dynamic range is constrained by the celestial field being observed, the efficiency of the telescope's hardware and the time spent observing. However, this theoretical dynamic range is limited further by imprecise models of instrumental effects and celestial sources used in the data reduction process, as well as by the quality of the radio environment.

The noise level in the final result of an observation can be set by several phenomena. In the ideal case, the noise level equals the thermal sky noise level, and the detection of sources or other features is limited by this noise level only. An image can also be limited by confusion noise when it does not provide enough resolution to distinguish sources. Sidelobes provide a third type of noise, which are generated by the point spread function (PSF) of the instrument, that convolve strong sources that are in or outside the field of interest. Finally, radio-frequency interference (RFI) can add additional noise to the final result of an observation. In this paper, we will aim at suppressing noise coming from RFI and sidelobe noise coming from off-axis sources, using similar techniques based on fringe theory.

Because we address two problems at once, we will introduce both problems individually. In the following subsection, we will introduce the problem of RFI and describe current techniques to deal with it. Thereafter, we will introduce the concerns of off-axis sources and approaches to deal with those as well.

\subsection{Radio-frequency interference}
While technical advances gave rise to better telescopes, different technical advances have ironically decreased the quality of the radio environment for radio astronomy. A potential problem that limits the effective dynamic range of modern telescopes such as LOFAR, the WSRT, the GMRT, the ATCA and the VLA, is radio-frequency interference (RFI). Fortunately, practically all RFI interferes within a limited amount of time or frequency channels, and can be flagged automatically in post-correlation. In \citet{post-correlation-rfi-classification}, the SumThreshold algorithm is described and is proven to be very accurate for that purpose. Further implementation of the method into the LOFAR pipeline has shown excellent results \citep{LOFAR-RFI-pipeline}.

Although reasonably strong temporal and spectral RFI can successfully be removed by flagging, it is not always a satisfactory solution. Sporadic continuous broad-band RFI for example poses a potential problem, since this type of RFI can not be removed by flagging. Doing so might affect considerable parts of the observation, potentially throwing away too much of the data. \citet{fringe-fitting-rfi-mitigation} has shown that the GMRT suffers from this type of RFI at low frequencies, for example caused by high voltage lines. \citet{fringe-fitting-rfi-mitigation} describes a method to remove this kind of RFI based on fringe fitting of RFI. This approach has been recently implemented in {\sc aips}\footnote{{\sc aips} is the Astronomical Image Processing System\\(\url{http://aips.nrao.edu/})} \citep{rfi-mitigation-in-aips}. This method will be analysed in \S\ref{sec:fringe-filtering}. Most other telescopes do not report such severe broad-band RFI: LOFAR, although build in a populated area, shows very little of this kind of RFI in the currently finished stations \citep{LOFAR-interference-results} and (E)VLA interference reports also mention spectral RFI affecting a few channels, but no broad-band RFI, although low frequency causes more problems \citep[\S4.6]{evla-observational-status}. Nevertheless, when approaching the thermal noise on low frequencies, such as LOFAR will do in the future, faint RFI might show up. The fringe fitting method is not so well applicable in these cases, because such RFI will be below the noise. By removing a spatial frequency component from (white) noise dominated data, a component from the noise will be removed instead of removing actual RFI. Work has been done to apply post-correlation RFI removal techniques for the (E)VLA, by ways of calibrating and removing the RFI source \citep{postcorrelation-lf-rfi-excision}, but this method is tedious and requires the RFI to be reasonably stable.

Another solution for removing continuous RFI is spatial filtering by eigenvalue decomposition \citep{multichannel-rfi-mitigation,hampson-spatial-nulling-2002, ellingson-spatial-nulling-2002}, which disentangles the contribution of sources from different directions, and subsequently removes the contributions from the direction of interference. Recently, this was implemented for the Parkes multibeam receiver \citep{spatial-filtering-parkes-multibeam}. However, the requirement of specialized hardware and/or having to configure the filter before correlation is a major disadvantage of spatial filtering techniques, in the context of interferometers. The latter requires the configuration to be fixed before the observation in most cases. This makes it hard to react to unanticipated RFI, and impossible to change the filter after observing if the filter has not worked correctly. RFI is often not stable enough to be removed during post-correlation processing.

Another technique for removing sporadic continuous RFI has been introduced in \citet{the-gmrt-eor-experiment}, which decomposes the time frequency data with a singular value decomposition (SVD). This method however was shown in \citet{post-correlation-rfi-classification} to potentially alter the astronomical data, making the method less attractive to use for data reduction without further research. In \citet{post-correlation-reference-signal}, the RFI is subtracted from the data after correlation by the use of a reference signal. Unfortunately, such a reference signal is not always available or practical to implement.

\subsection{Off-axis sources}
Signals from off-axis sources received in the sidelobes, like RFI, decrease the dynamic range of observations, or might even cause calibration to fail. New wide-field telescopes such as LOFAR see a large area of the full sky, and always have a few strong sources in their sidelobes. Examples of such sources are Cassiopeia~A, Cygnus~A and the Sun. These sources are often not of interest, but have to be removed accurately.

A common method to deal with off-axis sources is peeling \citep{peeling-calibration-for-lofar,lf-calibration-with-source-peeling}. Peeling is iterative, and changes the phase centre towards the source, optionally averages in time and frequency to suppress other sources, and self-calibrates and subtracts the source. This method has shown good results, but is very computational intensive --- too intensive to use by default on high resolution telescopes such as LOFAR. Demixed peeling is a variation on normal peeling, that is currently being tested for LOFAR observations. However, early results show similar computational requirements when the same removal quality is required \citep{demixing-lofar}.

Finally, in \citet{paper-ddr-filtering} a delay-delay rate (DDR) filter is proposed that disentangles the flux contribution into the different sky facets they originate from. The DDR-filter was used by Parsons \& Backer for first order calibration, but the idea of such a filter is also attractive for application in a later stage and over longer timescales, because the filter can be applied on post-correlated data without additional hardware. It is however unclear how accurate the filter will be for off-axis source removal. We will propose related filters, while trying to increase its application and accuracy.

\subsection{Outline}
In this paper we will describe and analyse new methods for filtering both RFI and off-axis sources, with the ultimate goal of reaching lower noise levels. We will start by analysing Athreya's fringe fitting method in \S\ref{sec:fringe-filtering} and describe why it is insufficient for e.g. LOFAR observations. In \S\ref{sec:novel-techniques}, several new methods will be introduced and analysed with the help of simulations. We will test our filtering approaches in \S\ref{sec:practical-applications} on a WSRT dataset at a frequency of about 140~MHz of the field centred on the radio galaxy B1834+62 \citep{double-double-radio-galaxies-schoenmakers}. At this low frequency, the WSRT is sensitive to very bright sources like Cygnus~A and Cassiopeia~A \citep{deep-wsrt-150mhz-observations}, which despite their large angular distance are not sufficiently attenuated by the primary beam. They therefore generate intense spurious sidelobes across the target field of view. We will discuss the results in \S\ref{sec:discussion}, where we will also discuss how time or frequency averaging and gridding may effect off-axis sources or RFI. Finally, we will draw conclusions based on our findings in \S\ref{sec:conclusions}. 

\section{Analysis of the Fringe filtering method}  \label{sec:fringe-filtering}
\subsection{Removing constant RFI}
\begin{figure*}
 \begin{center}
  \subfloat[Fit with constant fringe rate, amplitude found = $12.8$]{ \label{fig:rfi-constant-fringe-fit}
   \includegraphics[width=80mm]{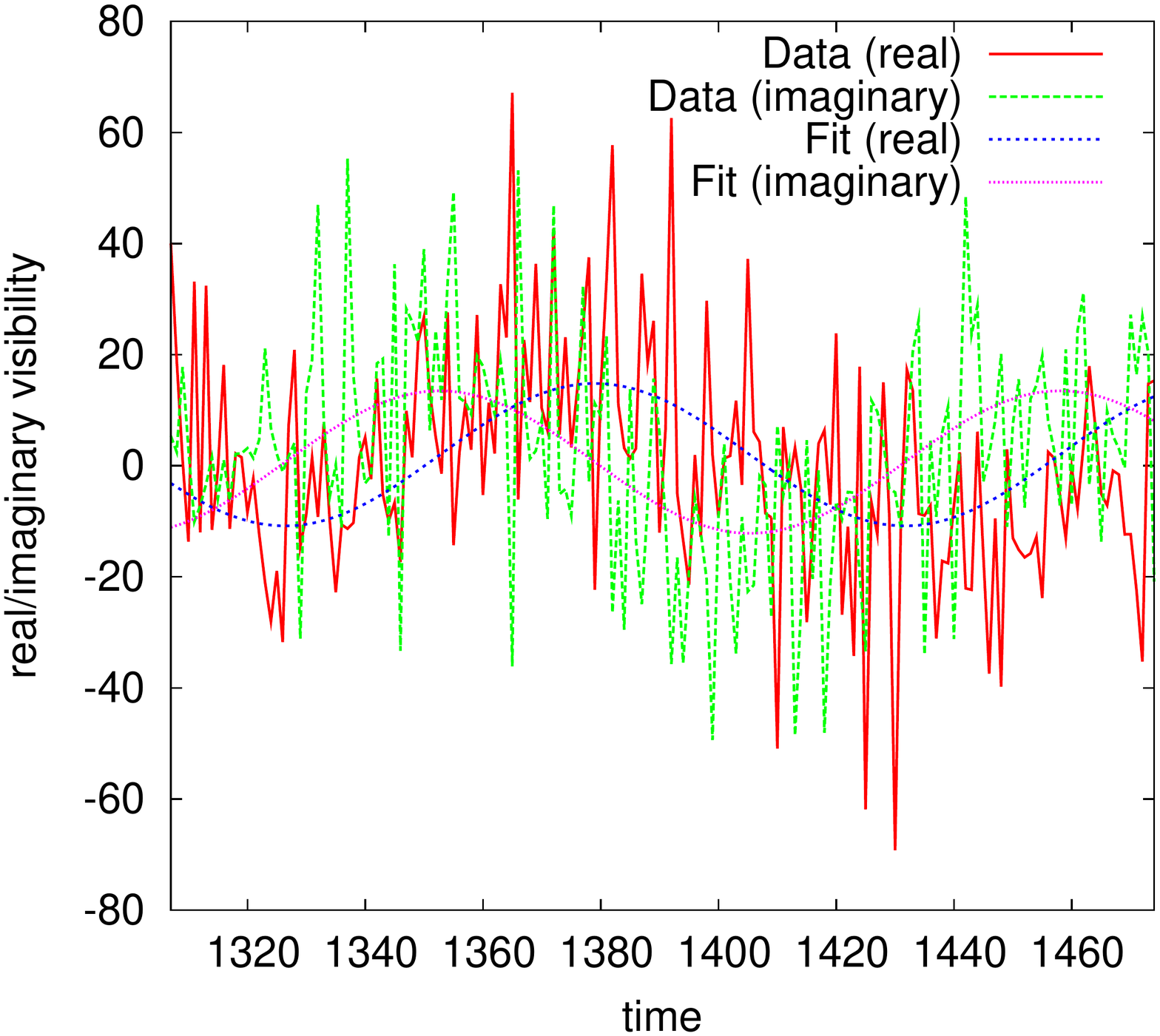}
  }
  \subfloat[Using fringe count, amplitude found = $14.7$]{ \label{fig:rfi-variable-fringe-fit}
   \includegraphics[width=80mm]{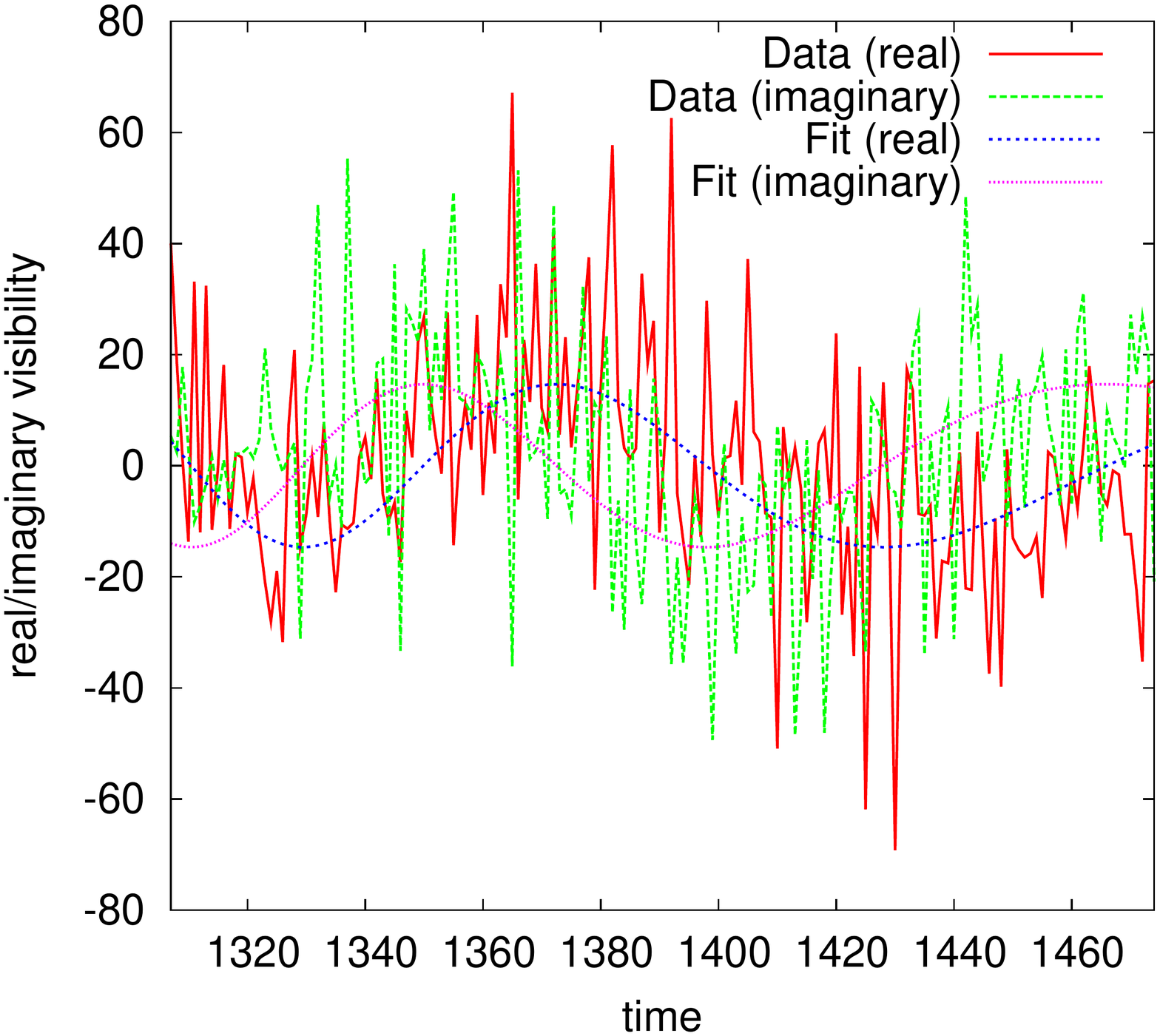}
   }
 \end{center}
 \caption[]{Comparison of fitting methods using simulated data: the original amplitude of the source is $16$. Only the shown data is used for the fit. Using a constant fringe speed (left panel) over this range produces a somewhat less accurate fit compared to using the fringe count for each sample in the fit (right panel). The $x$-axis is in time steps of 15 seconds from the start of the (simulated) observation. At time step 1570, the simulated baseline is orthogonal to the direction linking the target source and the phase centre and $\nu_F$=0. Hence, the fringe speed changes significantly over the displayed time range, which can be seen by the somewhat elongated fringes near the right. }
 \label{fig:rfi-fringe-fit}
\end{figure*}
\citet{fringe-fitting-rfi-mitigation} describes how geometrically stationary RFI can be removed from an observation by fitting out a sinusoid with a frequency opposite to the natural fringe rate. A stationary earth-bound RFI source receives a fringe rate opposite to the applied fringe stopping rate. Therefore, one can estimate its contribution. The natural fringe rate is given by:
\begin{align}
 \notag \nu_F(t) & = \frac{dw(t)}{dt} \\
 & = -\omega_{E} u(t) \cos \delta,
\end{align}
with $t$ the sidereal time, $\omega_E=1$ rotation/day, the rotation speed of the earth, $u(t)$ the component representing the standard $u$ position of the baseline in the $uv$-plane, $w(t)$ the standard $w$-component representing the applied phase delay and $\delta$ the declination of the phase centre. When a baseline is orthogonal to the direction of the phase centre, $\nu_F(t)$ is zero. A stationary source of RFI contributes to a correlation in the form of the complex function
\begin{equation} \label{eq:static-fringe-delay-solution}
 \textrm{RFI}(t)=\mathcal{A}e^{-i \nu_F t},
\end{equation}
with $\mathcal{A}$ the complex amplitude of the RFI at time $t$. The $2 \pi$ term is absorbed in $\nu_F$, such that its value is in radians/time unit. This amplitude is initially assumed to be constant over some period $\left[ t_0 , t_E \right]$, and $\nu_F$ is assumed not to change over this time interval. It is then possible to estimate $\mathcal{A}$ by performing a least square fit between the complex function $V(t)$, representing the observed visibilities, and the RFI signal by minimizing the error function
\begin{equation}
 \epsilon(\mathcal{A}) = \int\limits_{t_0}^{t_E} \left(\mathcal{A} e^{-i \nu_F t} - V(t)\right)^2 dt.
\end{equation}
Minimization of $\epsilon(\mathcal{A})$ results in
\begin{equation} \label{eq:rfi-gain-solution}
 \mathcal{A} = \int\limits_{t_0}^{t_E} V(t) e^{i \nu_F t} dt,
\end{equation}
which corresponds to $\mathcal{A}=\mathcal{F}(\nu_F)$, the frequency component $\nu_F$ of the Fourier transform $\mathcal{F}$ of $V$ over the time interval. Therefore, removing a Fourier component of a signal can be implemented as a standard frequency filter. Equation~\eqref{eq:static-fringe-delay-solution} corresponds to a single component of the delay-rate (DR) transform, creating a symmetry with the DDR filter proposed in \citet{paper-ddr-filtering}. An example of the application of Equation~\eqref{eq:rfi-gain-solution} on simulated data is given in Fig.~\ref{fig:rfi-constant-fringe-fit}. The two plots show the result of fitting a sinusoidal function to simulated data. We simulated a WSRT interferometer, correlating antennae RT0 and RT5: a 720m baseline. A single channel is simulated with a frequency of 147~MHz. The simulated observation has eight sources, seven of which are faint and in the primary beam, while the last source simulates an interfering source that is four times stronger. This off-axis source generates a visibility amplitude of 16 and is a 40$^{\circ}$ from the phase centre, hence far from the other sources. 

\begin{figure*}
 \begin{center}
  \subfloat[Original]{ \label{fig:rfi-fringe-fit-original-image}
   \includegraphics[height=45mm]{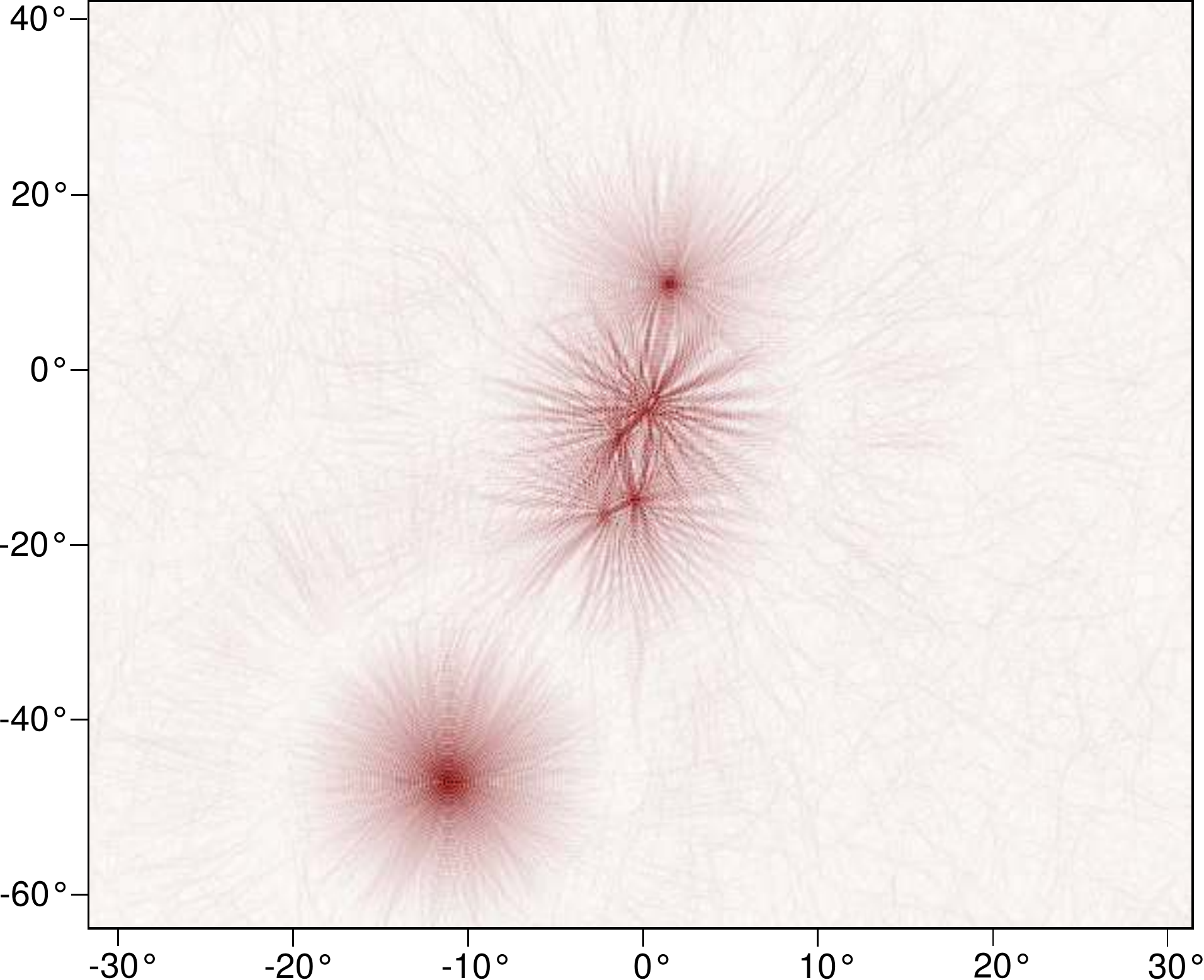}
  }
  \subfloat[Fringe fitting applied]{ \label{fig:rfi-fringe-fit-applied-image}
   \includegraphics[height=45mm]{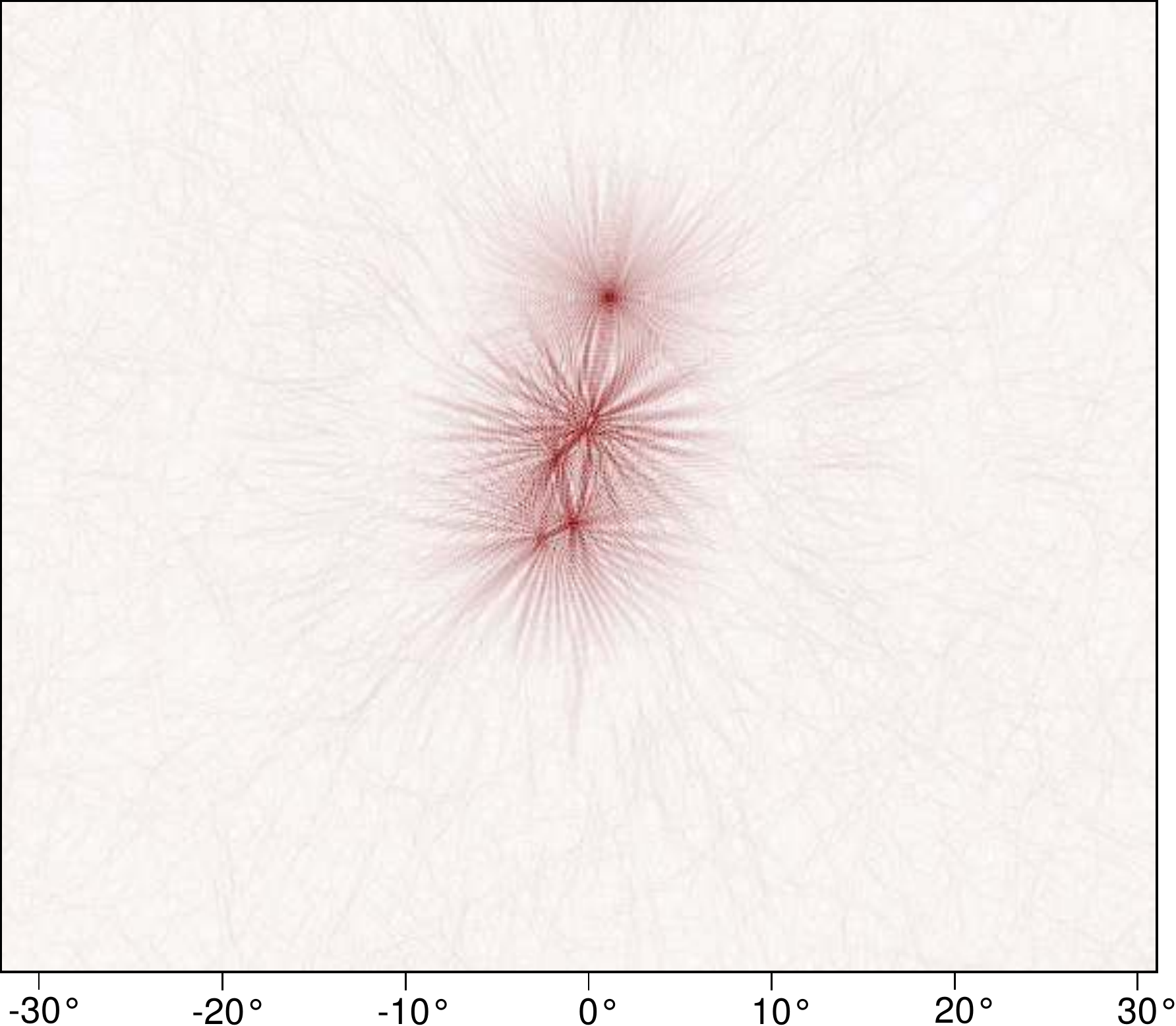}
   }
  \subfloat[Difference]{ \label{fig:rfi-fringe-fit-difference-image}
   \includegraphics[height=45mm]{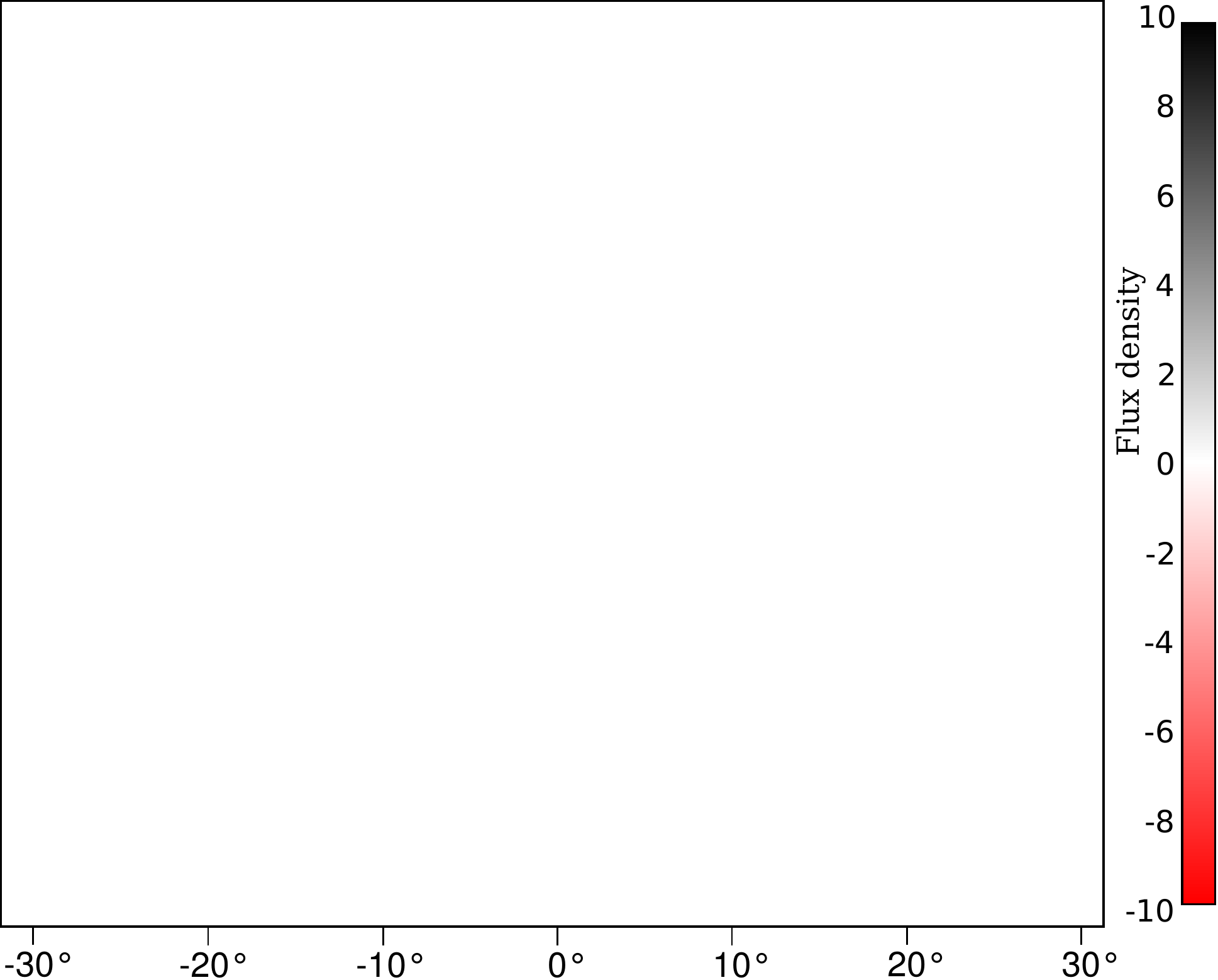}
   }
 \end{center}
 \caption[]{These images show the application of a fringe filter that takes out a hypothetical source with a constant amplitude (Equation~\eqref{eq:static-fringe-exact-solution}). The same 720m WSRT baseline and set-up as in Fig.~\ref{fig:rfi-fringe-fit} was simulated and imaged without deconvolving. The image in the left panel is the result of imaging without any filtering. The middle panel shows the result after application of the filter, while the right image shows the difference. The filter removes the source up to the sidelobe confusion noise of the other sources, which is over three orders of magnitude. The residual shows that it does not affect the sources of interest, again up to at least three orders of magnitude. This simulated situation is only hypothetical, since it is unlikely that the received power of distant sources remains constant.}
 \label{fig:rfi-fringe-fit-images}
\end{figure*}

Since $\nu_F$ changes slowly with time, Equation~\eqref{eq:rfi-gain-solution} will become inaccurate when increasing the time interval. Additionally, it can not be calculated near $\nu_F=0$. By observing that the number of wavelengths of delay caused by the geometrical delay corresponds to the number of rotations applied on the visibilities, we can replace $\nu_F t$ by $w(t)-w(t_0)$, where $w$ is the applied phase delay in radians/time unit as function of time. As $w(t_0)$ causes a constant phase shift, it can be absorbed in $\mathcal{A}$. By substituting $\nu_F t$ with $w(t)$ in Equation~\eqref{eq:rfi-gain-solution}, we get a more accurate solution for $\mathcal{A}$:
\begin{equation} \label{eq:static-fringe-exact-solution}
 \mathcal{A} = \int\limits_{t_0}^{t_E} V(t) e^{i w(t)} dt.
\end{equation}
An example of such a fit is given in Fig.~\ref{fig:rfi-variable-fringe-fit}. As long as the amplitude of the RFI source remains constant, this allows successful removal of the source when $\nu_F \gg 0$. As is visualized by Fig.~\ref{fig:rfi-fringe-fit-images}, it removes the strong source in the example without unwanted side effects on the area of interest.

\subsection{Removing variable RFI}
With the algorithm presented by Athreya, the RFI source is not only allowed to differ between baselines, but also in time. Since the beam rarely follows the RFI source, it is likely that the gain towards the RFI source will change. Athreya proposes tiling of the data, making separate fits on each tile, where each tile is approximately the size of a fringe. However, tiling the data and performing fits on each tile causes instability near the borders of the tiles.

A more accurate way is to perform individual fits for each sample, sliding a kernel of weights over the data that are used to perform the fit. Two trivial suggestions for a weighting function are the rectangular function and the Gaussian function. A rectangular function would result in a sliding window method, which has implementational advantages. However, a rectangular function produces a sinc response in delay space. Therefore, the fit will be affected by any other frequency in the data set that corresponds to non-zero values in the sinc function, which undesirably would remove part of the signal of interest. A Gaussian kernel would localize the frequency response somewhat better. A larger kernel or tile size would decrease the frequency response to other frequencies, but in order to remove the RFI it would require the received gain of the RFI to change less quickly.

Allowing the amplitude to change in time creates spirals in the complex plane. This kind of fitting has recently been implemented in the AIPS astronomical package as described by \citet{rfi-mitigation-in-aips}.\\

\subsection{Generalization of the fringe fitting method}
Up to now, the use of the method has been limited to the removal of a single (RFI) source that behaves like a point source at the celestial pole. It is common practice to peel and/or calibrate for sources that are outside the area of interest, because they need to be taken out carefully in order to avoid additional sidelobe confusion noise. In such a case, the off-axis source is similar to static RFI: the source itself is not of interest, but has to be taken out for calibration and imaging the field accurately. For this purpose, the fringe fitting method can be generalized to remove any point source. This requires a small change to Equation~\eqref{eq:static-fringe-exact-solution}, which now becomes:
\begin{equation} \label{eq:fringe-fitting-generalized}
 \mathcal{A} = \int\limits_{t_0}^{t_E} V(t) e^{i \left( w(t) - w_S(t) \right) } dt.
\end{equation}
Here, $w(t)$ is the standard $w$-component in the $uvw$ domain as before, while $w_S(t)$ is the $w$-component for an observation phase centred on $S$, the source to be removed. While the process is easier and faster than normal off-axis source calibration or peeling, in practice it will be of little use: it neglects information present in polarizations, as defined by the measurement equation \citep{understanding-radio-polarimetry-i}, and neglects the relations between baselines. Advanced calibration algorithms such as the SAGE calibration technique \citep{sage-calibration-i, sage-calibration-ii} solve for source parameters by combining this information at once, and will in general be more accurate, as long as the source is (coherently) seen in multiple polarizations or antennas.

\section{Novel filtering techniques} \label{sec:novel-techniques}
For high dynamic range, the source removal techniques as analysed in the previous section might not always suffice: the fringe fitting procedure can only remove a single unresolved source at a time. Also, since the fitting window has to be reasonably small, the fit will be slightly affected by the contribution of other sources. Therefore, the source has to be strong to be able to remove it, although the absolute error made will not depend on the strength of the source.

In the following sections, we will present several filters that are aimed to work when the fringe filter does not suffice. The key issues that these filter techniques share, are that they do not perform fitting on windows, but use the full data at once. They also remove high-frequency Fourier components that do not correspond with the fringe frequencies of sources of interest.

\subsection{A low-pass filter in time domain} \label{sec:time-domain-low-pass-filter}
\begin{figure}
 \begin{center}
  \includegraphics[width=70mm]{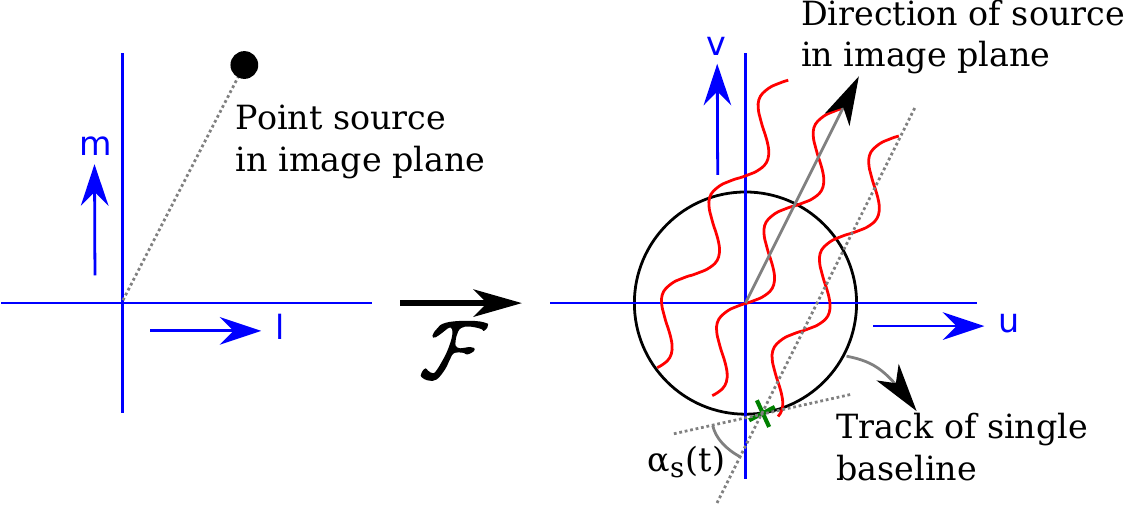}
 \end{center}
 \caption[]{Cartoon showing how a source in the image plane contributes fringes in the $uv$-plane. The further the source is from the phase centre (origin), the faster the fringe. Function $\alpha_S(t)$ is the angle between the direction of the source and the direction of a specific point in the uv-track as a function of time. The smaller $\alpha_S$, the faster the fringe speed in the track at that point.}
 \label{fig:explanation-fringe-speed}
\end{figure}

\begin{figure}
 \begin{center}
   \includegraphics[width=70mm]{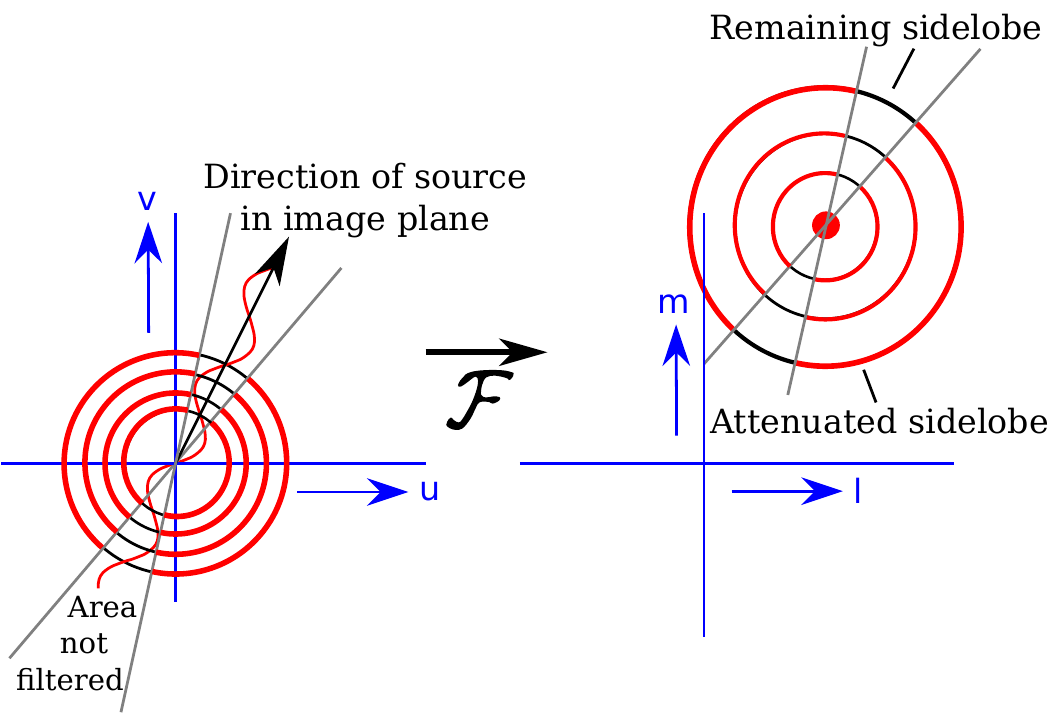}
 \end{center}
 \caption[]{Applying the low-pass filter on several baselines will filter parts of sources that exceed the frequency limit. For a particular source, this corresponds with multiplying the source with a hourglass shape in the $uv$-plane (left panel). Because of this multiplication, the sidelobes of the source in image plane (right panel) will be, relative to the phase centre, filtered in tangential direction. Sidelobes in radial direction will remain.}
 \label{fig:filter-convolution-explanation}
\end{figure}

The visibility of a single point source with strength $I_{lm}$ and coordinates $(l,m)$ is given by
\begin{equation}\label{eq:def-visibility}
 V\left(u,v,w\right) = I_{lm} e^{i2\pi(ul + vm + wn)}.
\end{equation}
Define $\boldsymbol{d}=(u,v,w)$ and $\boldsymbol{l}=(l,m,n)$. Since the source $I_{lm}$ is real, the phase $\phi$ of $V$ is given by
\begin{equation}
 \phi(\boldsymbol{d}) = 2\pi \boldsymbol{d}\cdot\boldsymbol{l}.
\end{equation}
The property that will be used in the filtering technique, is the implication of this formula that sources with large $|\boldsymbol{l}|$, i.e., that are far away from the phase centre, have a high fringe speed in the $uv$-plane.

\begin{figure*}
 \begin{center}
  \subfloat[Application]{ \label{fig:time-domain-low-pass-filter-applied}
   \includegraphics[height=45mm]{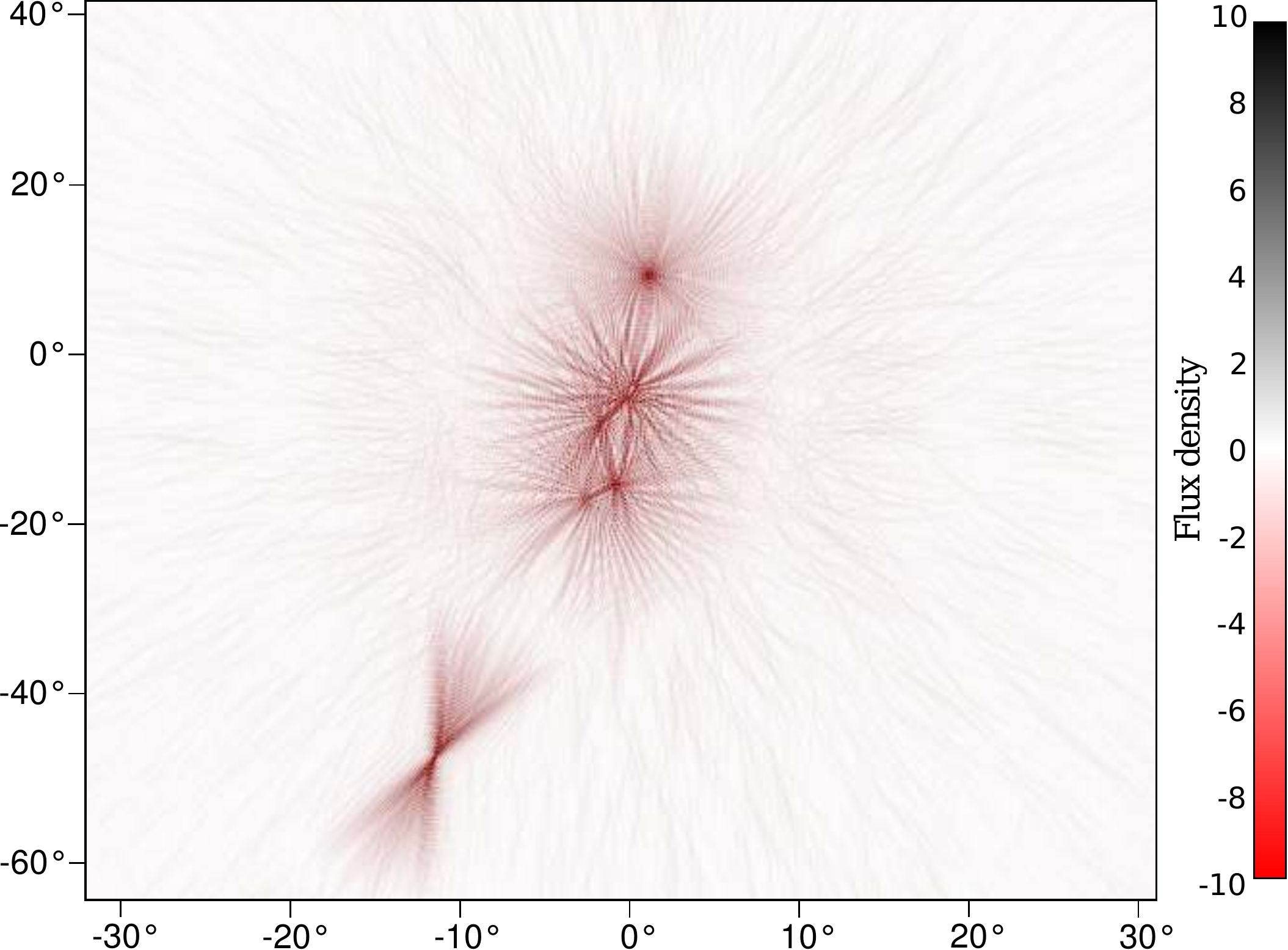}
  }
  \subfloat[Difference]{ \label{fig:time-domain-low-pass-filter-residual}
   \includegraphics[height=45mm]{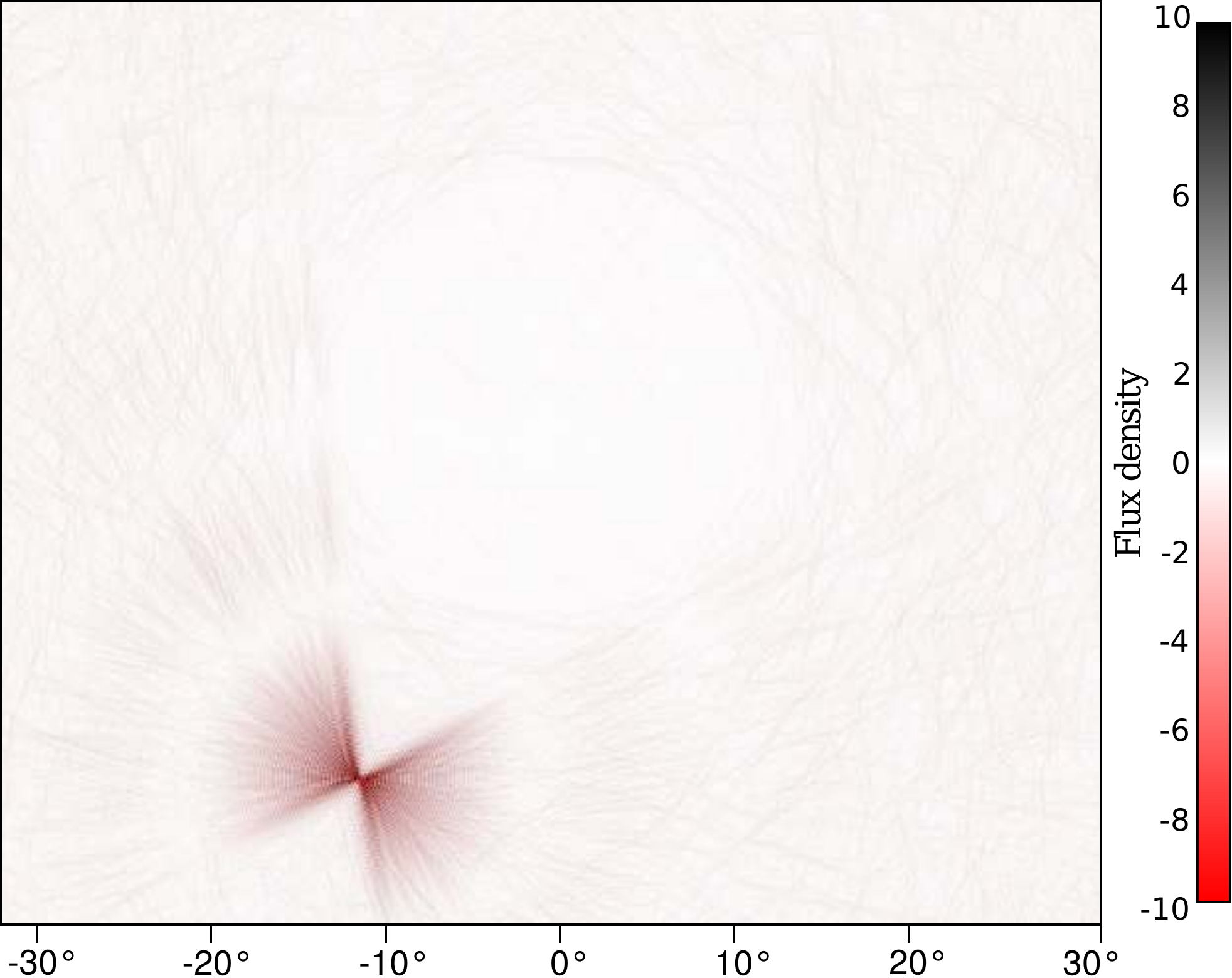}
   }
  \subfloat[Difference with model]{ \label{fig:time-domain-low-pass-filter-error}
   \includegraphics[height=45mm]{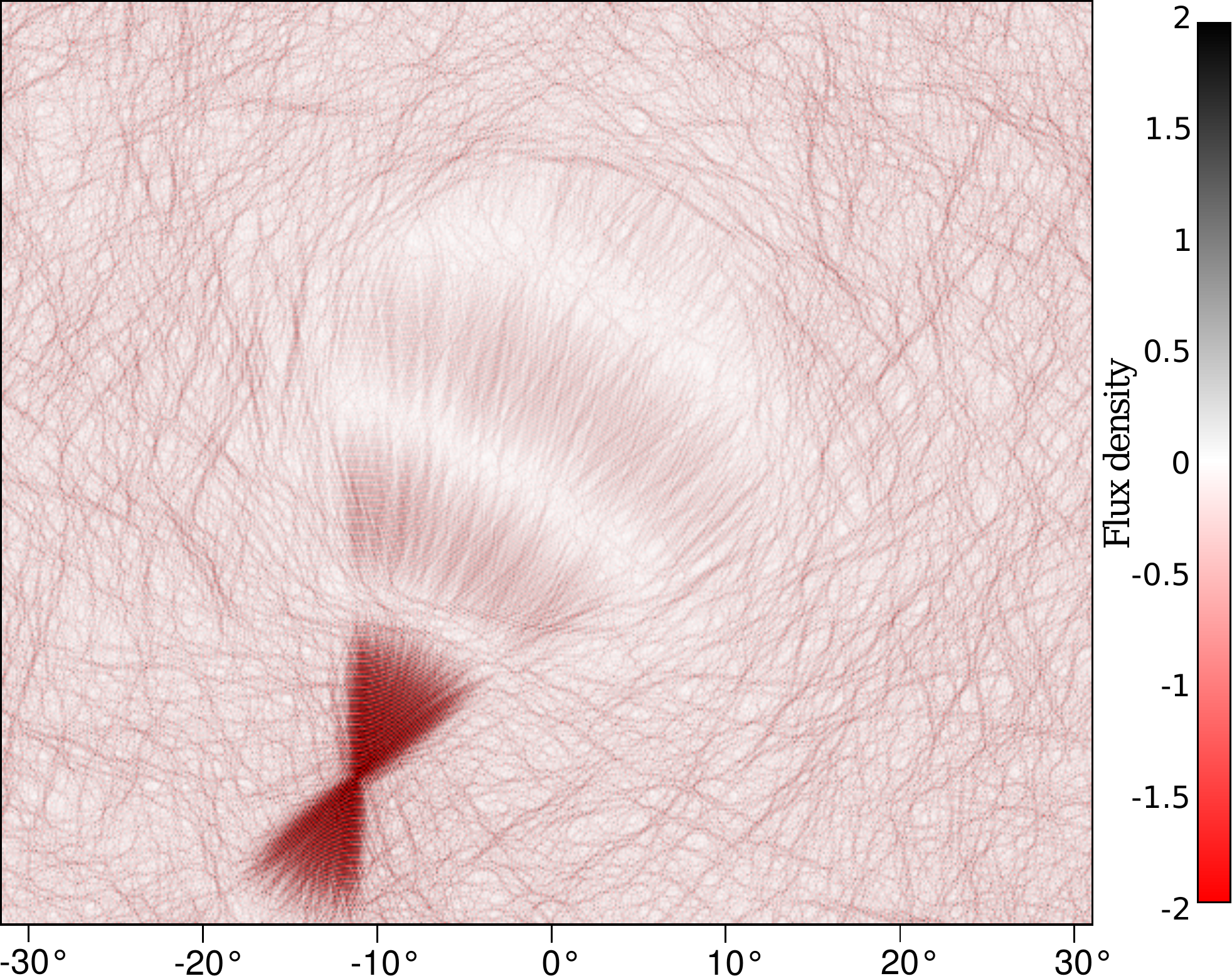}
   }
 \end{center}
 \caption[]{Application of a low-pass filter in the time domain (\S\ref{sec:time-domain-low-pass-filter}). The source has been attenuated by filtering (left panel), but some of the sidelobes have not been removed. This is because the fringe rate of the source does not always exceed the filtering frequency. The middle panel shows what has been removed and confirms that the sources of interest have not been attenuated (up to the 100 times lower noise level), the right panel shows with high contrast what has not been removed from the source. Note the different intensity scales.}
 \label{fig:time-domain-low-pass-filter}
\end{figure*}

Without loss of generality, we assume that our interferometer has a configuration such that its corresponding $uv$-track is a circle that is centred on the $uvw$-origin. This only occurs for an East-West Interferometer such as the WSRT. However, the technique can be straight forwardly extended to other interferometers that create possible elliptic tracks that might not be centred on the origin. In the assumed case, the $uv$-plane position $\boldsymbol{d}$ will be a function of time but have a constant size. If a time-sorted sequence of observed samples of a single correlation is considered, its fringe frequency is given by
\begin{equation} \label{eq:source-fringe-frequency-time}
 \nu_{S}(t) = \frac{d\phi}{dt} = \left|\boldsymbol{d}\right| \left|\boldsymbol{l}_S\right| \cos \alpha_S(t),
\end{equation}
where $\nu_S(t)$ is the fringe speed in fringes per second at time $t$ for source $S$, $|\boldsymbol{d}|$ is the radius of the $uv$-track, $|\boldsymbol{l}_S|$ is the distance of $S$ to the phase centre and $\alpha_S(t)$ is the angle between the $uv$-track and the line through $S$ and the phase centre as drawn in Fig.~\ref{fig:explanation-fringe-speed}. The fringe speed will be maximal at points where the corresponding $uv$-track is parallel to the direction of the source, and zero when the source direction and $uv$-track are orthogonal. The maximal fringe speed produced by a source is proportional to the distance between the source and the phase centre: $\nu_S(t) \propto \left|\boldsymbol{l}_S\right|$.

We will now consider low-pass filtering of the time-sorted visibility data with a filter frequency $\nu_F$, specified in fringes per wavelength. Such a filter will have the following two properties: First, sources with $\forall t: \nu_S(t) / |\boldsymbol{d}| < \nu_F$, will never be filtered. In image plane, the area corresponding to $\nu_S(t) / |\boldsymbol{d}| < \nu_F$ is a circle that is centred on the phase centre. The fringe speed in the $uv$-track is translation independent, hence it is not necessary for the track to be centred on the origin. In case the $uv$-track is an ellipse, the filtering area will be an ellipse as well, but we will continue to assume circularity. Second, sources outside the circle will be filtered during the periods in which $\nu_S(t) / |\boldsymbol{d}| \ge \nu_F$. The differential start and end angle, respectively $\alpha_S^s$ and $\alpha_S^e$, at which a source will enter the filtered area are given by
\begin{align}
\notag 
\alpha_S^s &= \arccos \frac{\nu_F}{\left|\boldsymbol{l}_S\right|}, \\
\alpha_S^e &= \pi - \arccos \frac{\nu_F}{\left|\boldsymbol{l}_S\right|}.
\end{align}
The area filtered is independent of the baseline length because $\nu_F$ is specified in fringes per wavelength. For a single baseline, the filter ratio can be calculated with $\left(\alpha_S^e - \alpha_S^s\right)/\pi$. Consequently, in an array with $N$ baselines with different sizes, the fraction of samples in which the source is filtered is given by
\begin{align}
\notag
\rho_s &= \frac{1}{N} \sum\limits_{i=0}^{N-1} \frac{\alpha_S^e - \alpha_S^s}{\pi} \\
\label{eq:time-lowpassfilter-attenuation}
       &= 1 - \frac{2}{\pi} \arccos \frac{\nu_F}{\left|\boldsymbol{l}_S\right|},
\end{align}
which is therefore the total attenuation of the source by the filter.

Although we have shown with Equation~\eqref{eq:time-lowpassfilter-attenuation} that the total attenuation of a source is known, the shape of the area that is filtered is important as well, as that defines the shape of the sidelobes. The effect of low-pass filtering is sketched in Fig.~\ref{fig:filter-convolution-explanation}: the filter removes the source fringes at two symmetric radial areas in the $uv$-plane. Subsequently, the application of this filter can be seen as an additional multiplication of the source in the $uv$-plane. Instead of a convolution with the nominal point spread function (PSF), sources in the image plane are convolved with a partly attenuated PSF. The side lobes that the source would normally have are not filtered in the direction of the phase centre, and can still increase the noise in the area of interest. This effect can be seen in Fig.~\ref{fig:time-domain-low-pass-filter}.

Although this filter does not directly suppress confusion noise, it does filter high frequencies that can increase aliasing effects during averaging or gridding (\S\ref{sec:averaging}). A more sophisticated filter will be presented in the next section, which utilizes the same theory about the fringe speed of sources.

\subsection{A projected fringe low-pass filter in time domain} \label{sec:projected-fringe-filter}
As was shown in Section~\ref{sec:time-domain-low-pass-filter}, in order to remove the side lobes of an interfering source from the area of interest successfully, the interferer has to be filtered over the entire length of the observation. We will now introduce a filter with the purpose of filtering out all sources in a certain direction beyond a minimum distance from the phase centre.
\begin{figure*}
 \begin{center}
  \subfloat[Fringe contribution of a source]{ \label{fig:uv-project-a}
   \includegraphics[width=40mm]{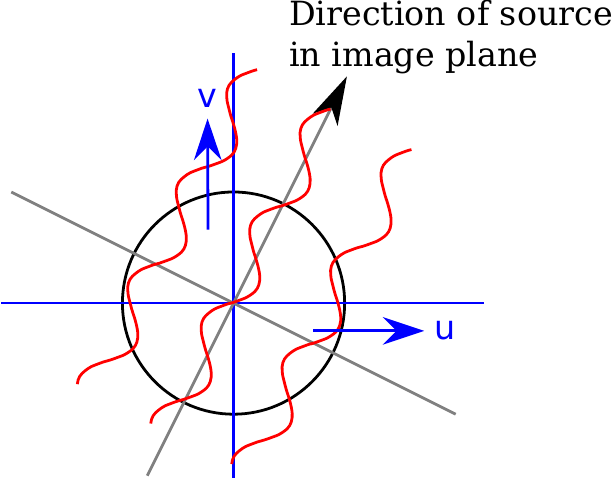}
  }
  \subfloat[Rotation]{ \label{fig:uv-project-b}
   \includegraphics[width=40mm]{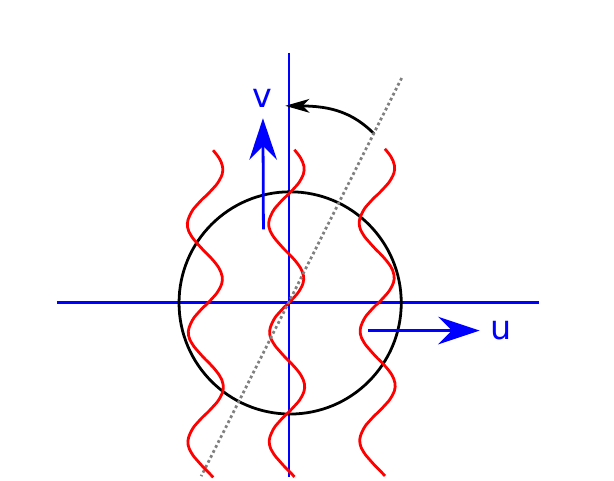}
  }
  \subfloat[Projection]{ \label{fig:uv-project-c}
   \includegraphics[width=40mm]{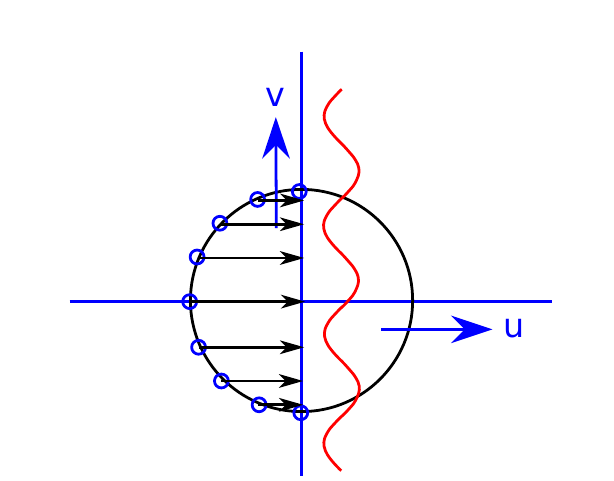}
  }
 \end{center}
 \caption[]{Creating a constant fringe rate towards a single direction. Panel~\subref{fig:uv-project-a}: A source with a certain direction from the origin in the image plane will cause a fringe in the $uv$-plane corresponding to that direction. Panel~\subref{fig:uv-project-b}: Rotating the direction of the source onto the $v$-axis will align its fringe with that axis. Panel~\subref{fig:uv-project-c}: Projection of the sample track onto the $v$-axis will make any source in the direction of rotation have a constant fringe rate.}
 \label{fig:uv-project}
\end{figure*}

\begin{figure}
 \begin{center}
   \includegraphics[width=80mm]{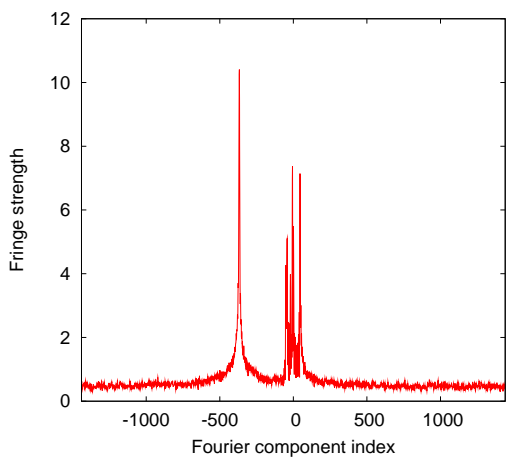}
 \end{center}
 \caption[]{Fourier transform of a $uv$-track that was rotated and projected, such that sources in a certain direction have a constant fringe speed. The model of Fig.~\ref{fig:time-domain-low-pass-filter} was used. Most of the contribution of sources near the centre collect near Fourier component index zero, while the contribution of the off-axis source shows up as a peak at an index away from zero.}
 \label{fig:projected-ft-plot}
\end{figure}

\begin{figure*}
 \begin{center}
  \subfloat[Applied]{ \label{fig:project-low-pass-filter-applied}
   \includegraphics[height=45mm]{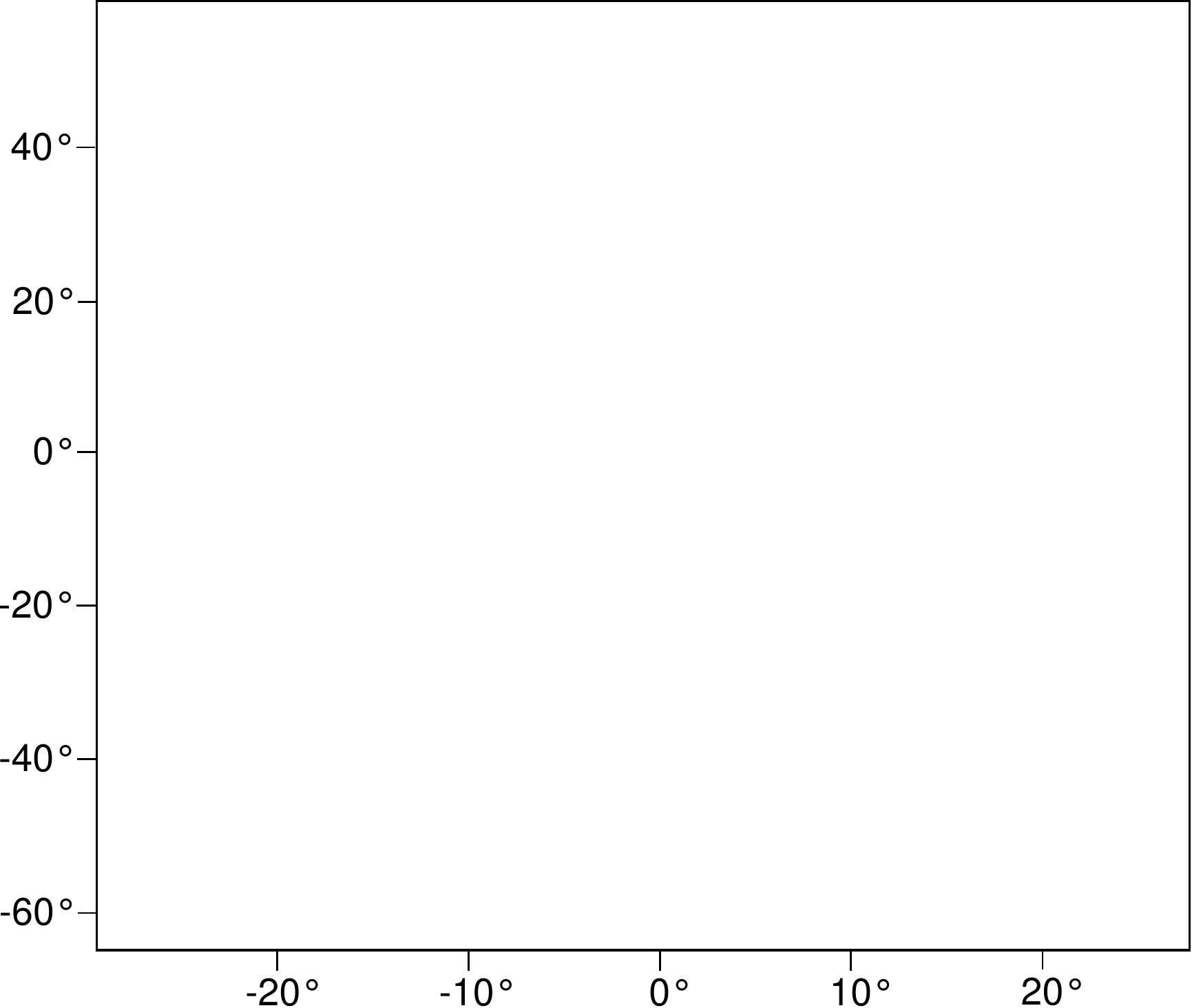}
  }
  \subfloat[Residual]{ \label{fig:project-low-pass-filter-residual}
   \includegraphics[height=45mm]{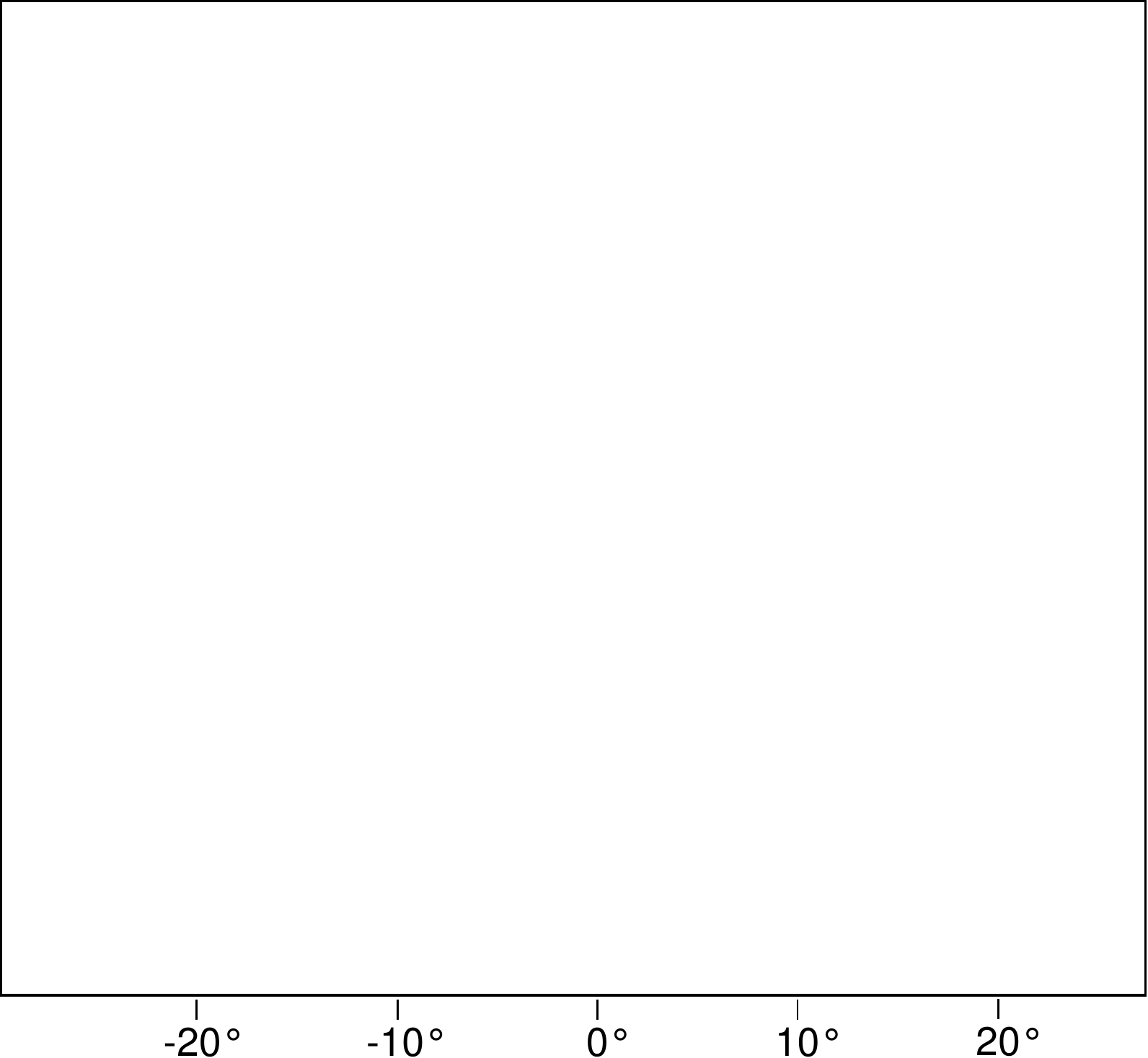}
  }
  \subfloat[Difference with model]{ \label{fig:project-low-pass-filter-error}
   \includegraphics[height=45mm]{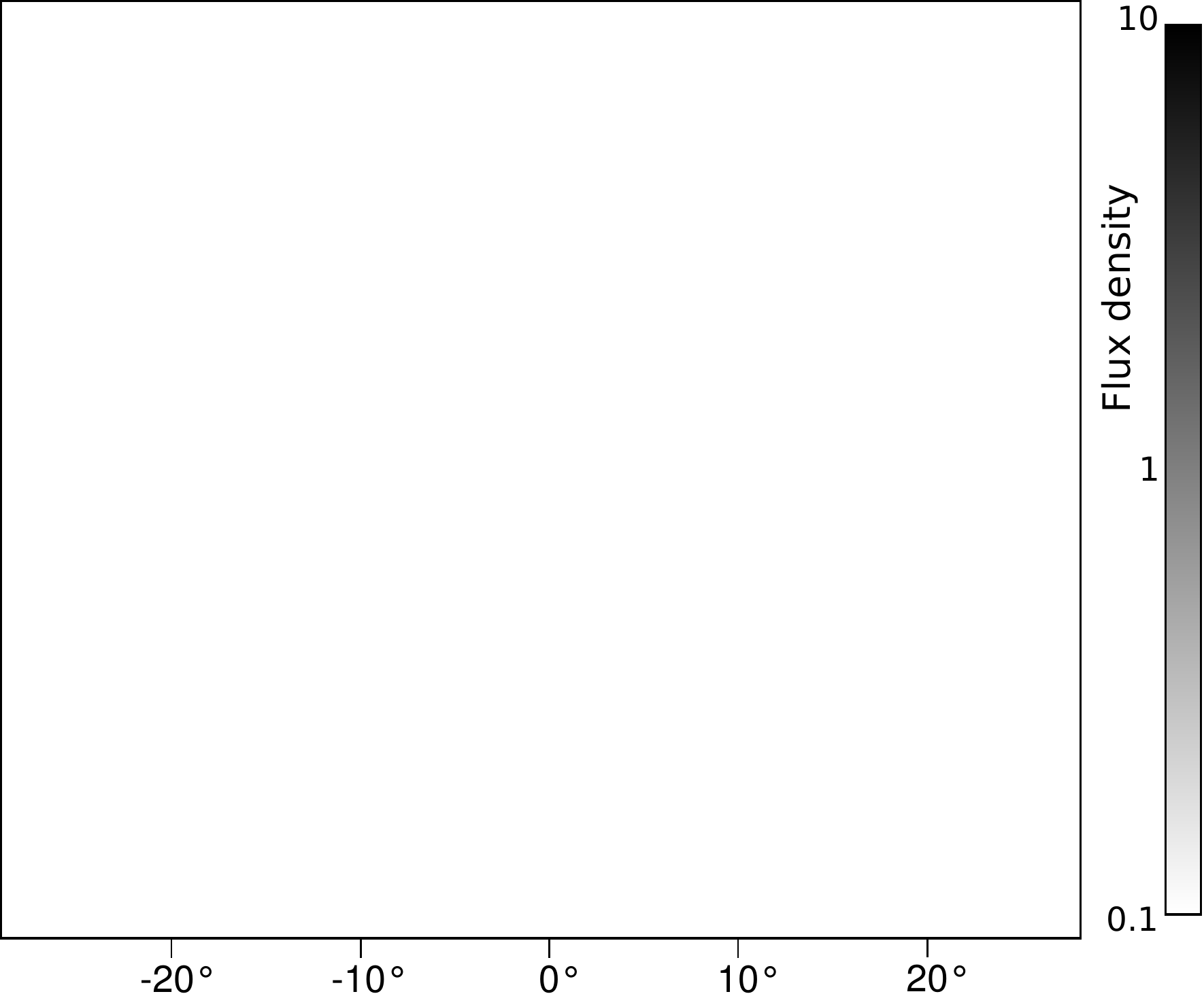}
  }
 \end{center}
 \caption[]{Application of the projected fringe low-pass filter (\S\ref{sec:projected-fringe-filter}) on simulated data. The projected fringe low-pass filter nulls a single direction starting at a certain distance, but does not preserve the phase centre well. In this simulation, the off-axis source has been removed completely up to the noise, two orders of magnitude lower. In \subref{fig:project-low-pass-filter-applied}, the filter is applied and the top source is removed. Panel~\subref{fig:project-low-pass-filter-residual} shows what has been removed from the image, while \subref{fig:project-low-pass-filter-error} shows what has been removed from the area of interest. }
 \label{fig:project-low-pass-filter}
\end{figure*}

The first step of the filter is to make the speed of fringes, coming from any source from a specific direction $\alpha_D$, constant in the time direction. This is done by rotating the $uv$-plane such that the fringes are parallel to the $v$-axis, and subsequently projecting the samples from the track onto the $v$-axis, thereby stretching the high-frequency fringes and pushing together the low-frequency fringes from sources from direction $\alpha_D$. Fig.~\ref{fig:uv-project} visualizes the transformation. At each point on the $uv$-track given by an angle $\alpha(t)$, the fringe frequency $\nu_S(t)$ of a source at time $t$ is multiplied by a factor due to the projection, resulting in a new fringe frequency $\nu_\textrm{projected}$ at angle $\alpha(t)$ on the circle, given by
\begin{equation} \label{eq:projected-speed}
 \nu_\textrm{projected} = \frac{\nu_S(t)}{\cos \left(\alpha(t)-\alpha_D\right)}.
\end{equation}
By substituting the definition of $\nu_{S}(t)$ from Equation~\eqref{eq:source-fringe-frequency-time} into this equation for a single source in the direction of the filter, i.e., $\alpha_S(t)=\alpha(t)-\alpha_D$, the result is $\nu_\textrm{projected} = \left|\boldsymbol{d}\right| \left|\boldsymbol{l}_S\right|$. Hence, the fringe speed becomes independent of time. Sources from other directions, however, will not become constant.

An example of this effect is shown in Fig.~\ref{fig:projected-ft-plot}, which shows the Fourier transform of a projected $uv$-track. The model of Fig.~\ref{fig:rfi-fringe-fit-original-image} was used as input. The projection is towards the direction of the strong source in the bottom. This source shows up as an isolated feature away from Fourier component index zero, because this source lies furthest away from the phase centre. Although the power of this source peaks in one component, it is distributed over several Fourier components, because the time series is finite. Therefore, the point is convolved with the Fourier transform of a windowing function. The sources near the phase centre collect at component indices around zero. 

By performing a low-pass filter with frequency $\nu_F$ on the projected samples, we will remove fringes from sources at time $t \in \left[t_0;t_e\right]$ for which
\begin{equation} \label{eq:projected-filter-test}
 \left|\boldsymbol{l}_S\right| \left| \frac{\cos \alpha_S(t)}{\cos \left(\alpha(t) -\alpha_D\right)} \right| > \nu_F
\end{equation}
holds.

Fig.~\ref{fig:project-low-pass-filter} visualizes the application of the filter. Its effect can be summarized by these three characteristics: (A) any sources at direction $\alpha_D$ that are further away than the limiting distance corresponding to $\nu_F$ will completely be removed; (B) sources at direction $\alpha_D$ within the limiting distance will not be removed at all; and (C) any sources from directions other than $\alpha_D$ will neither be removed completely nor stay untouched completely. The latter is because the denominator and the numerator in Equation~\eqref{eq:projected-filter-test} will have zero crossings at different $t$. Consequently, the left term in Equation~\eqref{eq:projected-filter-test} will become large when the denominator is near zero.

While incomplete filtering of sources in some directions that are not of interest is not very problematic, it is impractical that the only sources for which absolute preservation can be guaranteed, are sources that lie on the line going through the phase centre in the direction of the applied rotation. In the next subsection, we will present modifications that will solve this issue.

Despite this complication, this method might still be usable in practice. According to Equation~\eqref{eq:projected-filter-test}, the fringes of sources will all be filtered around the same angle $\alpha(t)$ in the $uv$ plane. This direction is known, and the area in the $uv$ plane that is affected is therefore known. Samples in this area can be removed from the data, causing a small loss of data. However, the source will successfully be removed without side effects.

\subsection{The iterative projected fringe filter in time domain} \label{sec:iterative-projected-filter}

\begin{figure}
 \begin{center}
   \includegraphics[width=80mm]{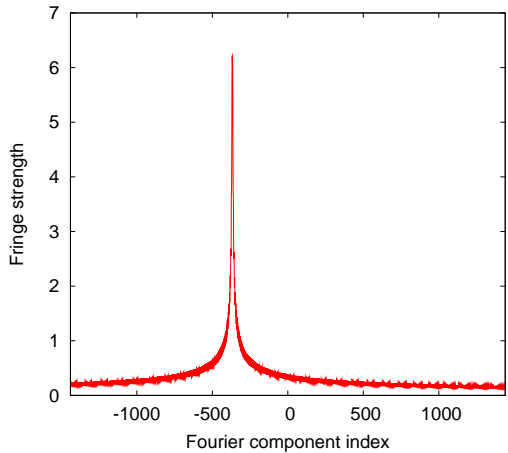}
 \end{center}
 \caption[]{Visualization of the first component in a one-dimensional CLEAN of the plot in Fig.~\ref{fig:projected-ft-plot}. }
 \label{fig:projected-ft-iteration-plot}
\end{figure}

\begin{figure}
 \begin{center}
   \includegraphics[width=80mm]{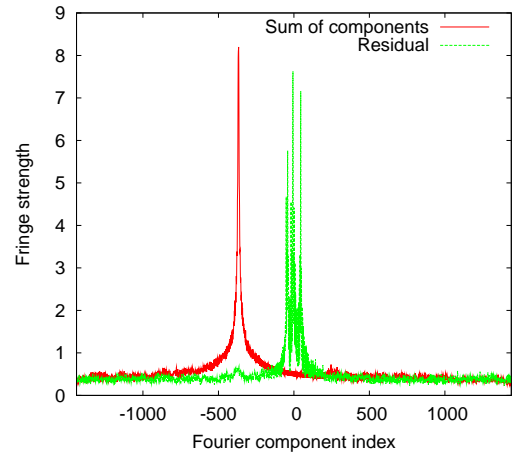}
 \end{center}
 \caption[]{In red, showing the sum of the first hundred components removed by the deconvolution and in green, showing the residuals that contain the data for the area of interest. In the Fourier transform similar to Fig.~\ref{fig:projected-ft-plot}, $\eta_\textrm{filter}$ part of the data around $\alpha_S(t) \approx \alpha_D$ was left out to make sure no sources in the area of interest map to higher components.}
 \label{fig:projected-ft-100iterations-plot}
\end{figure}

\begin{figure*}
 \begin{center}
  \subfloat[Applied]{ \label{fig:part-project-low-pass-filter-applied}
   \includegraphics[height=45mm]{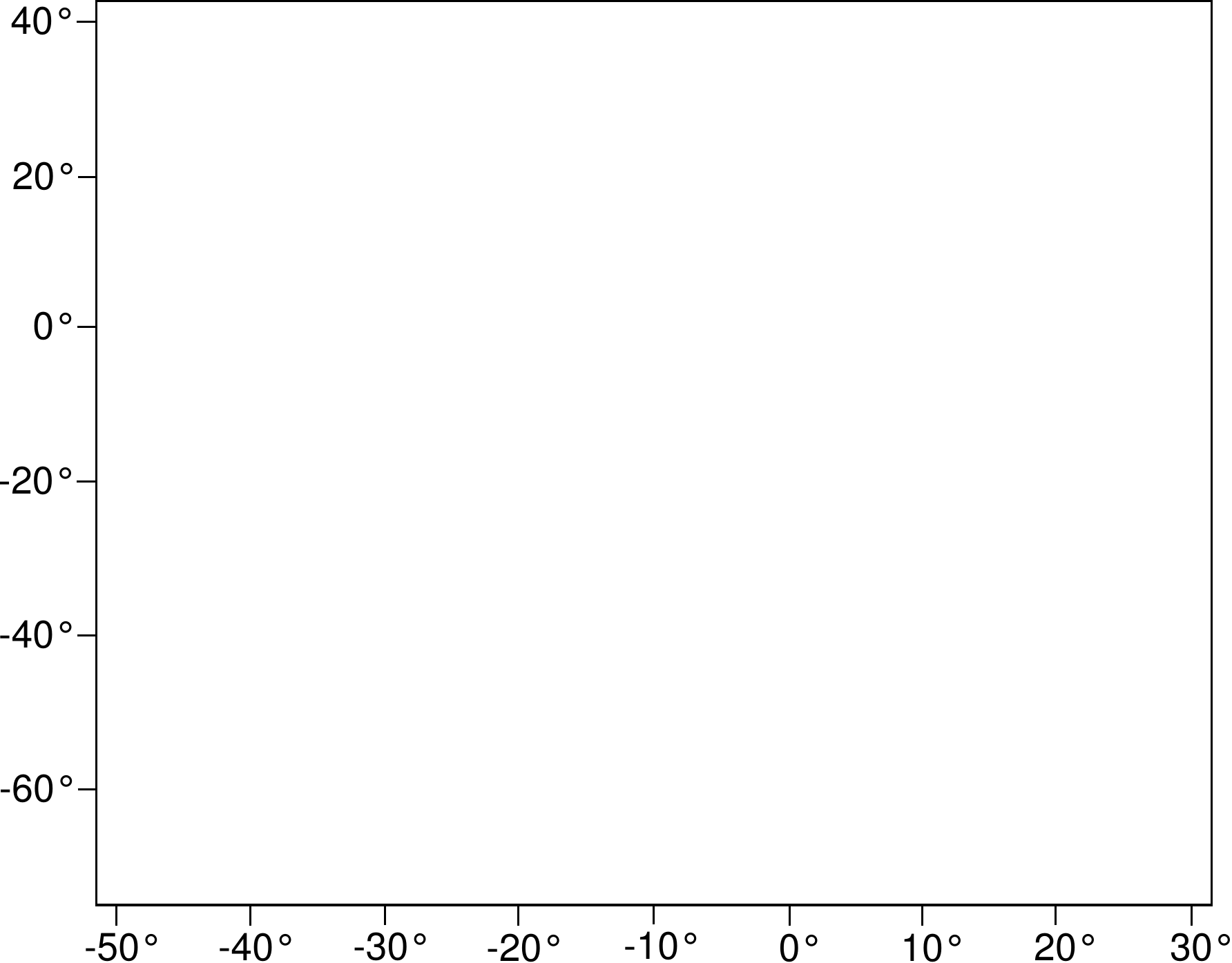}
  }
  \subfloat[Residual]{ \label{fig:part-project-low-pass-filter-residual}
   \includegraphics[height=45mm]{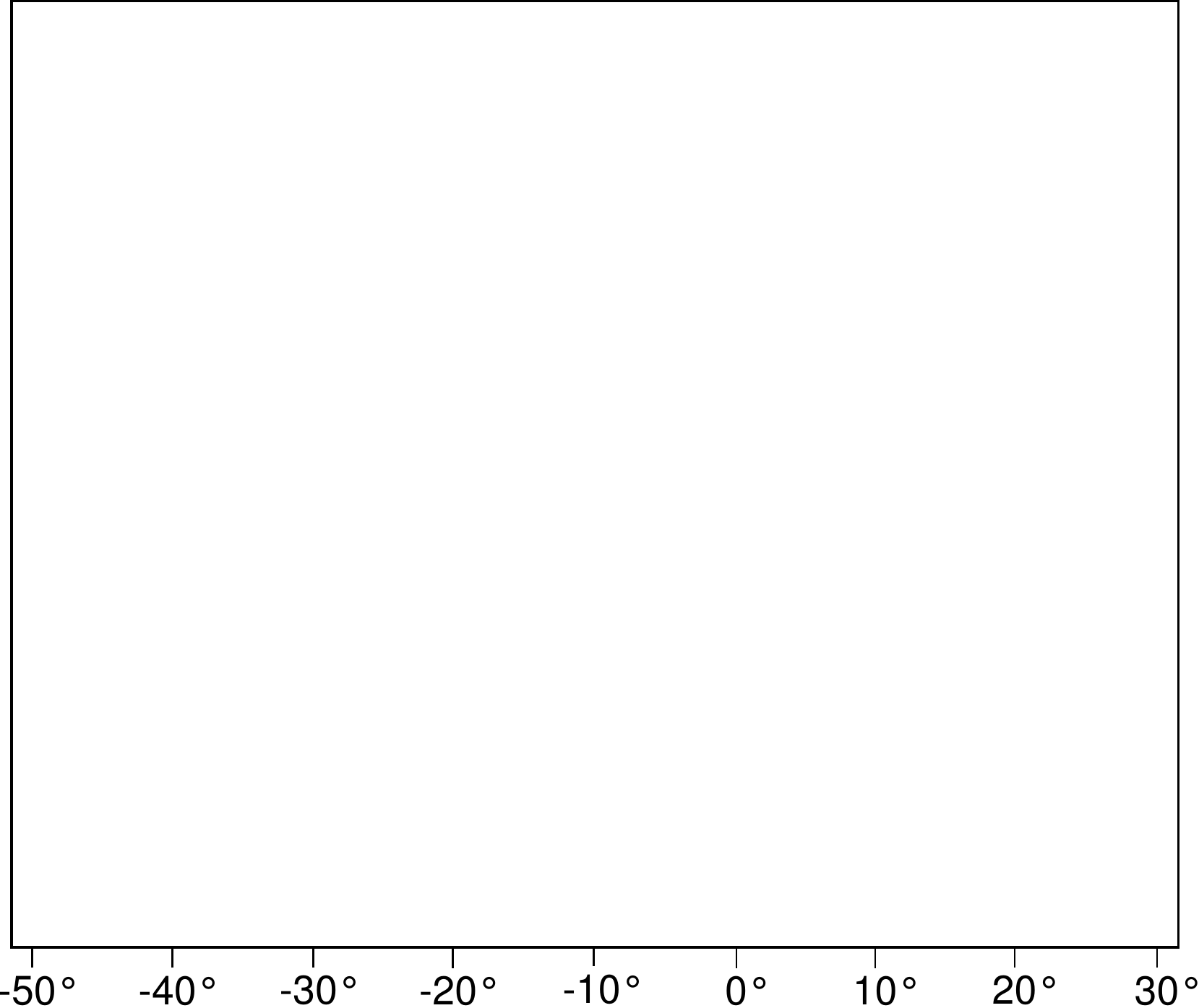}
  }
  \subfloat[Difference with model]{ \label{fig:part-project-low-pass-filter-error}
   \includegraphics[height=45mm]{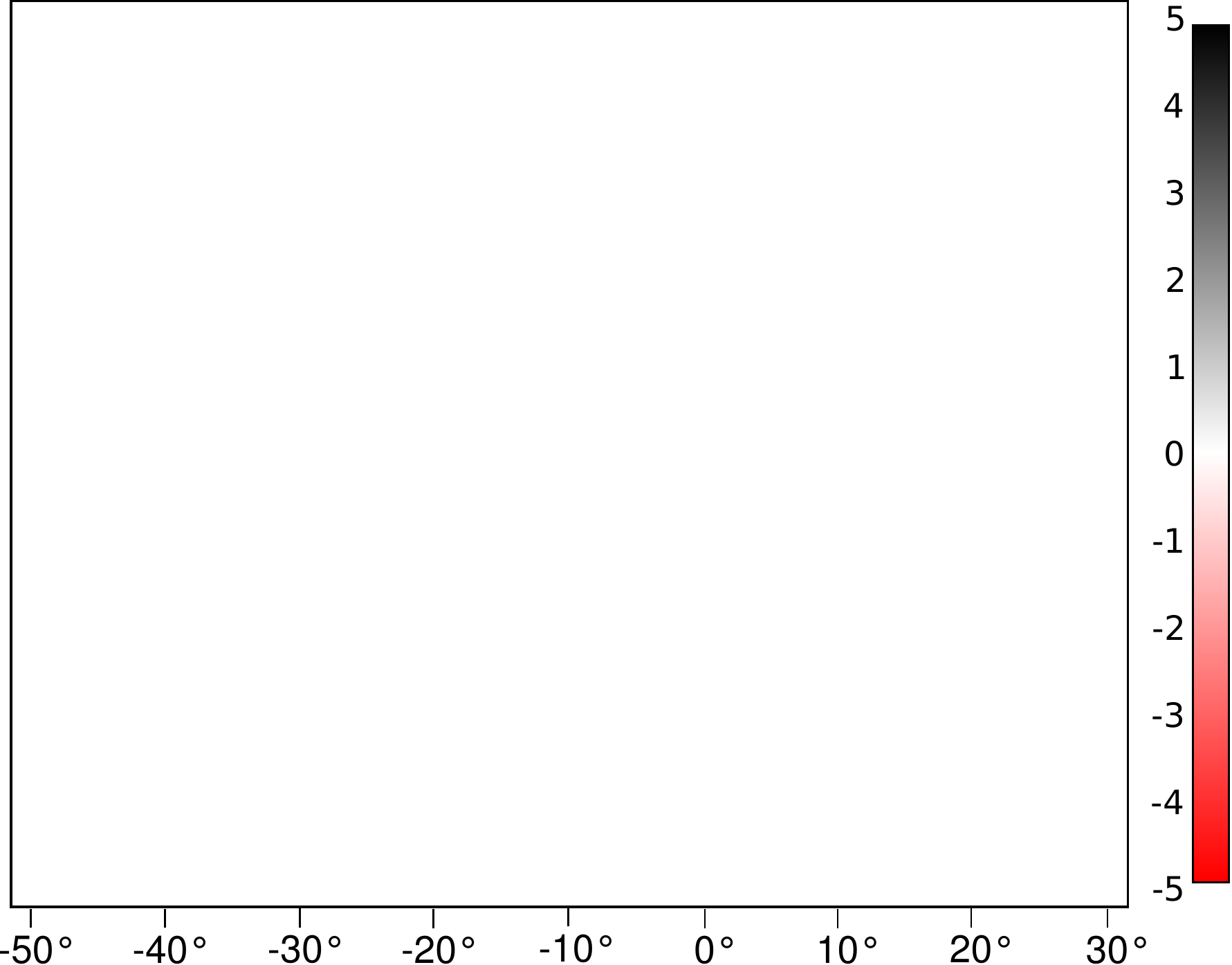}
  }
 \end{center}
 \caption[]{Application of the iterative projected fringe filter (\S\ref{sec:iterative-projected-filter}) on a single simulated baseline of 720 m as in Fig.~\ref{fig:rfi-fringe-fit-images}. The filter was aimed at the source in the bottom and iteratively removes fringes with high frequency. A value of $\eta_\textrm{filter}=0.2$ was used to preserve all of the centre sources, and 100 one-dimensional CLEAN iterations were performed in the projected fringe spectrum domain. Although this has attenuated the source without needing a model of the source, the sidelobes in the direction of the phase centre still remain. }
 \label{fig:iterative-project-low-pass-filter}
\end{figure*}

The projected fringe frequency of a on-axis source can exceed the filtering frequency when $\alpha_S(t) \approx \alpha_D$, i.e., when the uv-track is near parallel to the applied direction of the filter. To create an area of unfiltered sources in the image plane, one can leave this range out of the filter. This however, would create artefacts similar to the low-pass filter of \S\ref{sec:time-domain-low-pass-filter}, and would still not improve the dynamic range in the area of interest.

A solution is to perform a Fourier transform only on the part of the projected samples at which $\left|\alpha_S(t) - \alpha_D\right|>\eta_F$, for some small angle $\eta_F$, and use a deconvolution method to extrapolate the found frequencies to the area that has been left out. A one-dimensional CLEAN on the fringe spectrum can be used to remove and extrapolate fringes, taking fringes out one by one. Altogether, such a filter removes sources from a single direction $\alpha_D$ at a distance corresponding to $\nu_F$ and create a rectangular area around the phase centre which will be preserved. The width of this area is given by
\begin{equation}
\kappa(\nu_F, \eta_F)=\frac{\nu_F}{\left|\boldsymbol{d}\right|}\left|\sin \eta_F\right|.
\end{equation}
Off-axis sources from directions other than $\alpha_D$ will be partially removed and sources of interest will be fully preserved. We will discuss the results of practical application of this filter in \S\ref{sec:practical-applications}.

Fig.~\ref{fig:projected-ft-iteration-plot} visualizes the Fourier transform of the first component that will be removed by a one-dimensional CLEAN on the plot in Fig.~\ref{fig:projected-ft-plot}. In the Fourier transform, $\eta_\textrm{filter}$ part of the data was left out. Because of the finite time domain, the power in a single component is convolved with a function formed by the windowing function, which also depends on the angle between the source and the filter direction. Intuitively, one can think of this as the shape of the PSF in the projected fringe spectrum domain of a single baseline. 75~per cent of the power in the highest component are selected for subtraction in each iteration. Figs.~\ref{fig:projected-ft-100iterations-plot} and \ref{fig:iterative-project-low-pass-filter} show the resulting projected fringe domain and image domain respectively, after applying the iterative fringe filter with 100 iterations.

\subsection{Filtering in frequency direction} \label{sec:freq-lowpass-filter}
The filters that have been presented so far, have been applied in the time domain of correlations from a single baseline. If an interferometer observes several frequency channels over some limited bandwidth, a logical extension is to filter in frequency direction. The samples from different frequencies in the same baseline at the same time form a straight line in the $uv$-plane. A source $S$ produces a fringe speed $\mu_S$ in frequency direction given by
\begin{equation} \label{eq:frequency-fringe-rate}
 \mu_S(t, \lambda) = |\boldsymbol{d}(\lambda)| |\boldsymbol{l}_S|\sin \alpha_S(t),
\end{equation}
and $|\boldsymbol{d}(\lambda)| \sim \frac{1}{\lambda}$.

A low-pass filter in the frequency direction removes fringes of off-axis sources at which $\mu_s(t, \lambda) < \mu_f$.
In contrast to filtering in time, the situation differs on some points:
\begin{itemize}
  \item The use of the $\sin$ function in Equation~\eqref{eq:frequency-fringe-rate} implies that sources produce a high fringe rate in frequency direction when the $uv$-track is orthogonal to the source direction in the image plane. The result is that the source sidelobes in direction of the phase centre, which is the area of interest, will be removed. Therefore, a low-pass filter in frequency direction would complement a filter in time direction, which depends on the cosine of the source angle and the uv-track (Equation~\eqref{eq:source-fringe-frequency-time}). Therefore, the part that is not filtered by the latter can be further attenuated with a frequency direction low-pass filter.
  \item While most radio sources are constant over the observation time, they vary over frequency. Low-pass filtering in frequency would low-pass filter the sources variation over frequency. Because the primary beam is smaller at higher frequencies, an off-axis source can have a steep apparent spectral index.
  \item In the frequency direction, the number of fringes is limited by the observing bandwidth, and the bandwidth might be limited such that the fringes of a source rotate too little for filtering. For example, if a bandwidth-frequency ratio of 2.5~MHz/100~MHz is assumed for a 100~m baseline (approximately the shortest WSRT baseline observing with a single band), a source needs to be at a distance of about 8\degree~from the phase centre to create a single fringe within the bandwidth.
\end{itemize}

Due to these characteristics, the use of a frequency filter can complement a low-pass filter in time, but might be limited to the longer baselines or large filter radii. To be effective, sufficient bandwidth is required. The available bandwidth for filtering might be further limited if the apparent spectral indices of the off-sources are steep.
\section{Practical applications} \label{sec:practical-applications}

\begin{figure*}
\begin{tabular}{rp{22mm}p{22mm}p{22mm}p{22mm}p{22mm}p{22mm}}
   \begin{sideways} Position error \end{sideways} &
   \includegraphics[width=25mm]{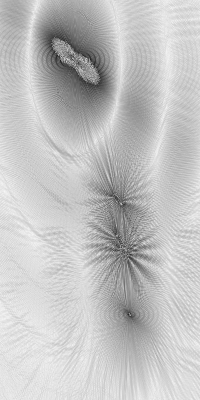} &
   \includegraphics[width=25mm]{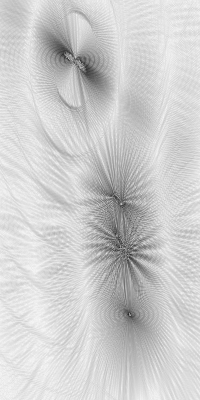} &
   \includegraphics[width=25mm]{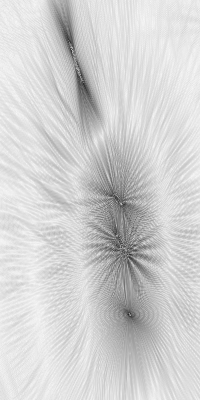} &
   \includegraphics[width=25mm]{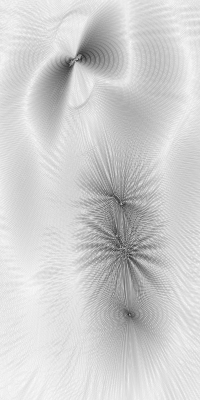} &
   \includegraphics[width=25mm]{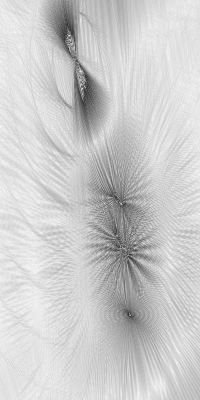} &
   \includegraphics[width=25mm]{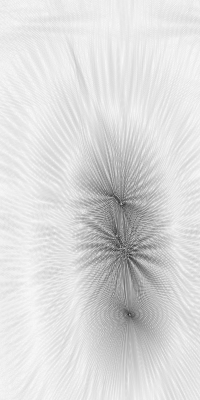} \\
   \begin{sideways} Faint source \end{sideways} &
   \includegraphics[width=25mm]{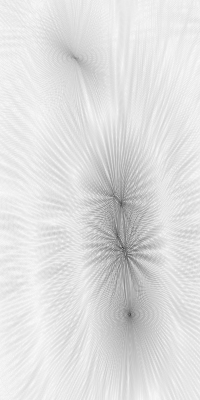} &
   \includegraphics[width=25mm]{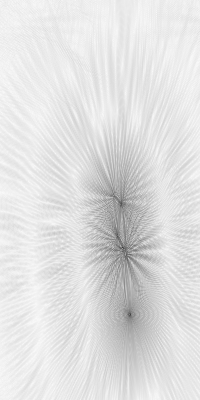} &
   \includegraphics[width=25mm]{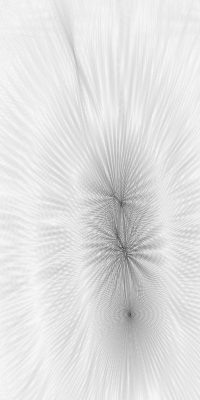} &
   \includegraphics[width=25mm]{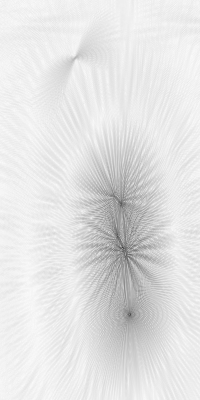} &
   \includegraphics[width=25mm]{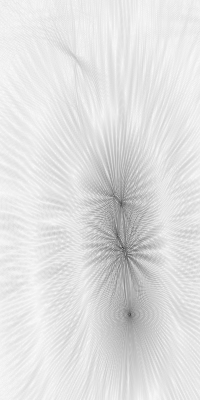} &
   \includegraphics[width=25mm]{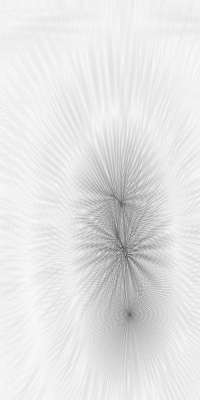} \\
   \begin{sideways} Variable source \end{sideways} &
   \includegraphics[width=25mm]{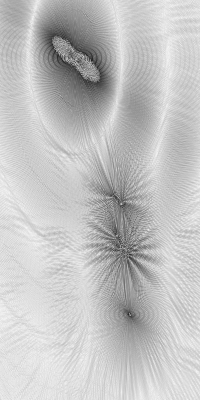} &
   \includegraphics[width=25mm]{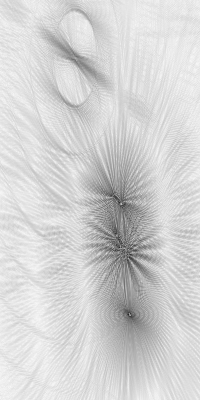} &
   \includegraphics[width=25mm]{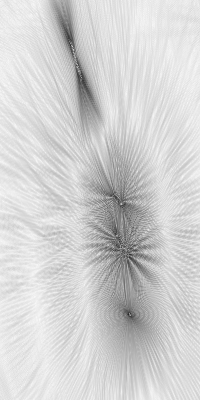} &
   \includegraphics[width=25mm]{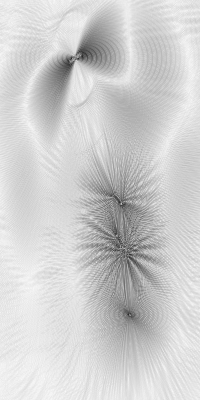} &
   \includegraphics[width=25mm]{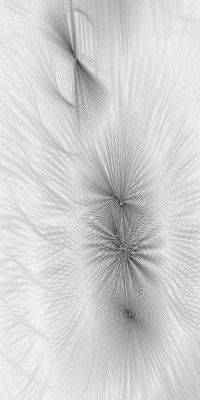} &
   \includegraphics[width=25mm]{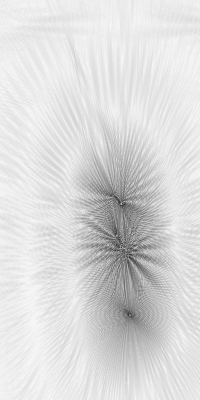} \\
   \begin{sideways} Constant source \end{sideways} &
   \includegraphics[width=25mm]{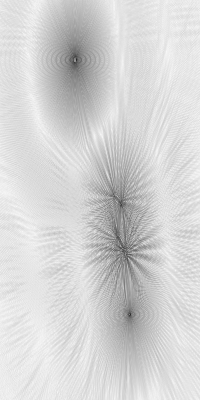} &
   \includegraphics[width=25mm]{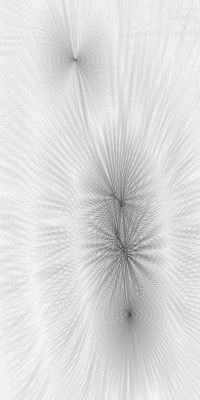} &
   \includegraphics[width=25mm]{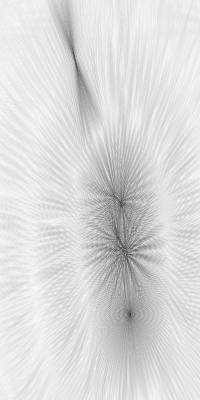} &
   \includegraphics[width=25mm]{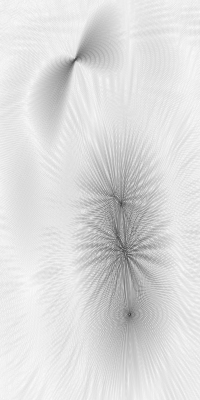} &
   \includegraphics[width=25mm]{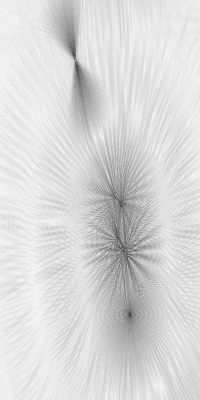} &
   \includegraphics[width=25mm]{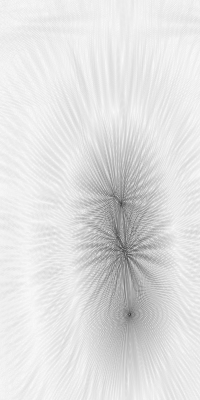} \\
   &
   Original &
   Single~fringe filter &
   Time low-pass &
   Frequency low-pass &
   Projected fringe filter &
   Time \& frequency low-pass
\end{tabular}
\caption[]{Simulated test sets with various types of off-axis sources that need to be removed. On its own, the single fringe filter removes the largest part of the source and its sidelobes, and only becomes inaccurate when the source changes in time or when the the model is inaccurate. The time and frequency low-pass filter complement each other, and together can remove everything outside a certain radius, if bandwidth allows. The projected fringe filter seems not to work very well -- it removes a part of the source, but leaves artefacts in the image in every test case.}
 \label{fig:all-filters}
\end{figure*}

\begin{figure}
 \begin{center}
   \includegraphics[width=80mm]{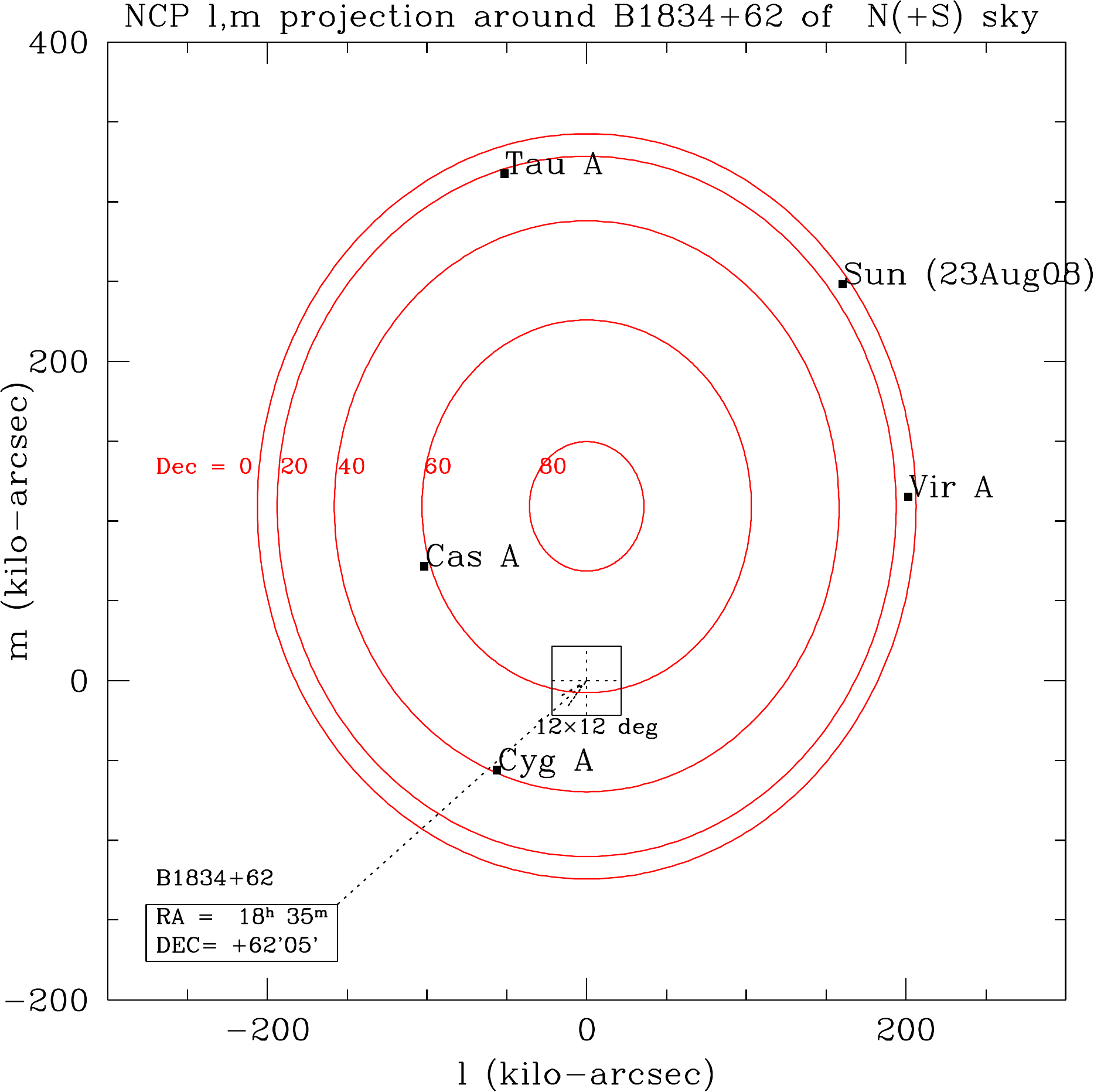}
 \end{center}
 \caption[]{Position in the sky of B1834 relative to other strong sources.}
 \label{fig:wsrt-b1834-sky-map}
\end{figure}

\begin{figure*}
 \begin{center}
  \subfloat[Original]{
   \includegraphics[height=48mm]{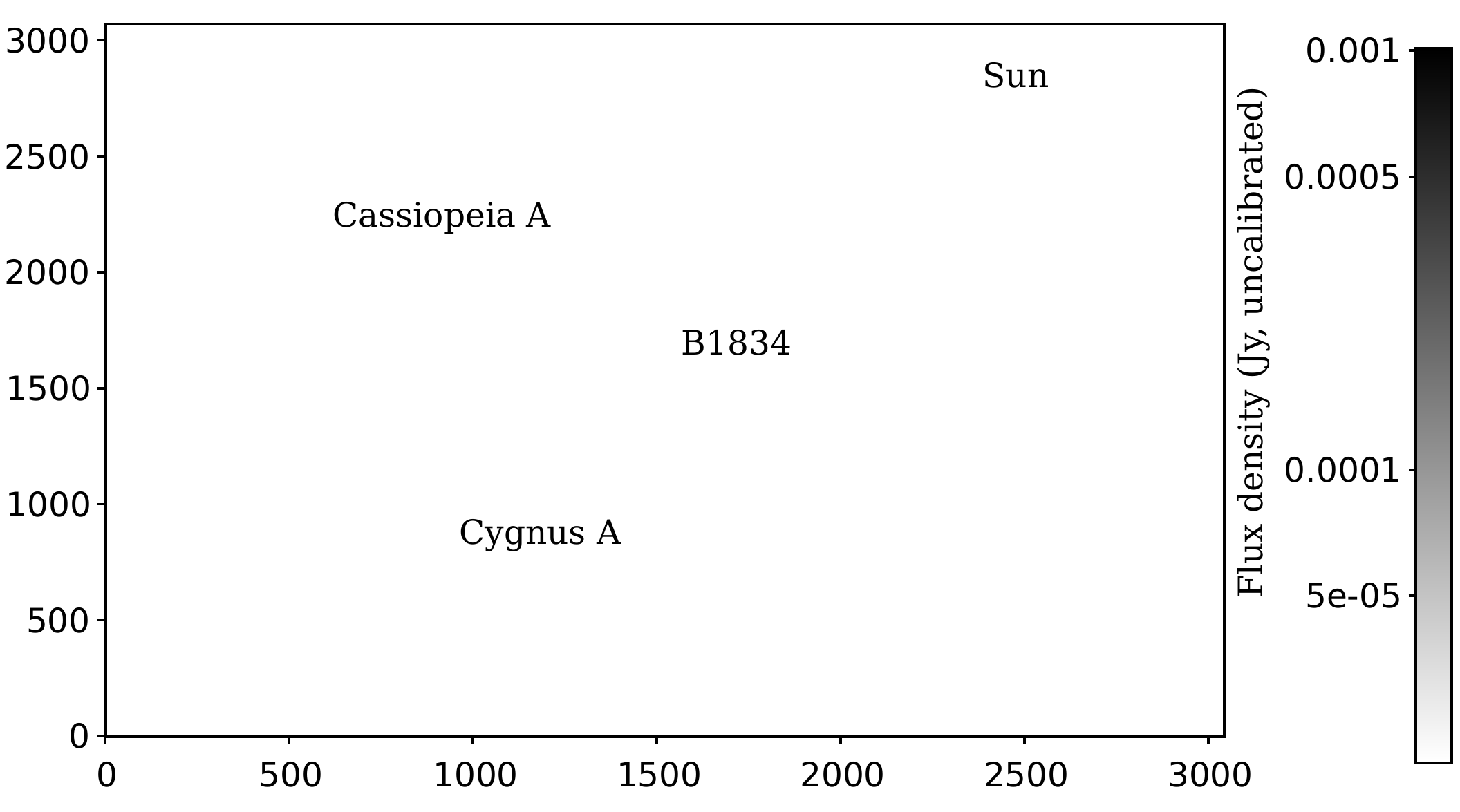}%
  } \\
   \large{\emph{Filtered}} \hspace*{6.5cm}%
   \large{\emph{Difference}}\hspace*{3cm}\\%
  \vspace*{-4mm}%
  \subfloat[Low-pass filter in frequency direction]{%
   \includegraphics[height=48mm]{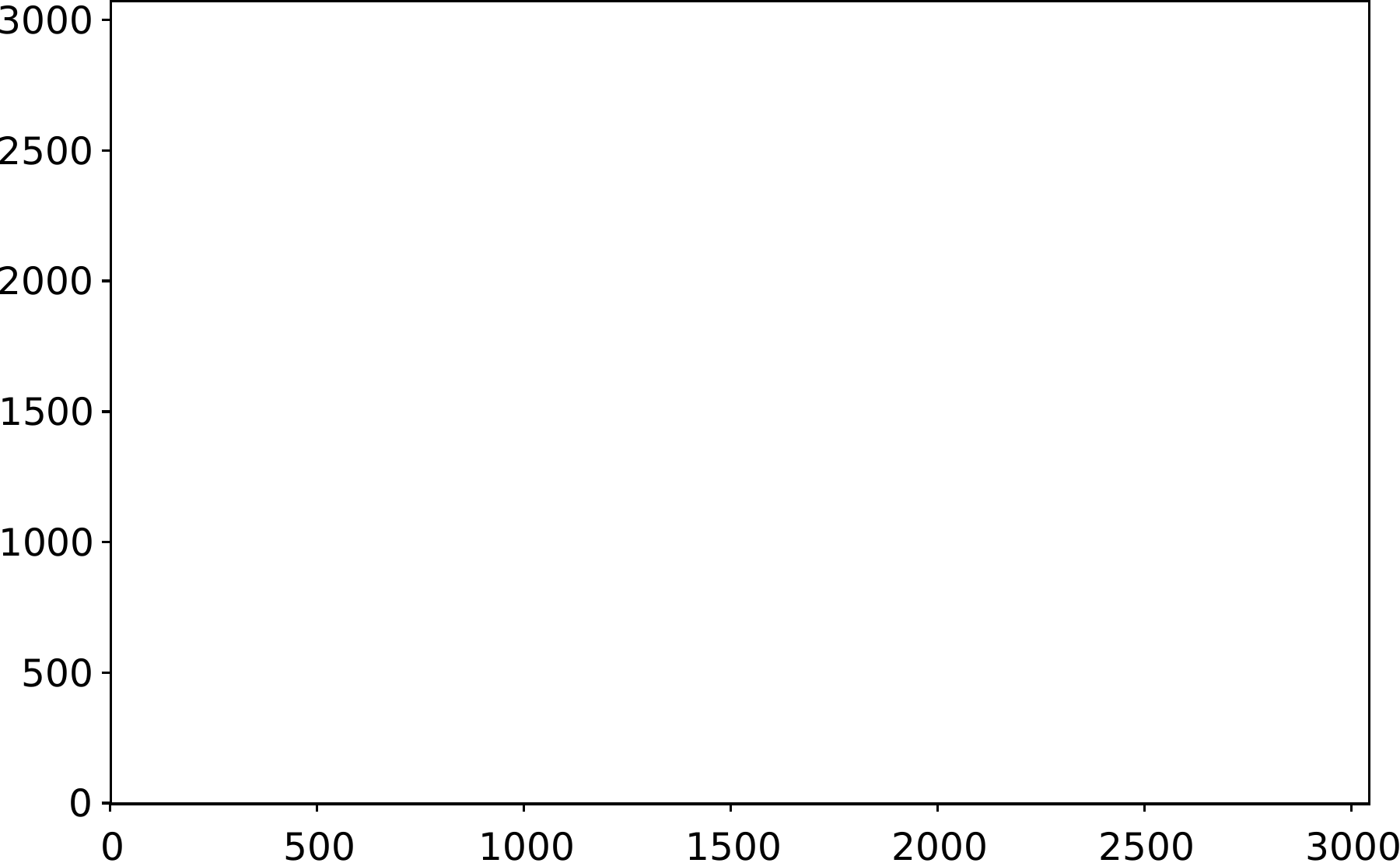}%
   \hspace{3mm}
   \includegraphics[height=48mm]{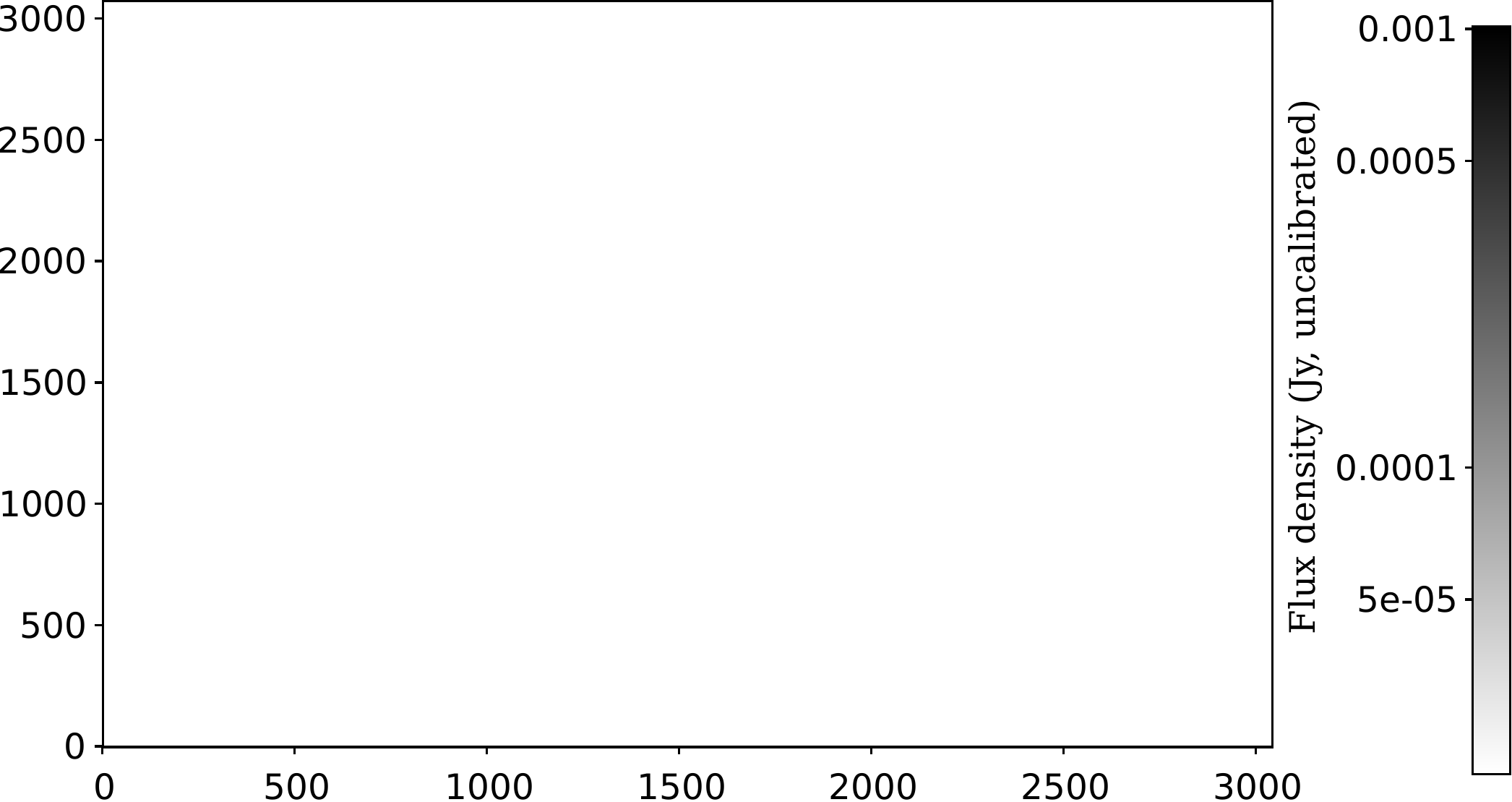}
  } \\
  \subfloat[Low-pass filter in time direction]{
   \includegraphics[height=48mm]{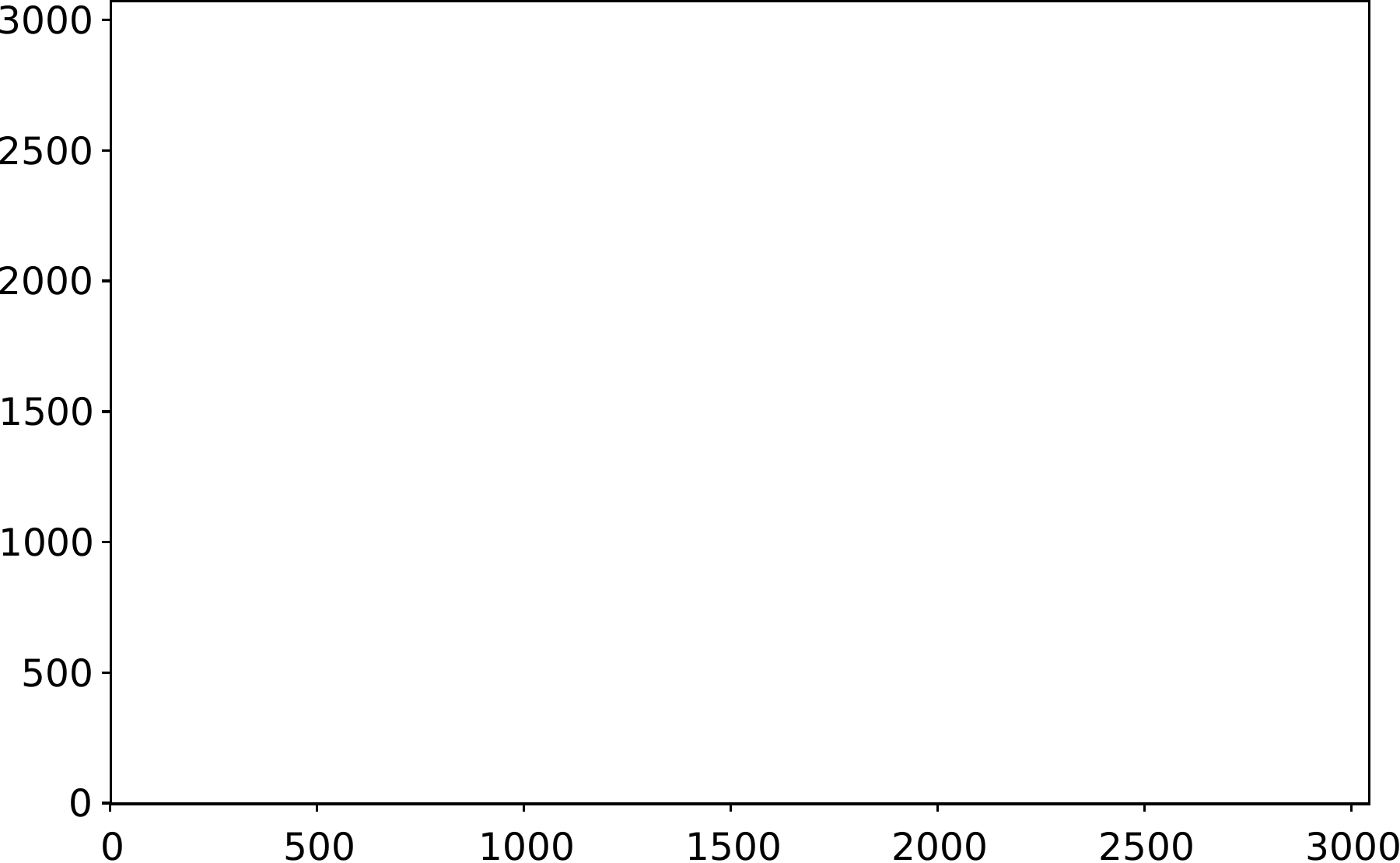}%
   \hspace{3mm}
   \includegraphics[height=48mm]{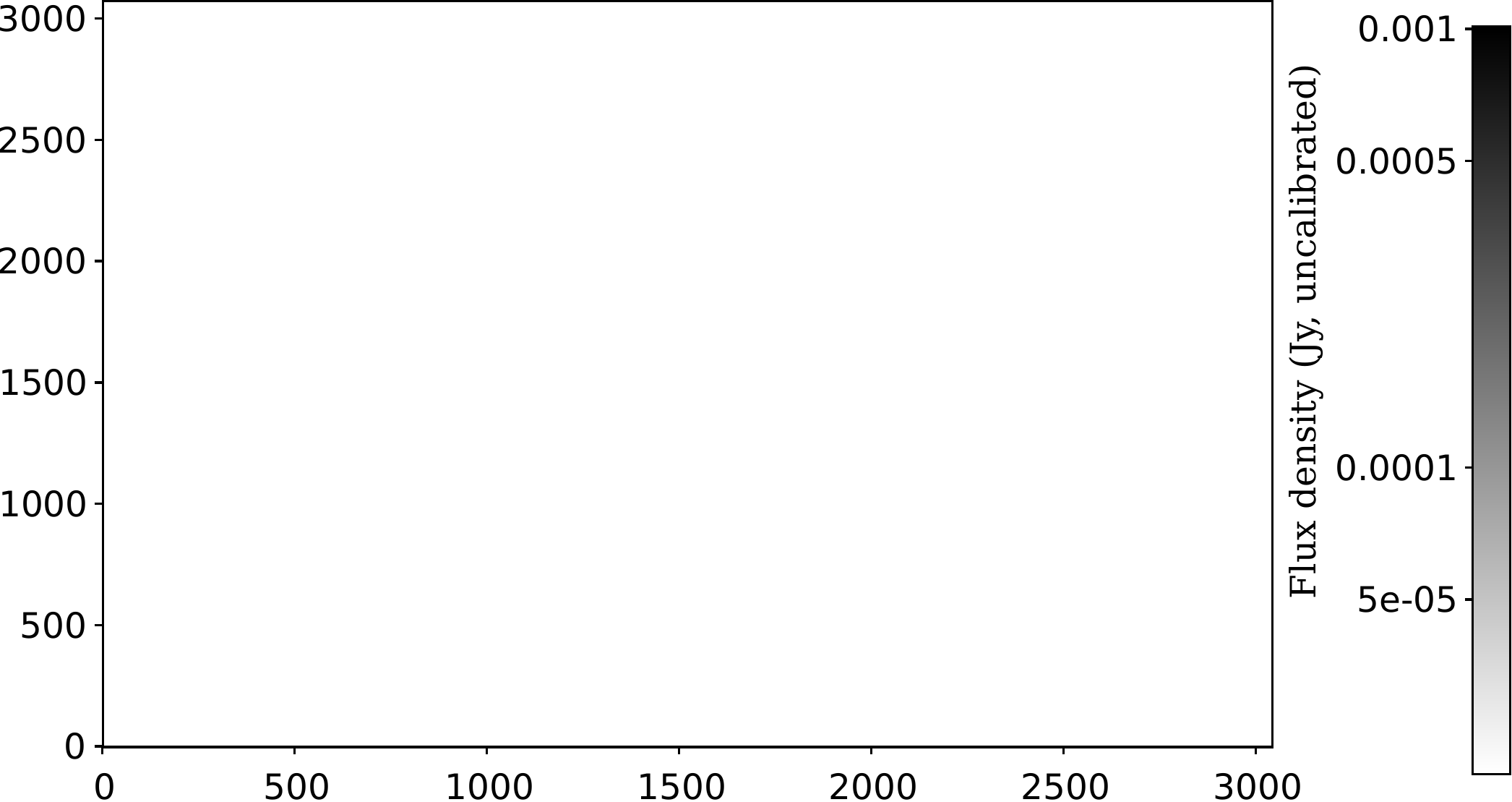}
  } \\
  \subfloat[Low-pass filter in both directions]{
   \includegraphics[height=48mm]{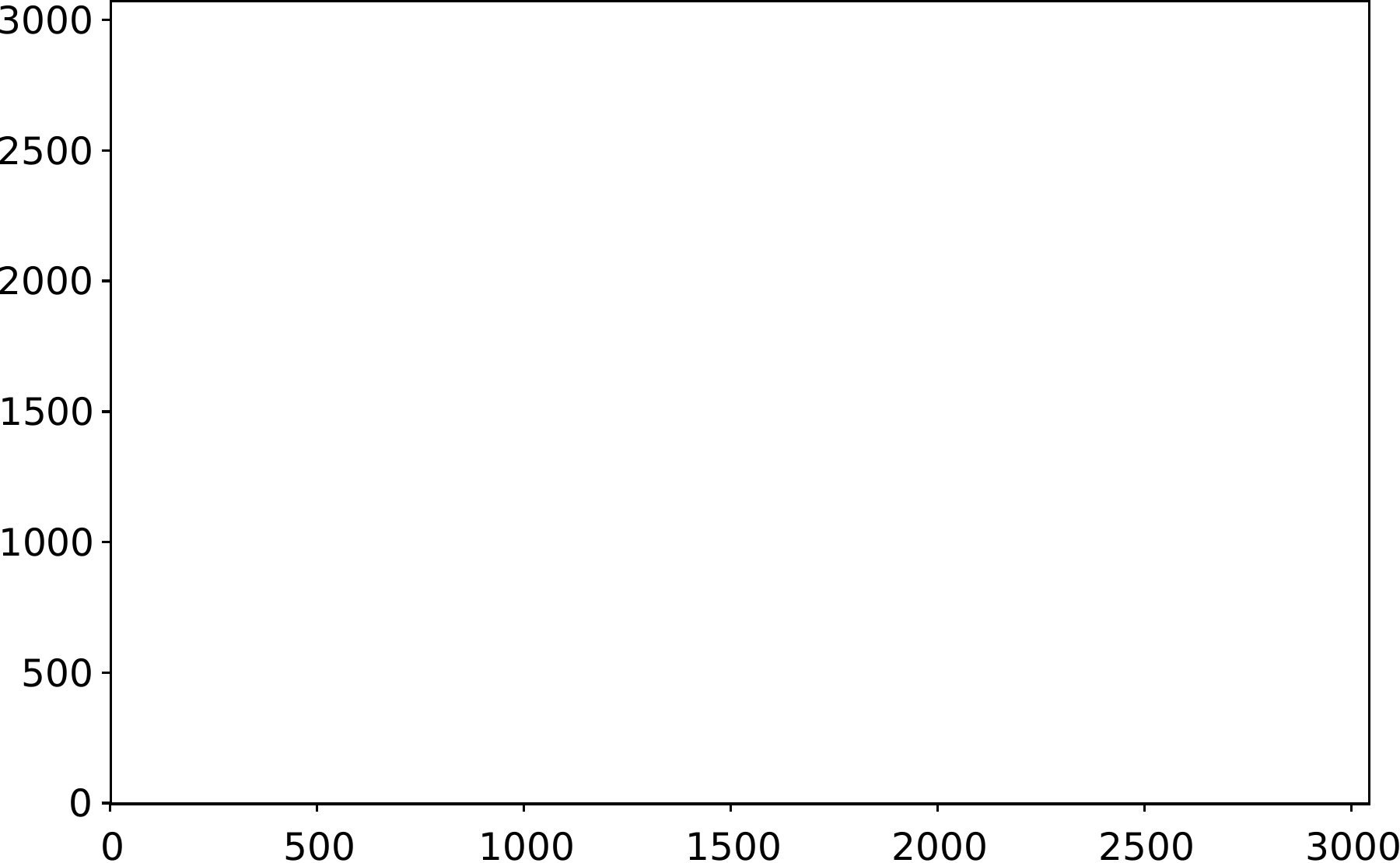}%
   \hspace{3mm}
   \includegraphics[height=48mm]{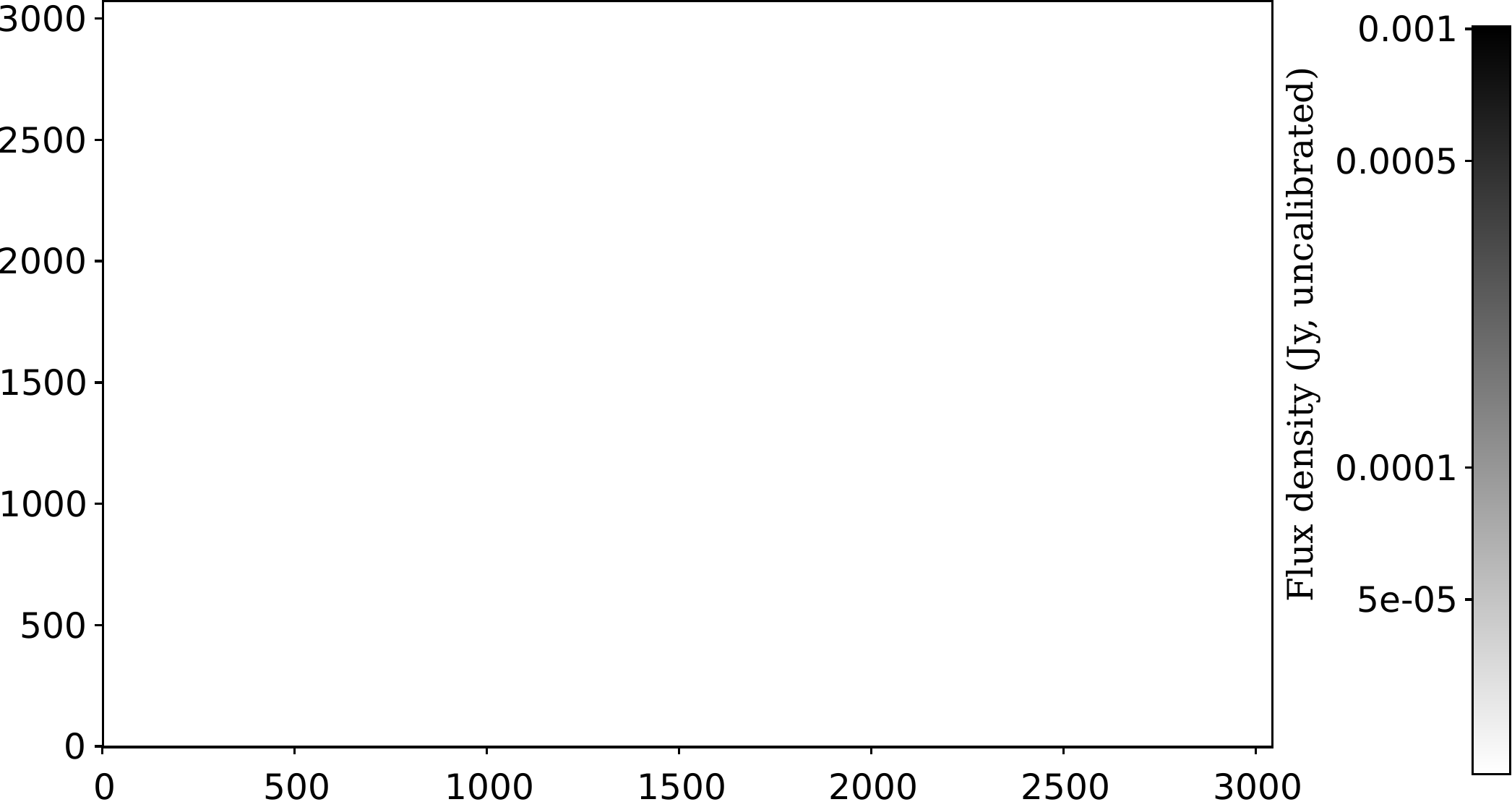}
  }
 \end{center}
 \caption[]{Application of the low-pass filters on a single 1.3~km baseline of an actual WSRT observation of the B1834 area, observed partially in daytime. Frequency filtering removes the Sun down to the noise, including its sidelobes in the area of interest. The filter is less effective near the circular filter edge. The rings are aliasing effects.}
 \label{fig:wsrt-single-baseline-0xa}
\end{figure*}

Several filters for off-axis sources were described in the previous chapters. Fig.~\ref{fig:all-filters} shows an overview of all the filters, applied on several classes of simulated off-axis sources. The fringe filter works well, as long as an accurate model of the source exists, and the received strength of the source does not change much in time. The low-pass filters in time and frequency direction together remove the off-axis source quite well. The projected iterative fringe filter in time direction can only attenuate the off-axis source moderately, even though it requires to know the direction to filter rather accurate. Application of the method on real data shows comparable results.

\subsection{Attenuation efficiency}
To test the level to which sources can be removed, we have simulated a single 40 degrees off-axis source in an otherwise empty field, i.e., without any on-axis sources, and also without noise. We simulated a single 2.5~MHz band at 130~MHz with a standard WSRT configuration and compared the level of the sidelobes before and after source filtering. The single fringe filter shows 40~dB of sidelobe attenuation on a constant source, but only attenuates up to 3~dB of a varying source, which provides a more realistic setting.

The frequency direction low-pass filter can remove 10~dB of a source, which can be varying. Because the low-pass filters are less effective near the borders of the band and the start and end of the observation, we have tried flagging 5~per cent of the border channels in the time frequency plane after filtering. This leads to 20~dB of attenuation. The low-pass filter in time direction does in theory not remove sidelobe noise in the direction of the source. However, in practice, it attenuates the RMS in areas around the phase centre by zero to 3~dB. This is because of a property of gridders: high fringe frequencies are mapped back to the area of interest, i.e., resampling causes aliasing effects. Therefore, removing the high frequencies before imaging lowers the noise as well. The RMS decrease in the radial direction due to low-pass filtering in time is around 25 dB. The large difference between attenuation of the tangential direction of time low-pass filtering versus the radial direction of frequency low-pass filtering is due to the limited bandwidth: in time, the observation contains lots of fringes which can be accurately filtered, but only a few fringes appear in frequency direction.

In the same test, the projected fringe low-pass filter shows 25~dB of attenuation around the phase centre. Finally, the projected iterative fringe filter attenuates only up to 3~dB.

Obviously, these results are highly dependent on many parameters, including the distance of the source to the phase centre, the amount of available bandwidth and its central frequency, the time and frequency resolutions and, for the single fringe filter, the speed of change of the source due to instrumental effects and the number and size of the interferometers.

\subsection{Low-pass filtering a WSRT observation}
\begin{figure*}
 \begin{center}
  \subfloat[Original]{
   \includegraphics[height=42mm]{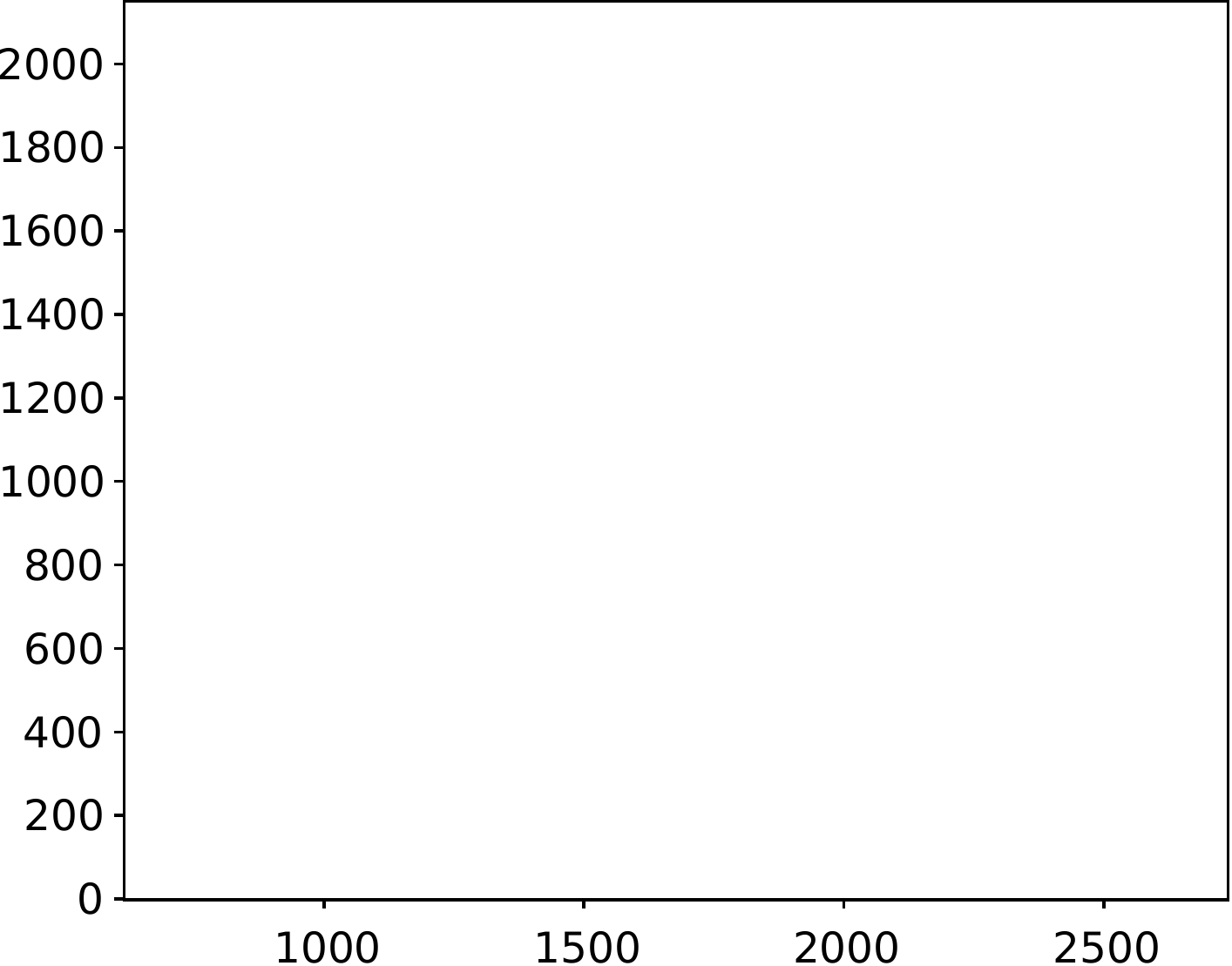}
  }
  \subfloat[Both filters applied]{
   \includegraphics[height=42mm]{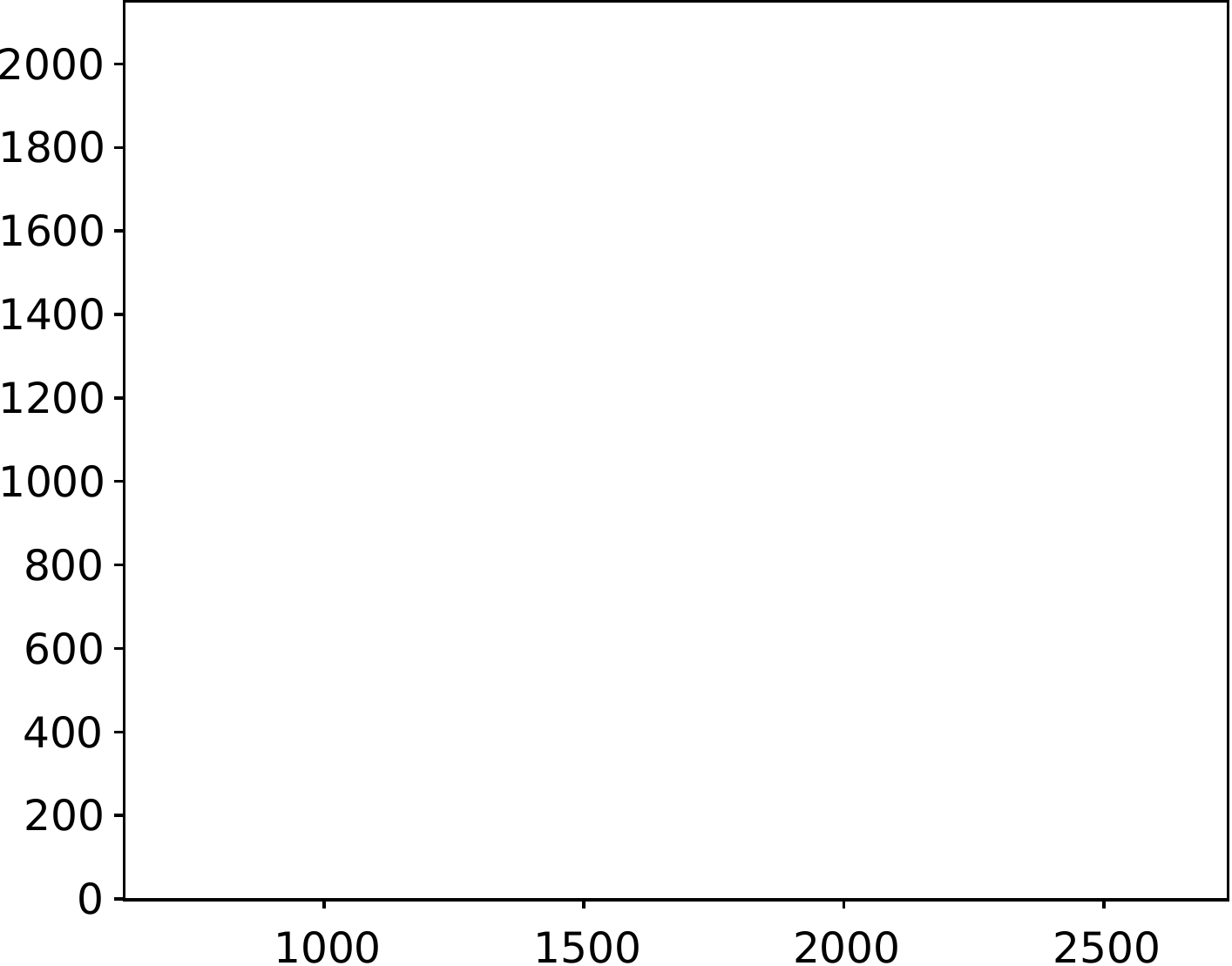}
  }
  \subfloat[Difference]{
   \includegraphics[height=42mm]{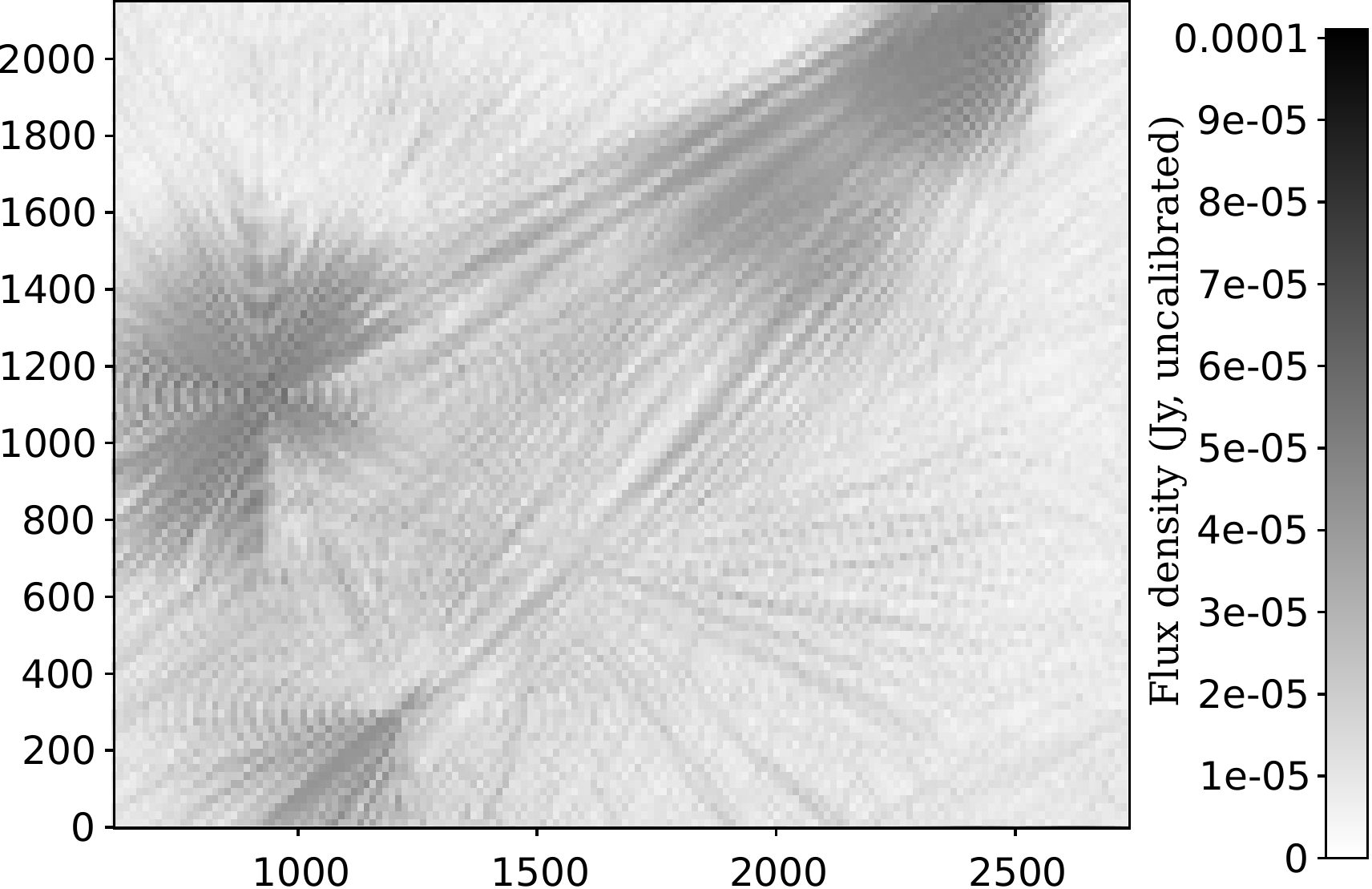}
  }
 \end{center}
 \caption[]{Application of low-pass filters in both directions as in Fig.~\ref{fig:wsrt-single-baseline-0xa}, but on a shorter baseline of 288 meters. The Sun is successfully attenuated, but the filter has been less effective on Cygnus~A and Cassiopeia~A.}
 \label{fig:wsrt-single-baseline-0x2}
\end{figure*}

We will now apply the filtering approaches on a WSRT dataset of the field centred on the radio galaxy B1834+62. This field was observed to search for polarized emission in this double double radio
galaxy \citep{double-double-radio-galaxies-schoenmakers} at very low frequencies. The observations were done in August 2008 and lasted for 12~h. The backend was configured to observe 8 frequency bands, each 2.5 MHz
wide and covered in 512 spectral channels, at frequencies ranging from 115 to 163 MHz. Here we will only use data from the band at 139 MHz. The integration time was 10~s, the spectral resolution, after Hann tapering, was 10~kHz. At this time and spectral resolution even sources more than 1 radian from the phase tracking centre were not significantly smeared. The field was affected by sidelobes from Cygnus~A, Cassiopeia~A and the Sun (for about 8 hours). An image of the locations of these sources, in the NCP projection of the whole sky suitable for the WSRT --- an East-West array --- is shown in Fig.~\ref{fig:wsrt-b1834-sky-map}.

Although each of these three sources is not in the primary beam, each of them is strong enough to lower the dynamic range of the observation considerably because of their sidelobes in the image plane. It is hard to remove these sources from the observation, because they are in the sidelobes of the beam and, especially in the case of the Sun, they are complex and their apparent strength varies over time. Because we do not have accurate models of the sources in our observation, the low-pass filters are a good choice, and we will show that the low-pass filters prove to be quite effective for attenuating the three sources.

Fig.~\ref{fig:wsrt-single-baseline-0xa} shows a single baseline of the B1834 observation. The baseline used is RT0~$\times$~RTA, a 1.3~km East-West baseline, and only data from a single 2.5~MHz band at 140~MHz was used. The displayed images correspond to several tens of degrees of the sky. The observation is limited by confusion noise of the Sun (right top corner, also aliased to the bottom), Cassiopeia~A (left top) and Cygnus~A (left bottom). The observation takes 12~hours and the (resolved) contribution of the Sun moves through the image and sets halfway. Consequently, the Sun and its sidelobes would be very hard to remove with traditional methods. The two low-pass filters together remove the Sun down to the noise: in the filtered image, its peak value is 1~per cent of the original value. It is hard to remove more, i.e., make the filter circle smaller, since only a small bandwidth is available. Because of this, the edge of the filter border is blurred in the frequency filtering cases. For the same reason, Cassiopeia~A should have been filtered but is removed only 95~per cent, and Cygnus~A should not have been filtered, but is attenuated 25~per cent. These errors occur because these sources are too close to the filter border. Other sources within the filter radius have been attenuated less than 1~per cent.

The application of the low-pass filters on this baseline shows the practical effectiveness of the filters: filtering in time direction removes the tangential components of the sources, while the frequency direction removes the radial components. The frequency filter is not as accurate as the time filter, because of the limited 2.5~MHz bandwidth available. This causes the circular ``filtered'' area not to have a sharp edge that a perfect sinc function would produce. Instead, the edge is somewhat blurred. As a consequence, a part of Cassiopeia A has been removed, although it did not exceed the theoretical cutting frequency.

In Fig.~\ref{fig:wsrt-single-baseline-0x2}, a shorter baseline was processed with the filtering techniques. Baseline RT0~$\times$~RT2 was used, which is only 288 meters long. Because of the combination of a short baseline and the small available frequency bandwidth, the frequency filter is only able to filter out 80~per cent of Cassiopeia A on this baseline. The Sun is still successfully attenuated over 99~per cent, up to the noise. Cygnus A is 10~per cent attenuated. No other sources in the area of interest have been visibly attenuated. Because the off-axis sidelobe noise RMS is around 10~per cent of the peak of strong on-axis sources in the area of interest, one can conclude from this image only that the on-axis sources have been preserved for at least 90~per cent.

\begin{table}
 \begin{center}
 \caption[]{Fringe speed in time and frequency directions as a function of scale, looking at zenith with a 1~km baseline. }
 \label{tbl:fringe-frequency-examples}
 \begin{tabular}{|r||r|r|r|}
 \hline
 & \multicolumn{2}{c|}{1~km} & \multicolumn{1}{c|}{$\lambda=$21~cm} \\
 \hline  \hline
 \multicolumn{1}{|l||}{Scale} & Time & Freq & Time \\
 \hline
 & $\lambda$/h & MHz$^{-1}$ & h$^{-1}$ \\
  45\degree
    & 2.9  & 2.4
    & 140     \\
 10\degree
    & 0.72 & 0.58
    & 34    \\
 1\degree
    & 0.073 & 0.058
    & 3.5   \\
 \hline
  & $\lambda$/d & GHz$^{-1}$ & d$^{-1}$ \\
 10~arcmin
    & 0.29 & 87
    & 14  \\
 1~arcmin
    & 0.029 & 8.7
    & 1.4  \\
 \hline
 \end{tabular}
 \end{center}
\end{table}

\begin{figure*}%
\begin{center}%
\hspace*{-4mm}\includegraphics[height=100mm]{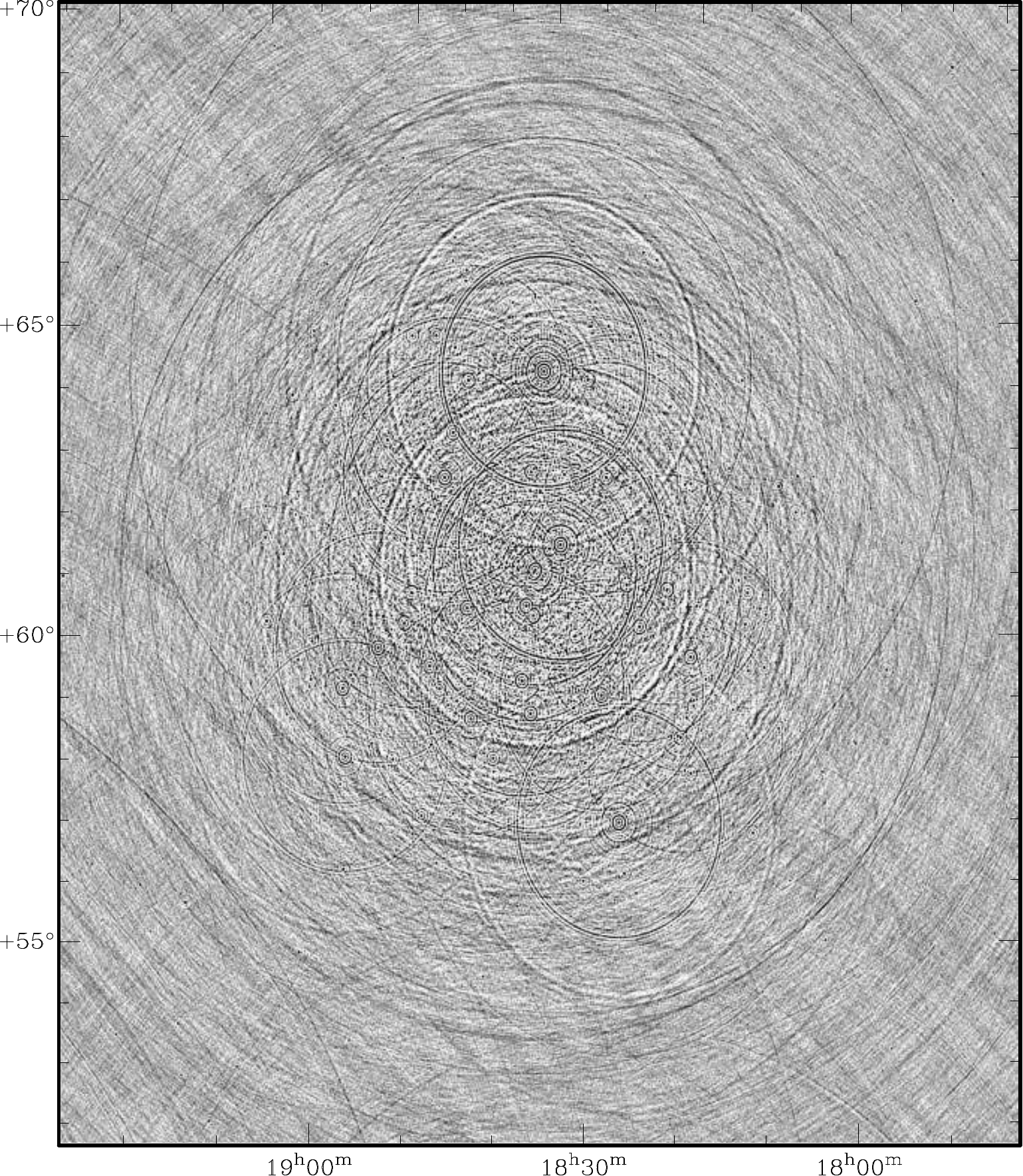}\nolinebreak\hspace{1mm}%
\includegraphics[height=100mm]{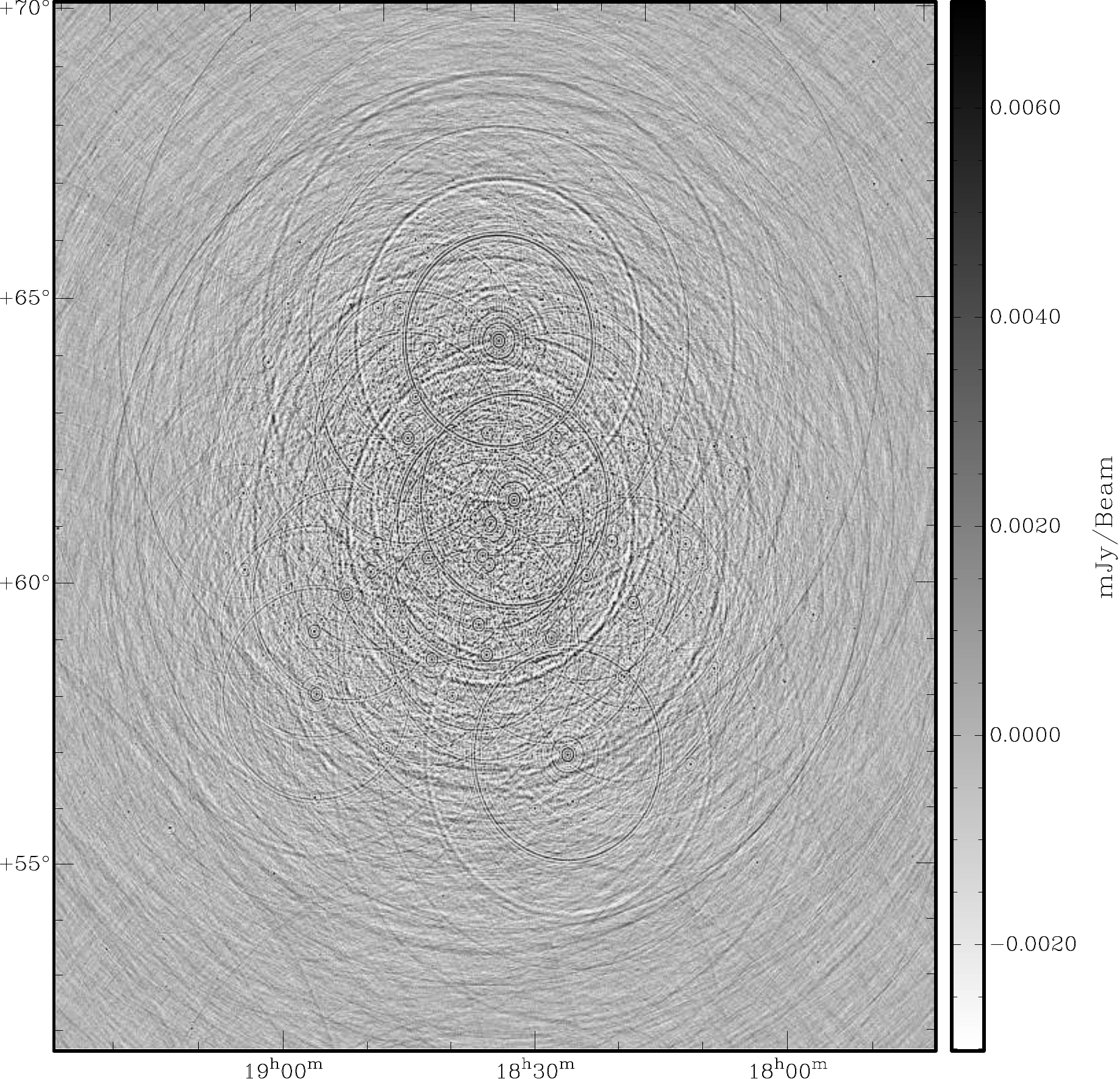}\\\vspace{0.3cm}\noindent%
\hspace*{-4mm}\includegraphics[height=100mm]{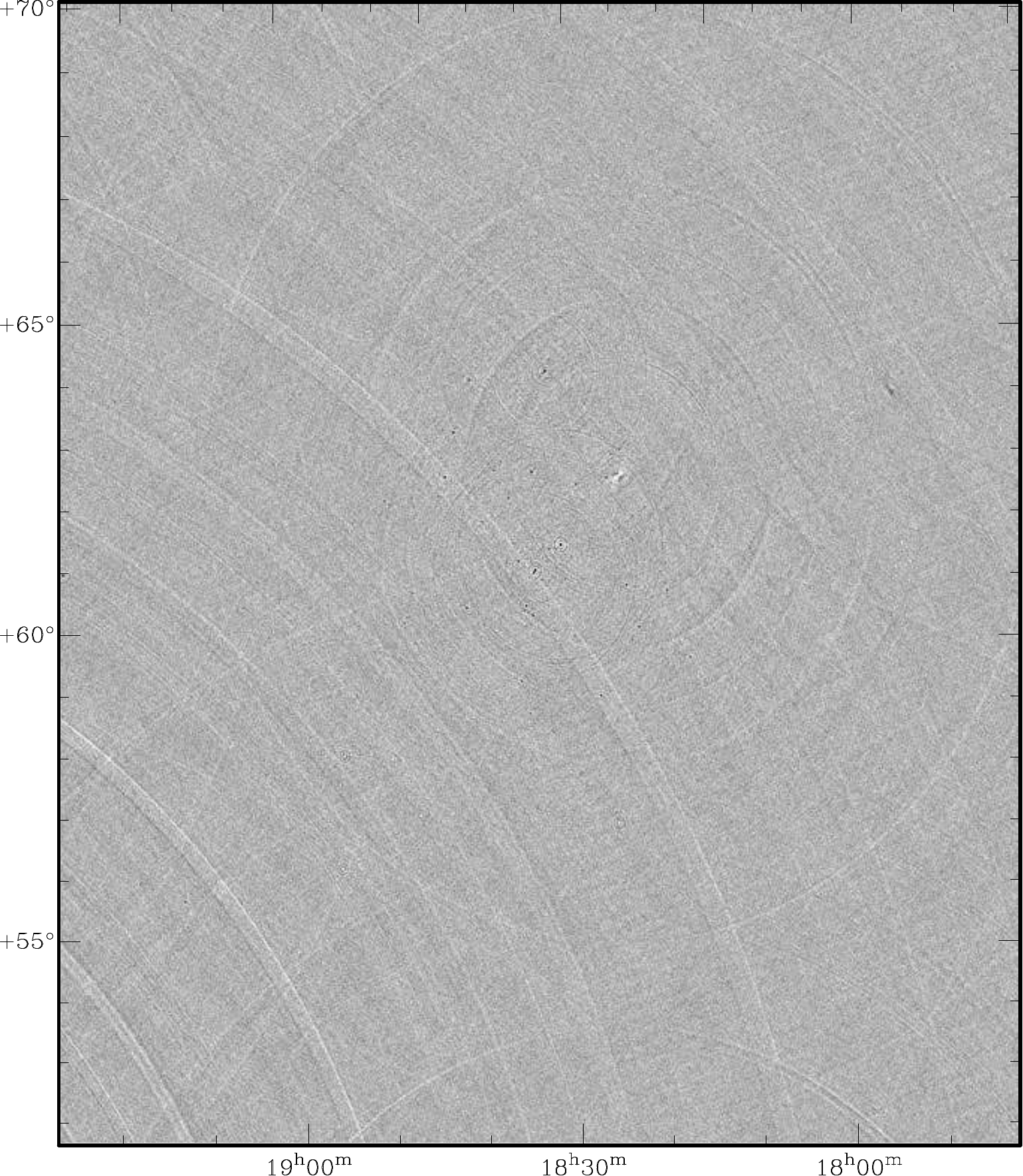}\nolinebreak\hspace{1mm}%
\includegraphics[height=100mm]{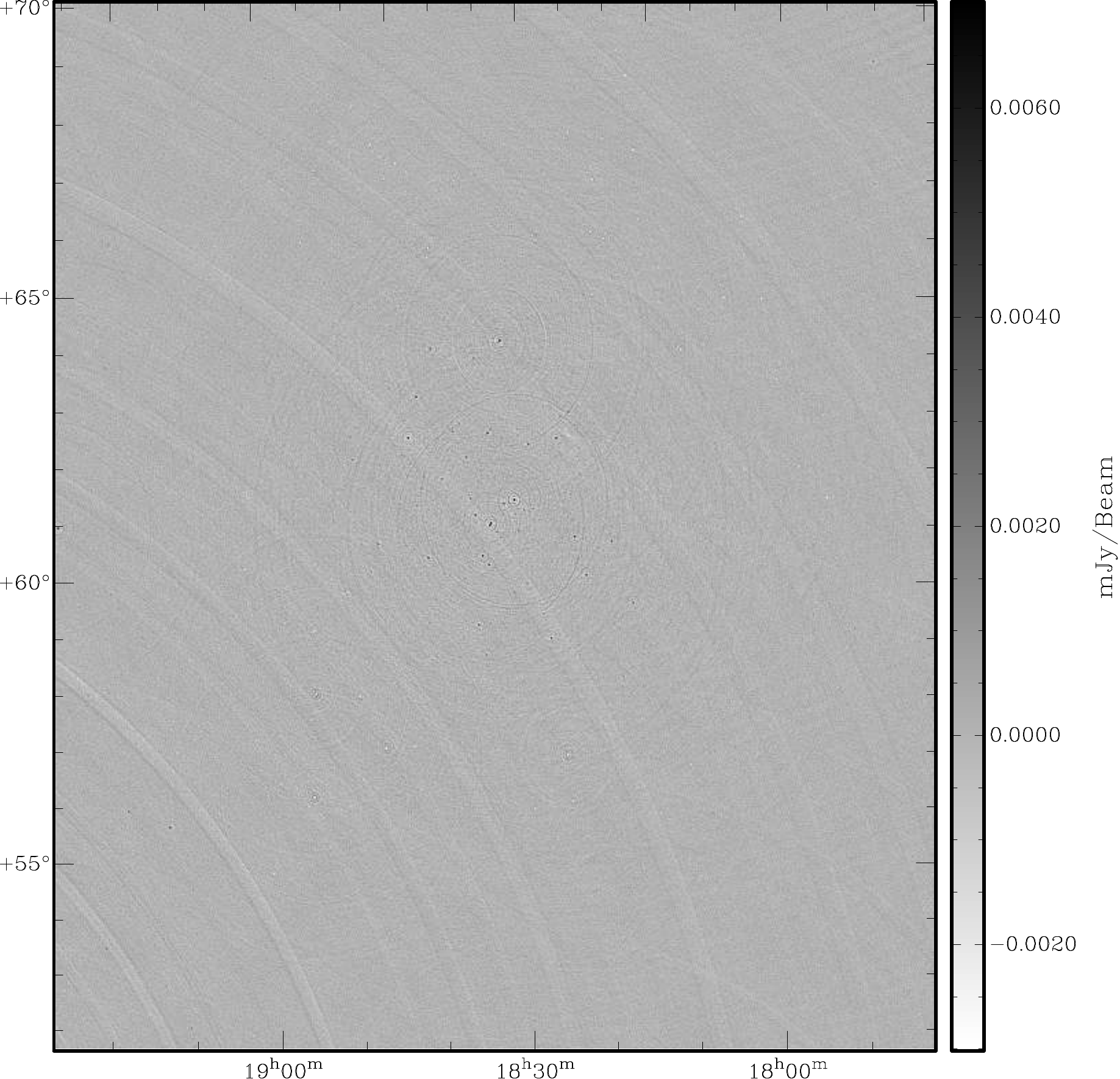}\\%
 \end{center}%
 \caption[]{A WSRT observation of field B1834 at 140~MHz containing three strong off-axis sources (see Fig.~\ref{fig:wsrt-b1834-sky-map}). WSRT can observe eight bands with 2.5~MHz bandwidth at this frequency, however, for this image, only one of the eight bands is used. The top and bottom figures show Stokes I and Stokes Q respectively. The left images are from the raw data, the right images shows the same data after low-pass filtering the set in both time and frequency directions. Even though the filter is limited by the small bandwidth, the suppression of the confusion noise of off-axis is significant. The effect is more detectable in the polarized images. Depending on which area is used for RMS calculation, the Stokes I and Q images show a noise reduction by a factor of 1.5-2 and 2-3 respectively. Moreover, a ghost of one of the off-axis sources (Cyg A) is strongly attenuated (see Fig.~\ref{fig:wsrt-b1834-ghost}). }
 \label{fig:wsrt-b1834-full}\label{fig:wsrt-b1834-full-original-stokesq}
\end{figure*}

As discussed, the filter frequency scales linearly with the baseline size: on long baselines, the fringe speed of sources is fast in both the frequency direction and the time direction. On short baselines, a source might cause only a few fringes or less in the frequency direction. It is therefore more difficult to filter short baselines, and Fig.~\ref{fig:wsrt-single-baseline-0x2} visualizes this problem. While the tangential contribution of Cygnus A has been removed effectively in the figure, only a small part of its radial contribution has been removed. The filter was able to remove the Sun because it is further away. On very short baselines, the real and imaginary components produced by a source are almost constant, and applying a low-pass filter in frequency direction on such a baseline will perform similar to averaging the frequency channels together. In such cases, the filter will not affect the astronomical data, but only average the noise out. If the fringe speed does not exceed the filtering frequency sufficiently on all baselines, the source will appear in the shorter baselines, hence the large scale structures of the source sidelobes will remain. In general, the combination of bandwidth, filter area and baseline length define the success of the frequency filter. Table~\ref{tbl:fringe-frequency-examples} shows a few configurations and their corresponding fringe speed for a particular baseline size and distance to the phase centre.

In Fig.~\ref{fig:wsrt-b1834-full}, all baselines were imaged together. The unfiltered Stokes I image is quite severely affected by sidelobes coming from off-axis sources. Moreover, because the off-axis sources come in through the far side of the primary beam, they appear in the polarized images as well. After filtering, the confusion noise is reduced significantly. Depending on which empty region is selected to calculate the RMS over, the noise has gone down by a factor of 1.5 to 2 in Stokes I, while the polarized images show a factor of 2 to 3 decrease in noise. Because the short baselines could not be filtered correctly in the frequency direction due to the limited bandwidth, the low-frequency components of the sidelobes remain. With sufficient bandwidth, such as for LOFAR, the results will be even more significant. CLEANing the images of Fig.~\ref{fig:wsrt-b1834-full} removes some of the bright sources in the centre, but the strong sources in the sidelobes can not be removed by CLEANing. As one can expect, the CLEAN algorithm is able to CLEAN deeper and find more sources in the filtered image. 

\begin{figure}
 \begin{center}
  \subfloat[Original (Stokes I)]{
    \includegraphics[height=37mm]{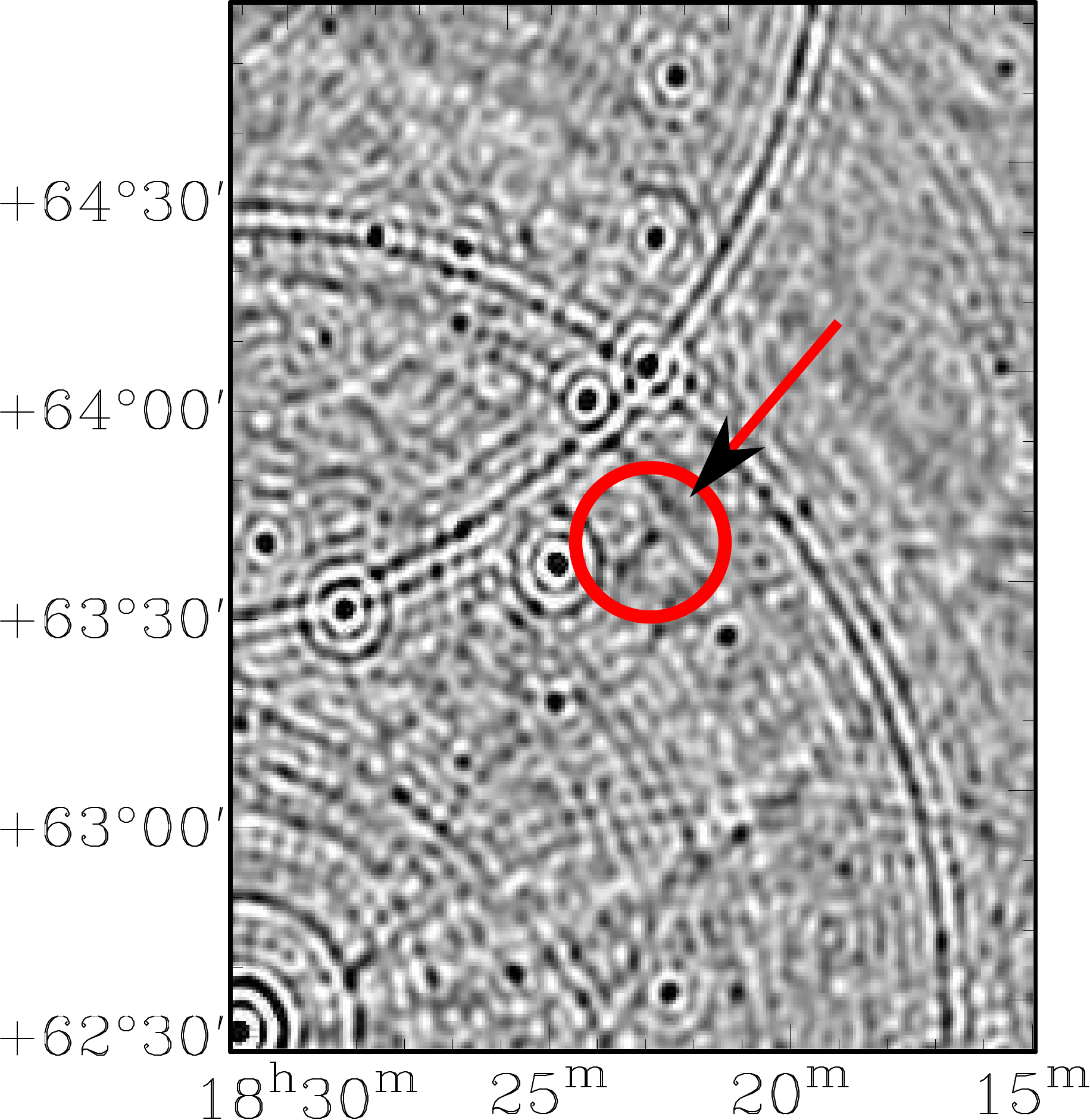}%
  }%
  \subfloat[Filtered (Stokes I)]{
   \includegraphics[height=37mm]{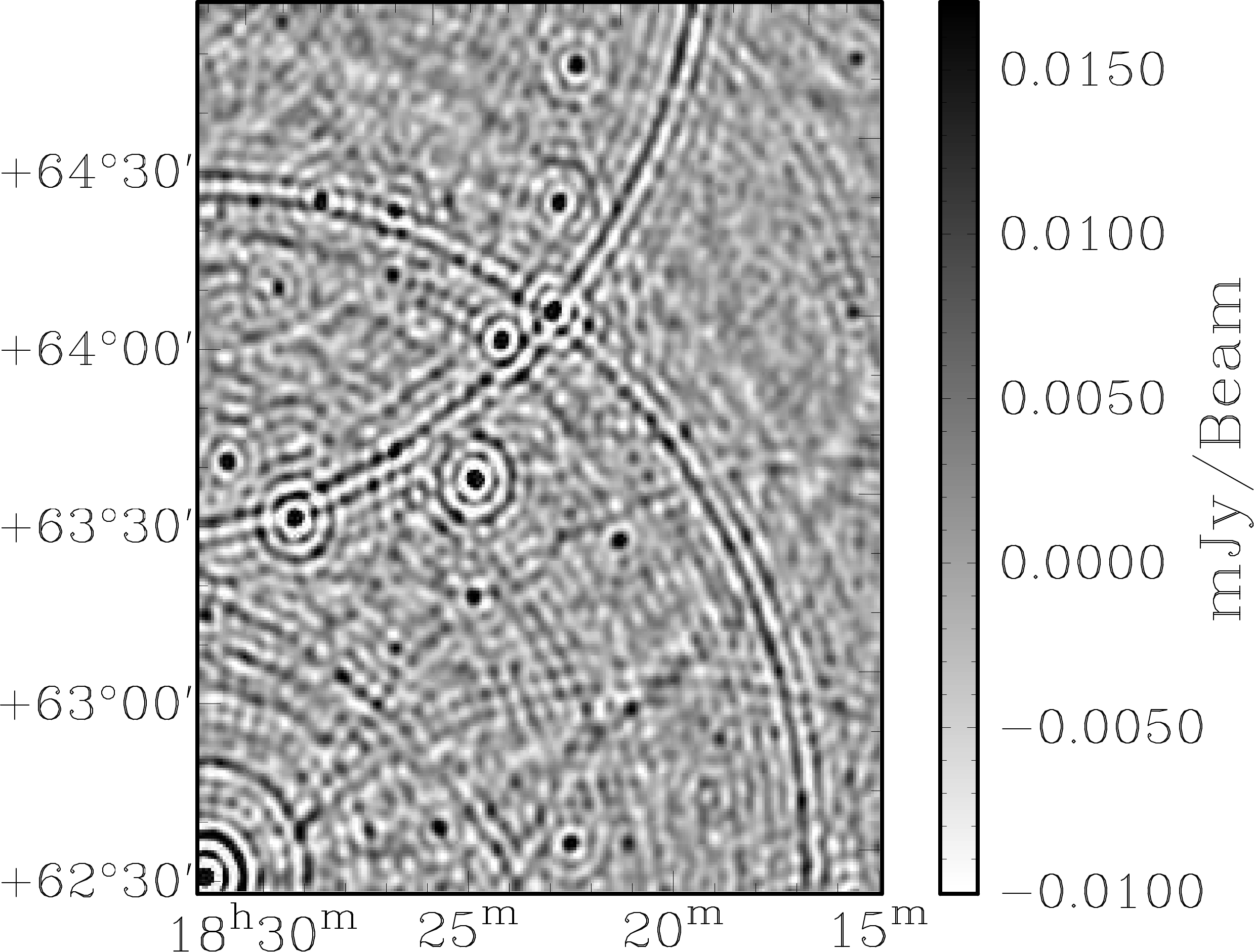}
  }\\
  \subfloat[Original (Stokes Q)]{
    \includegraphics[height=37mm]{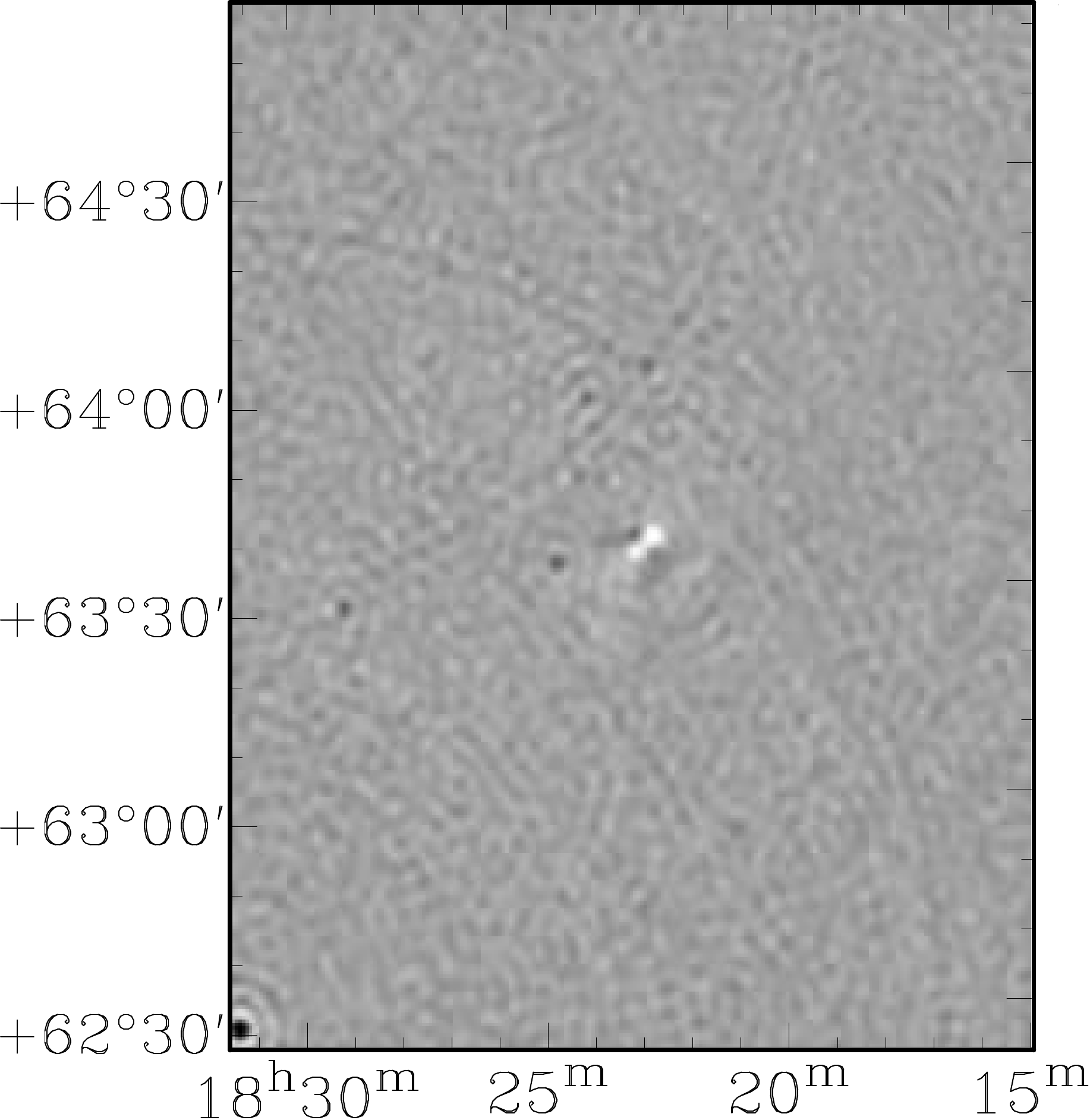}
  }
  \subfloat[Filtered (Stokes Q)]{
   \includegraphics[height=37mm]{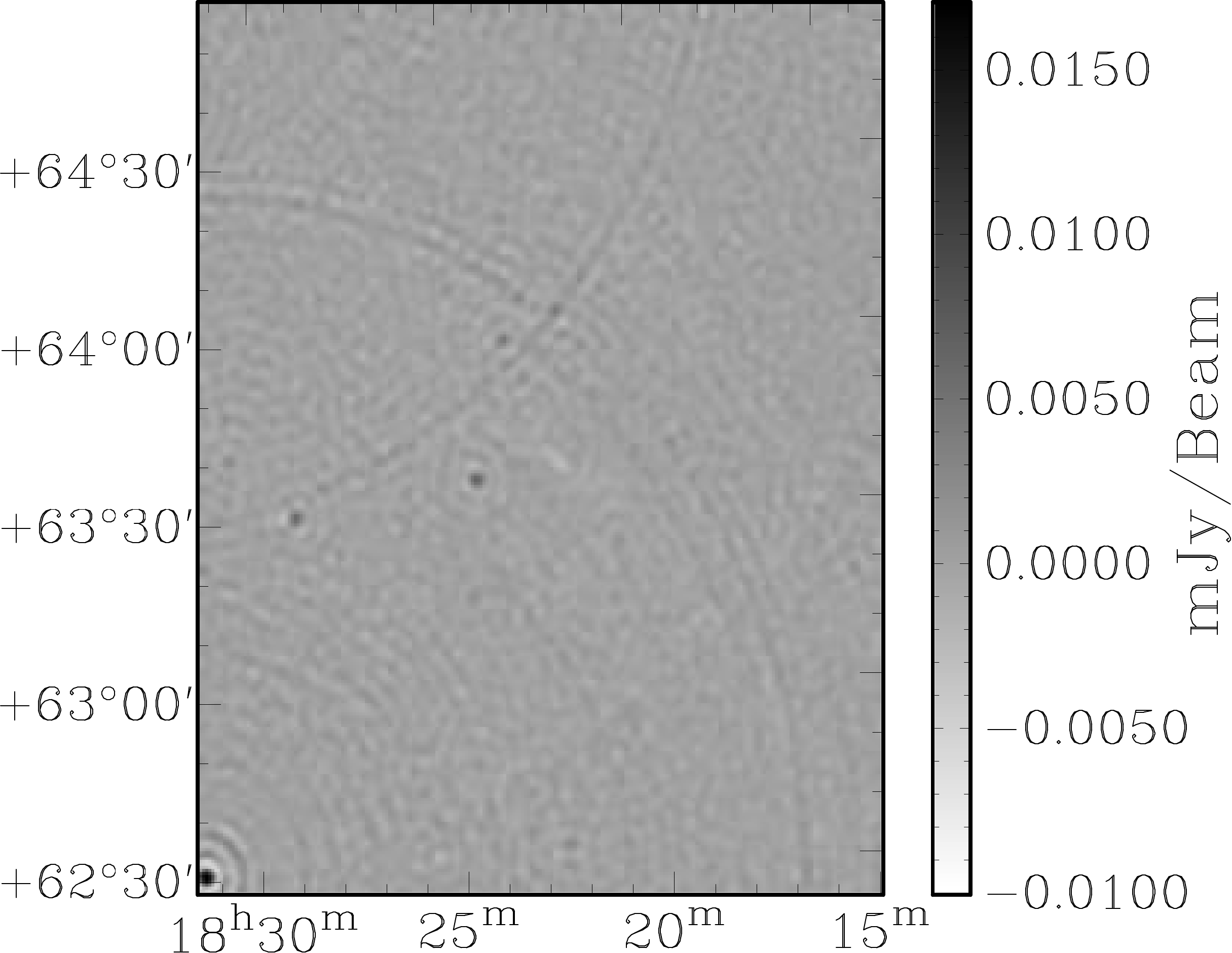}
  }
 \end{center}
 \caption[]{Enlargement of the central area of Fig.~\ref{fig:wsrt-b1834-full}: Aliasing of off-axis source causes a ghost in the primary field, which is attenuated by the low-pass filter.}
 \label{fig:wsrt-b1834-ghost}
\end{figure}

Another less obvious effect of the filter is suppression of ghost sources that are caused by aliasing of the off-axis sources. When looking at Fig.~\ref{fig:wsrt-b1834-full-original-stokesq}, it appears that there is one strong polarized source near the centre of the field. However, when performing the low-pass filters, the source disappears. The reason for this is that the source is not a real source, but a low frequency projection of an off-axis source: a ghost. A zoom in on this ghost as in Fig.~\ref{fig:wsrt-b1834-ghost} shows that the ghost is also present in Stokes I. This ghost is an aliasing artefact caused by the gridding in the imager. It appears as a normal source and contains regular sidelobes, as can be seen in Fig.~\ref{fig:wsrt-b1834-full-original-stokesq}. Low-pass filtering in time and frequency attenuates the ghost, as will any other method that attenuates the original off-axis source. The aliased ghost is caused by baselines which are gridded just below the Nyquist rate of the source. If the source is sampled correctly, its ghost will not appear at all. On the other hand, if the source is badly undersampled, its contribution will average out.

\subsection{Dealing with flagged samples}
\begin{figure*}
 \begin{center}
  \subfloat[Original]{
    \includegraphics[width=86mm]{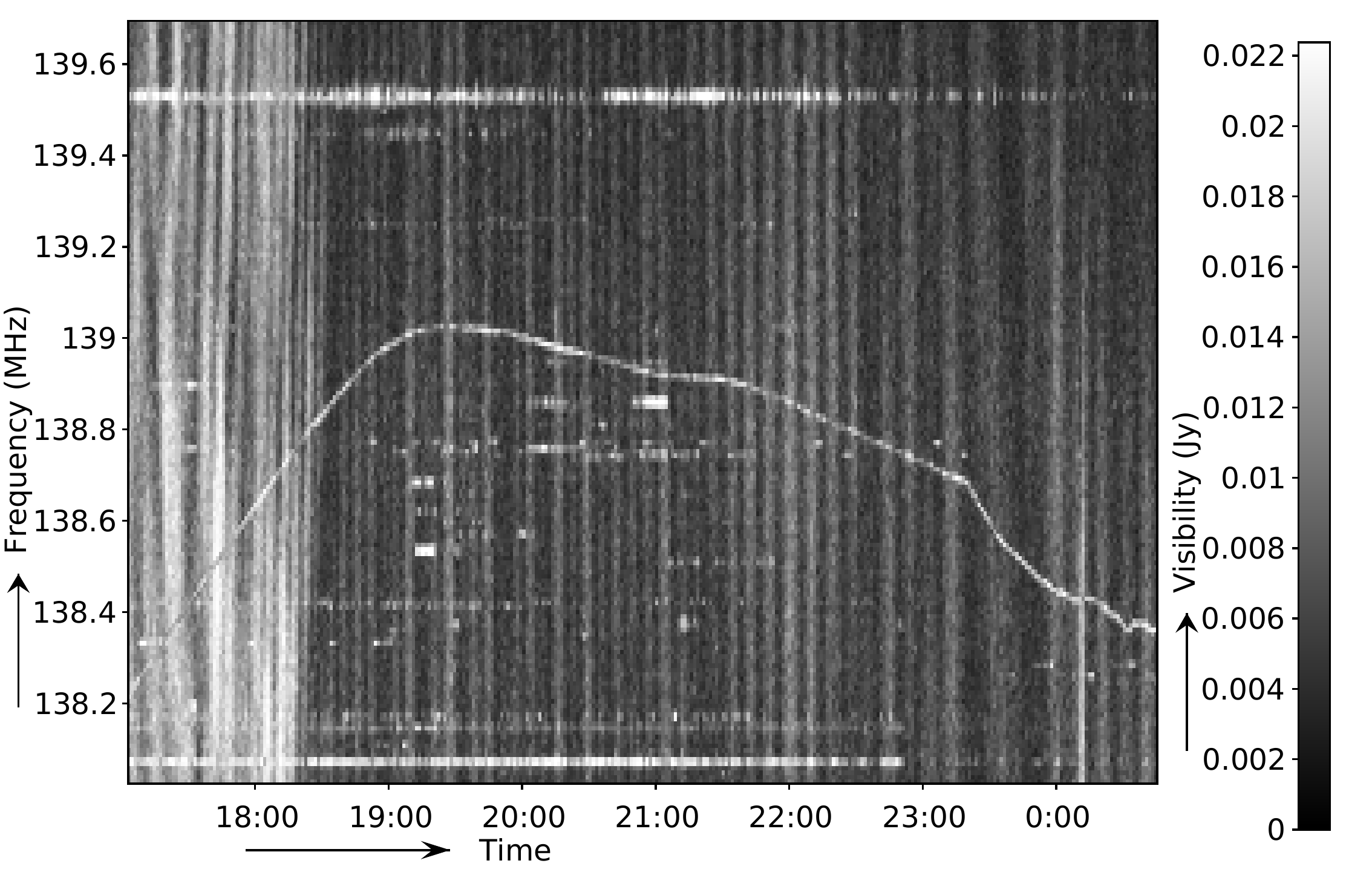}
  }
  \subfloat[After automated flagging]{\label{fig:flags-interpolated-flagged}
   \includegraphics[width=86mm]{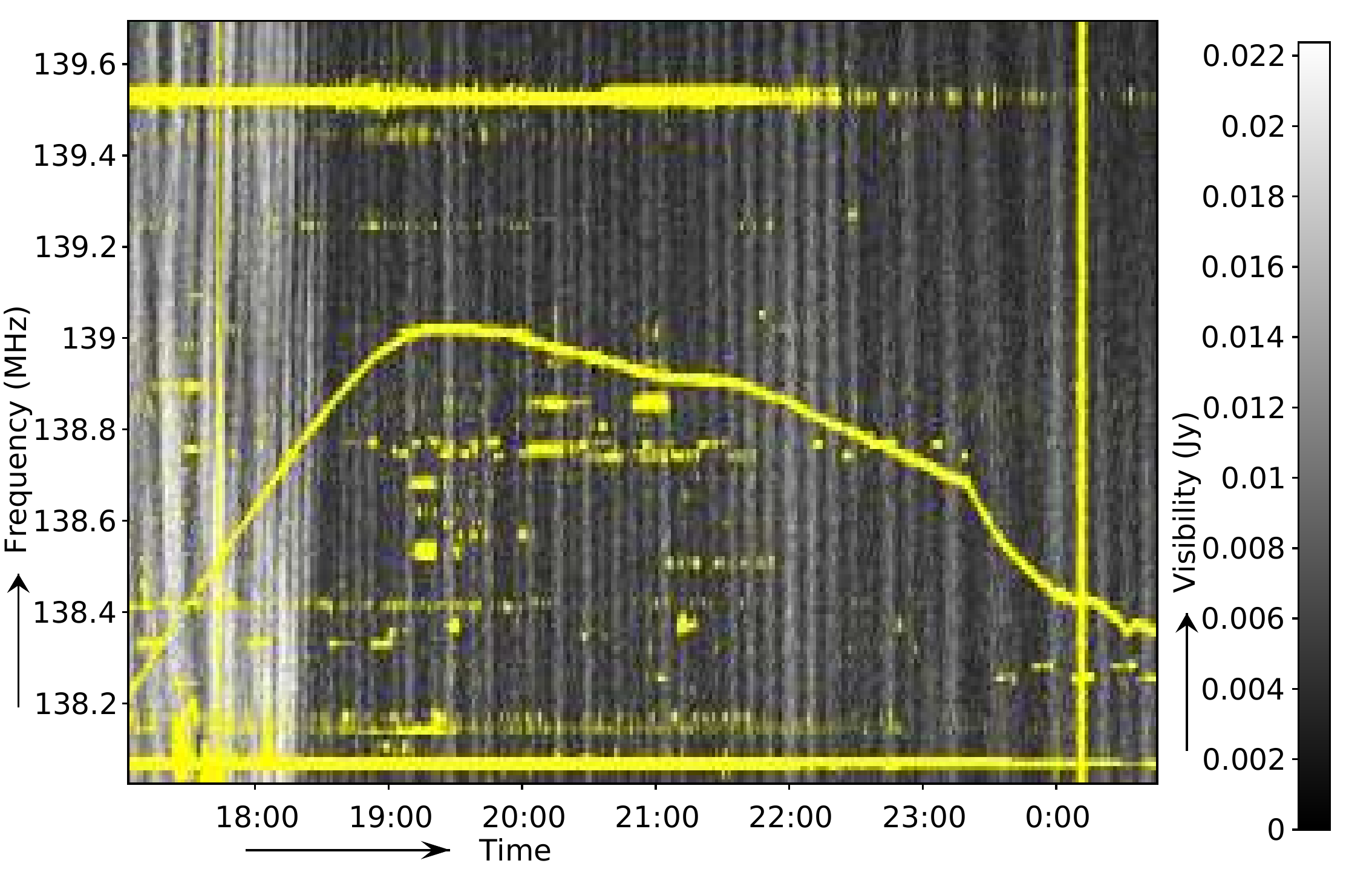}
  }\\
  \subfloat[After interpolation of flagged areas]{\label{fig:flags-interpolated-sub}
    \includegraphics[width=86mm]{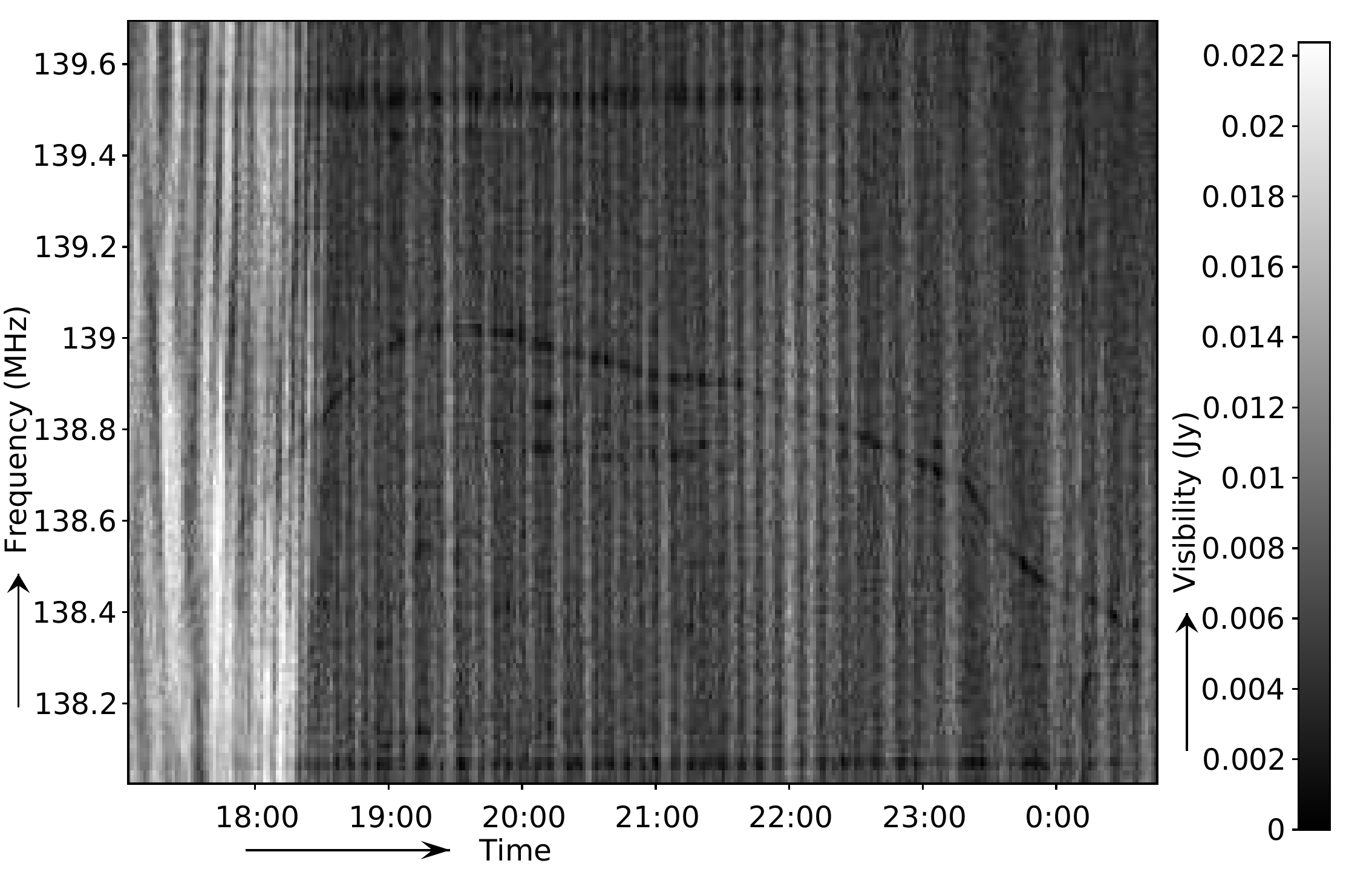}
  }
  \subfloat[After low-pass filtering in time and frequency directions]{
   \includegraphics[width=86mm]{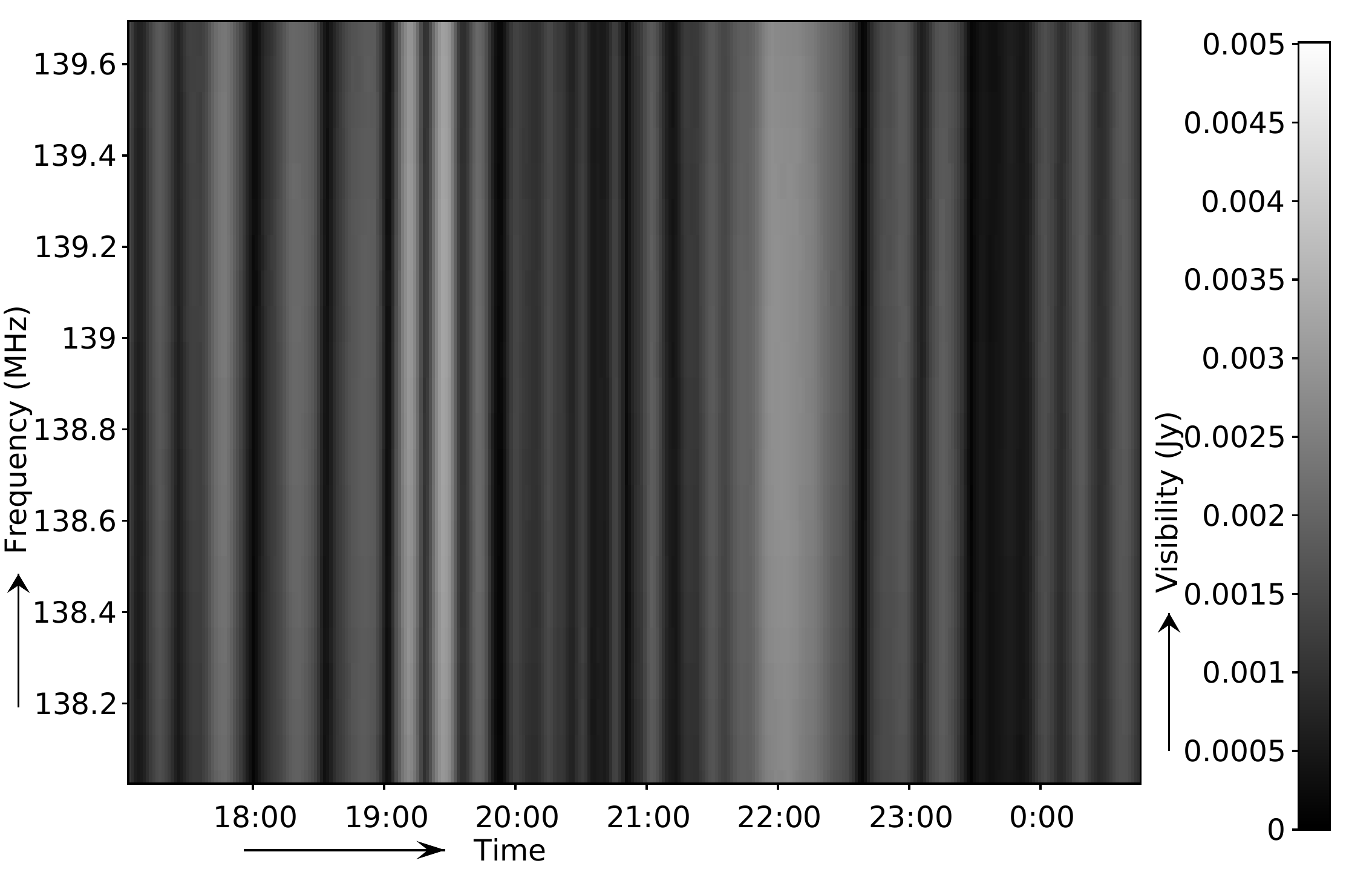}
  }
 \end{center}
 \caption[]{A baseline for which the flags have been interpolated and filtered. Note that in panel~\subref{fig:flags-interpolated-sub}, the RFI is still visible in the interpolated time frequency plot by eye, because those areas have a lower variance compared to the original.}
 \label{fig:flags-interpolation}
\end{figure*}

A complicating factor for low-pass filtering the time-frequency domain is the fact that the time-frequency plane contains flagged data due to RFI contamination. This has to be taken into account before convolving the data with a sinc function. To solve the problem, we will mimic how flags are handled during other stages of reduction. Two techniques for solving flagged samples are commonly used. The first is to set flagged samples to zero and account for the missing samples when deconvolving. The second is, if the samples are flagged before either correlation, further averaging or gridding, to only average over unflagged samples. The latter is similar to linear interpolation of the flagged samples, albeit the uv-position should be changed slightly because of the change of the centroid, to prevent bandwidth or time smearing. Before correlation or at high time and frequency resolutions, the difference between neighbouring samples is small enough that the error due to linear interpolation is small.

Since these methods have shown sufficient accuracy in practice, we have used a similar linear interpolation scheme: the data is interpolated by performing a Gaussian convolution on the unflagged data. The flagged samples in the original image are subsequently replaced with values from the convolved image. The result of this procedure on one of the WSRT B8134 set is given in Fig.~\ref{fig:flags-interpolation}. Normally, only data that are not flagged are used for imaging. For Fig.~\ref{fig:flags-interpolation}, this is the data from panel~\subref{fig:flags-interpolated-flagged}. To be able to filter the set, the flagged samples are interpolated as in panel~\subref{fig:flags-interpolated-sub}. Tests using all baselines of the WSRT B1834 set show that the difference between imaging of the flagged set and the interpolated set in which all samples are used are small, as sources in the area of interest are changed less than 1~per cent. After low-pass filtering, we reapply the old flags. The rationale for this is to make sure that ranges that contain RFI are not used during further reduction, and the interpolated data is only used for filtering.

\subsection{Computational requirements}
For filtering the observation of B1834, we have used a regular desktop with a dual core Intel Core2 CPU running at 2.13~GHz and 2~GB of memory. Filtering the measurement set to create Fig.~\ref{fig:wsrt-b1834-full} in time and frequency direction, including interpolating the RFI samples, takes on the order of an hour on this machine, while we have been performing the filtering step with a non-optimized proof-of-concept script. This time is comparable with the time it takes to image the data set with the {\tt lwimager}\footnote{The {\tt lwimager} or Light Weight Imager is part of the casarest program, a subpackage of the Common Astronomy Software Applications package (\url{http://casa.nrao.edu/})} that was used to create the images. The measurement set contains 91 baselines with 4 polarizations, 4300 time steps and 512 channels, and is 8 gigabytes in size. The IO takes about 15~per cent of the time. Hence, the computational requirements for filtering are not excessive. The method performs around an order of magnitude faster than demixed peeling as implemented in the LOFAR pipeline.

One complicating factor is that observations with a large number of frequency channels are often split up in many (sub-)bands. This is for example the case for LOFAR observations. Since the total data can become large, the sub-sequences are divided over several nodes on a cluster. Efficient synchronisation of the data between the nodes is not trivial, but by using a few nodes concurrently, we have been able to successfully filter a high resolution LOFAR observation within a few hours.

\section{Discussion} \label{sec:discussion}

\subsection{Comparison of filter methods}
The filters discussed were the single fringe filter (\S\ref{sec:fringe-filtering}), the low-pass filter (\S\ref{sec:time-domain-low-pass-filter} and \S\ref{sec:freq-lowpass-filter}) and the projected fringe filters (\S\ref{sec:projected-fringe-filter} and \S\ref{sec:iterative-projected-filter}).

The single fringe filter as proposed by Athreya and the introduced projected fringe filter can be applied before ionospheric calibration. We have shown that the single fringe filter is acceptable accurate for removing stable RFI sources, as long as the source to be removed is strong and reasonably constant. The filter should include the change in fringe frequency within the window as in Eq~\eqref{eq:static-fringe-exact-solution} for maximum accuracy. We do not observe stable, broadband RFI in LOFAR or WSRT that can be dealt with this method. To remove off-axis sources with the single fringe filter, an accurate model of the source is needed. In practical situations with non-constant sources, the fitting error exceeds 10~per cent and is therefore highly inaccurate in comparison to common ways to remove sources. It is therefore too inaccurate to be useful for off-axis source fitting.

One of the reasons for a projected fringe filter to be useful is that it requires no model, except for a direction to filter towards. However, the iterative projected fringe filter was shown not to be accurate enough and will in general remove little more than 50~per cent of the source's power. Hence, the iterative projected fringe filter provides little benefit when removing (celestial) off-axis sources. The projected fringe low-pass filter can remove a source completely, but has the unwanted effect of filtering part of the area of interest. However, this unwanted effect only occurs on a small part of the data; the further the source that is to be removed is from the area of interest, the smaller the area. A possible approach might therefore be to exclude the part of the data on which the fringe speed of the area of interest exceeds the filter speed. Subsequently, the data can be calibrated on first order, and the calibration solutions can be extrapolated to the excluded data. The method is about an order of magnitude faster than peeling and demixed peeling. This approach needs further research.

In contrast to the single and projected fringe filters, the use of the introduced low-pass filter lies mainly in removing off-axis sources. The low-pass filter in frequency will low-pass filter any structure in frequency direction, thus is probably only useful for multi-frequency synthesized imaging. In this situation, the frequency low-pass filter is an ideal tool to improve the signal to noise ratio of the area of interest after all calibration and subtraction of modelled sources has taken place, because it attenuates radial sidelobes. When structure in frequency direction is important, e.g., when performing spectrography, the method can not be applied. The frequency low-pass filter is not necessarily limited to application after calibration. Because the phases and amplitudes are reasonably stable in frequency direction, it can be assumed that filtering in frequency direction will not remove information essential for calibration -- as long as all modelled sources are within the unfiltered area in image plane.

The low-pass filter in time might be less applicable for uncalibrated data, because it removes the high-frequency components introduced by quick phase or amplitude changes such as ionospheric changes. This problem is less relevant on longer baselines, because of the faster fringe speed: at $\lambda$\nolinebreak{}=\nolinebreak{}21~cm, a single degree off-axis source has a fringe duration of 17~min on a one kilometre baseline. The low-pass filter in time removes tangential sidelobes of off-axis sources, which implies that the sidelobe confusion noise in the area of interest is not directly attenuated. Nevertheless, this filter can be useful to reduce aliasing effects, such as removing an aliased ghost, where it is complementary to the frequency low-pass filter.

In case the low-pass filter in the time or frequency direction is applied before calibration, one should make sure that the filter does not introduce baseline-specific errors (closure errors), because these might cause self-calibration to fail. Since all presented filters are applied on individual baselines, this holds for all the filters. Although \citet{fringe-fitting-rfi-mitigation} argues that fringe fitting does not introduce closure errors, that only holds if the fit is perfectly accurate. It is unclear if this is generally true, because the accuracy of the fit is dependent on the fringe rate, and therefore baseline dependent. However, as long as the baseline-dependent error is small, self-calibration will benefit from the removal of the RFI source. We have not yet looked at calibrating filtered data, and this requires further research.

For low-pass filtering we have only looked at applying a rectangular windowed sinc convolution (truncated sinc), naturally imposed due to the finite time/frequency range. Especially when the window is small in comparison to the size of a fringe rotation, non-rectangular windows might improve efficiency. Different window functions can provide different trade-offs between the sidelobes and the steepness of the filter edge in the image plane: functions with a small resolution bandwidth, such as the rectangular function, will create a sharp edge that has ripples. On the other hand, functions with high sidelobe fall-off will create a smoother edge and will suppress the ripples better. An example of such a function is the Hann function \citep{harmonical-window-functions}.

It is harder to distinguish off-axis sources from on-axis sources in data that corresponds to specific areas in the $uv$-plane. The $uv$-areas for which this is the case, are areas at which the rotation angle of the $uv$-track is near the rotation angle of the off-axis source in the image plane. The reason for this is that the fringes of off-axis sources are slow in time direction in these $uv$-areas, and cannot be distinguished from the slow fringes of sources near the phase centre. Any method that tries to separate off-axis sources from on-axis sources, will consequently be less accurate in these areas. Unfortunately, off-axis sources cause sidelobes that interfere with the phase centre in these same areas, hence it is important to accurately remove the off-axis sources from these areas in order to achieve high dynamic ranges. Using frequency bandwidth to distinguish sources is necessary in these ranges. Many algorithms look at small bandwidths at a time. For example, most algorithms currently applied for LOFAR, such as demixed peeling or self-calibration, currently only use information from one or a few subbands at a time, while a LOFAR subband is only 200~KHz. To accurately separate off-axis sources with these algorithms, multiple subbands have to be combined together.

Low-pass filtering is an implicit effect of integrating and averaging that occurs in the standard pipeline of interferometers. The implications of that will be discussed in the next section.

\subsection{Adverse effects of time and frequency averaging} \label{sec:averaging}
\begin{figure}
 \begin{center}
   \includegraphics[width=70mm]{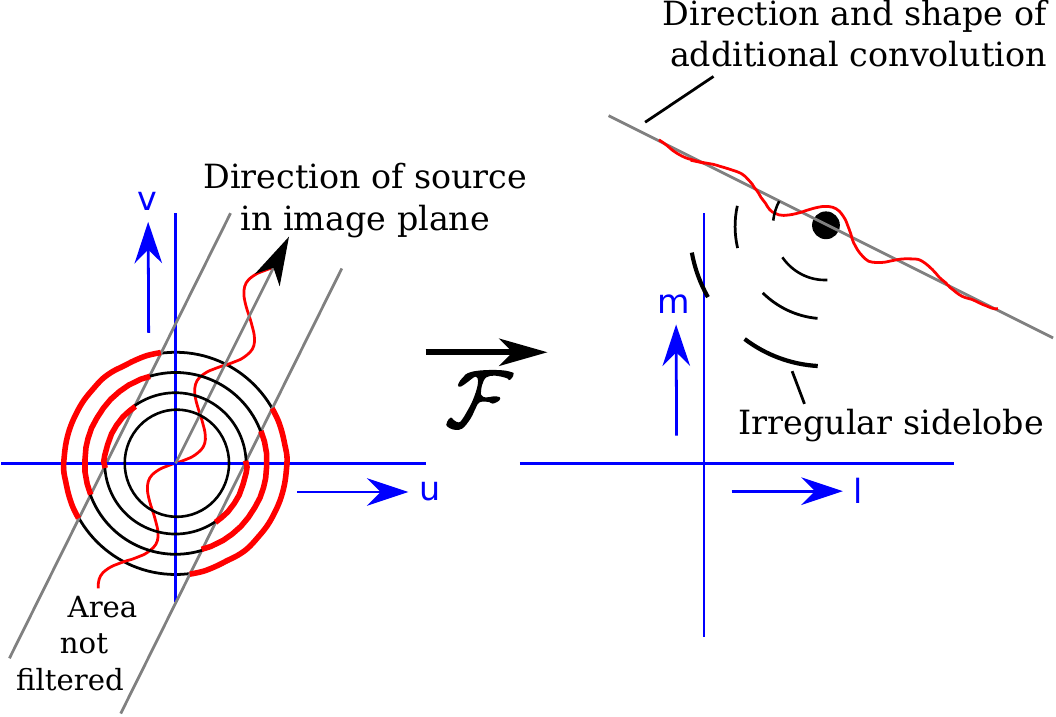}
 \end{center}
 \caption[]{The effect of over-averaging an observation in time direction, causing off-axis sources to be partly filtered in the long baselines. The left and right panel show respectively the $uv$-plane and the image domain.}
 \label{fig:averaging-convolution-explanation}
\end{figure}

To reduce the data volume, the correlation coefficients are integrated over time directly after correlating, and are sometimes further time averaged, for example after a RFI flagging procedure has detected corrupted samples, as is the default for LOFAR. When imaging, the visibilities are once more averaged for gridding, to be able to apply a fast Fourier transform (FFT). Nyquist's theory states that the original signal can be reconstructed as long as the sampling frequency is at least two times the highest frequency. Hence, in order not to lose information, the sampling frequency in time and frequency should be twice the fringe frequency of the source given by respectively Equation~\eqref{eq:source-fringe-frequency-time} and \eqref{eq:frequency-fringe-rate}. In this section we will discuss two side effects of averaging: (1) the effect of low-pass filtering and (2) the effect due to aliasing.

When data is averaged, the highest frequency components can no longer be presented, and the data is therefore filtered of high frequencies. The corresponding side effects of time and frequency averaging can be deducted from the low-pass filtering results. Since the amount of averaging is normally independent of the baseline size, i.e., all baselines will be averaged equally, an off-axis source will only be filtered in long baselines. This has been sketched in Fig.~\ref{fig:averaging-convolution-explanation} for over-averaging in time direction. Over-averaging in frequency is similar, but in radial direction. For these reasons, the effect of time and frequency averaging is baseline dependent and will contribute to closure errors. It is also a direction-dependent effect (DDE), since the distance of the source to the phase centre defines its fringe speed, and therefore the amount of attenuation. Therefore, different positions on the sky will be differently attenuated. Finally, averaging in time and frequency directions only complement each other partly: even by over-averaging the time and frequency directions significantly, the shorter baselines will still contain the source.

In an over-averaged set, a source will appear at its original location, but the source is fully present only in a subset of the baselines, which will cause it to have irregular sidelobes. Therefore, the source can not perfectly be removed with CLEAN, unless CLEAN is performed baseline by baseline or on smaller ranges of baselines, which is harder due to the low signal-to-noise ratio and dirtier point spread function of fewer baselines. Direction-dependent calibration might help, but directions that have been attenuated might still cause problems, e.g., in some antennas they will generate high gain solutions and therefore introduce noise. For these reasons, it is important to remove strong sources with fast fringe rates before time or frequency averaging in order to avoid their side lobes or added noise in the area of interest. This effect is most prominent in interferometric elements with a large field of view --- a small element beam will naturally attenuate off-axis sources.

\begin{figure}
 \begin{center}
   \hspace{-1cm}
   \includegraphics[width=90mm]{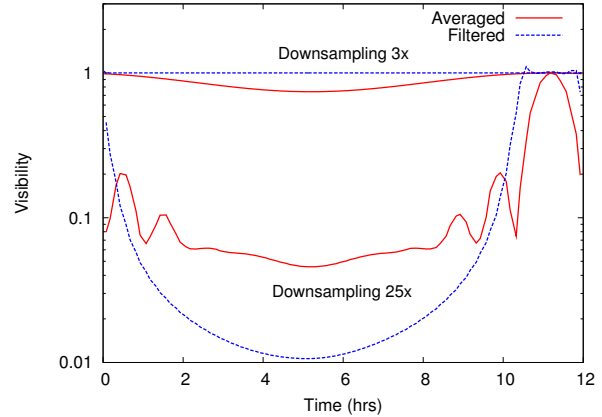}
 \end{center} \vspace{-1cm}
 \caption[]{Simulated effect of decreasing the time resolution with a factor of 3$\times$ and 25$\times$, on one single baseline with a single source, using two different methods: (A) averaging the data; and (B) low-pass filtering the data followed by nearest neighbour interpolation.}
 \label{fig:avg-vs-filtering}
\end{figure}

A second side effect of averaging comes from the fact that averaging is not a perfect low-pass filter, and will cause aliasing effects of high frequencies in the lower fringe frequencies. This will increase the noise generated by off-axis sources because they will not be filtered as much as possible. Time averaging can also distort sources of interest and can even generate ghost sources if off-axis sources have not been removed beforehand, as was seen in Fig.~\ref{fig:wsrt-b1834-ghost}. To remove these effects, a low-pass filter can be used before down sampling the visibilities.

Fig.~\ref{fig:avg-vs-filtering} shows the difference on a simulated observation between these two methods of changing the time resolution: (A) averaging the data; and (B) low-pass filtering the data followed by nearest neighbour interpolation. The down sampling factor was 3 and 25 for respectively the top and the bottom lines. The source is 30$\degree$ from the phase centre and the simulated WSRT baseline is 720~m, observing at 140~MHz and 62$\degree$ declination. The maximum fringe speed is 30~Hz and the correlator integration time was 5 seconds. The figure demonstrates the non-ideal effect of averaging: sources which fringes beat with half the (new) Nyquist speed are attenuated up to 25~per cent, which does not occur in the filtered case. Moreover, a source that beats faster than the Nyquist speed (bottom lines) is better attenuated with less aliased sidelobes by the filtering compared to averaging. The attenuation effect of averaging quickly decreases when the source is closer to the phase centre, but is still on the order of one percent at one degree when three times averaged.

Time averaging has been used to average out RFI or other sources that have a high fringe rate. \citet{fringe-fitting-rfi-mitigation} describes that RFI can be attenuated because of fringe stopping, although it is said that this is less effective at low frequencies. In \citet{rfi-mitigation-in-aips}, the authors also describe averaging out RFI. As this article has shown, although the source itself is attenuated by averaging, and therefore helps calibration, we have shown it is better to perform an explicit low-pass filter before downsampling. The fringe frequency expressed in fringes/sample is almost always higher in time direction compared to frequency direction. Hence, if one relies on fringe stopping and correlator averaging to suppress RFI or off-axis sources, the noise in the area of interest is still affected by the source, since time averaging does not remove sidelobes in the direction of the phase centre (Fig.~\ref{fig:time-domain-low-pass-filter}).

Time and frequency averaging are also part of the peeling algorithm, where it is used to filter off-axis sources. From the perspective of maximum attenuation, the baselines should be filtered with a filter size relative to the baseline length, instead of the de facto method of uniform averaging. This would suppress off-axis sources as much as possible, and equal in all baselines. However, care should be taken not to remove small temporal changes due to the ionosphere, that are needed for calibration. Fortunately, the ionosphere is typically stable in timescales of several minutes.

It is well known that data averaging can cause tangential and radial smearing when averaging respectively the time and frequency dimension \citep{big-book-on-synthesis-imaging}. The symptoms of bandwidth and time smearing can be intuitively explained with the results of this paper. As we have seen, the tangential and radial smearing happens because the longer baselines attenuate the source in a particular area of the $uv$-plane.

By using appropriate resampling techniques such as described in the paper, instead of time or frequency averaging which is used de facto, it is possible to reduce a data set to a smaller size with fewer artefacts. This might especially become important for arrays with a large field of view, long baselines and high data rates, such as LOFAR and the SKA, or high frequency interferometers such as ALMA. In the future, it might be interesting to resample short baselines to lower resolutions, as these baselines contain the slowest fringe rates. This could further reduce the size of a measurement. However, operations such as calibration currently can not handle irregularly sampled data.

\subsection{Relation to gridding}
To perform the two-dimensional FFT transform used for imaging the data, the uv-tracks are normally gridded onto a uniform grid. Like averaging, this has the side effect of low-pass filtering the data: the maximal fringe speed in any direction is defined by the grid resolution. In contrast to time or frequency averaging, the filter size is relative to the length of the baseline: long baselines are gridded with a finer resolution compared to short baselines. The filtering effect of gridding is therefore equal to low-pass filtering in time and frequency: off-axis sources will be attenuated equally in all baselines. The somewhat counter-intuitive fact is that coarsely gridding the uv-plane will suppress sidelobes of off-axis (RFI) sources in the image plane, and might increase the signal-to-noise in the area of interest. Furthermore, frequencies that can not be represented in the UV-plane, correspond with sources that fall outside the image plane. Therefore, imaging only the area of interest is an efficient way of filtering off-axis sources not of interest.

Analogues to time and frequency averaging, the down-sampling before gridding is performed in a non-ideal way, for example by averaging\footnote{Most software packages do use more elaborate ways of sampling the data on the grid, for example by using prolate spheroidals.}. From the conclusions in this work, we think aliasing effects are the reason why off-axis source that are not visible in the image plane, still produce sidelobes when performing regular gridding. The side effects are similar to the effects presented in Fig.~\ref{fig:avg-vs-filtering}, which shows that sources both faster and slower than the Nyquist frequency are inefficiently attenuated. To solve this, the high fringe frequencies should be removed before gridding the data on the uv-plane. Again, the best way to do this is to low-pass filter the time and frequency directions before gridding.

\subsection{Relation to other techniques}
Although we have not tried combining this method with techniques such as (demixed) peeling, it is likely that the presented low-pass filters can complement these. There are two reasons for this:
\begin{itemize}
 \item During calibration, the solutions are constrained by solving for antenna gains and by using the measurement equation. Calibration normally assumes solution constantness over short time intervals and small bandwidths, and does not assume relations over the full time or frequency range. The low-pass filter uses the full time-frequency domain of a single baseline to disentangle sources. Therefore, it uses information that is complementary to the information used in standard removal techniques.
 \item The low-pass filtering techniques are not model-based. On one hand, this allows direct and unbiased removal with less chance of inadvertently biasing towards an incorrect model, but on the other hand implies that there might not be enough data to separate sources in certain cases. Another difference with model based fitting, is that model based fitting can fail to converge due to an insufficient signal to noise level. Low-pass filtering is not limited by the signal to noise: due to the linearity of the Fourier transform, the result of low-pass filtering two time or frequency streams separately followed by averaging is equal to filtering the average of the two streams.
\end{itemize}

Because the low-pass filtering techniques do not involve non-linear fitting, they are much faster. If the filter techniques can be used for first order removal of off-axis sources, they might save a considerable amount of processing time. Investigation of the relation between the filter methods and other techniques will be the focus of further research. The LOFAR telescope provides a good test case for further research. Because of its large data volumes, its processing power is a considerable limitation, and it could potentially benefit a lot from faster source subtraction algorithms.

\section{Conclusions \& Outlook} \label{sec:conclusions}
We have shown that several filters can be used on individual baseline correlations to attenuate both off-axis sources and RFI sources in radio observations, thereby increasing the dynamic range of the observation. Because of the high performance of the filters, they are suitable for modern high-resolution observatories and can offer a complementary or alternative way to remove the sources. Especially the low-pass filter in the time and frequency directions are attractive, as they effectively attenuate all sources and their sidelobes outside a certain radius from the phase centre. However, they work less well on shorter baselines, and need a considerable bandwidth to remove sources effectively.

The next step is to further test the methods on other data, preferably with larger bandwidths, to see if the methods work in practice as well as in theory in other cases as well. Applying the filter on LOFAR data is attractive, because the off-axis source removal methods currently used are computationally intensive. With the large bandwidth of LOFAR, it would in theory be possible to, e.g., filter all sources outside 10 degrees even on baselines as short as 100 meters.

\section*{Acknowledgements}
The Westerbork Synthesis Radio Telescope is operated by ASTRON (Netherlands Institute for Radio Astronomy) with support from the Netherlands Foundation for Scientific Research (NWO). We also would like to thank Wim Brouw, Ue-Li Pen and Ronald Ekers for discussions and Panagiotis Labropoulos for help with the EoR-cluster.

\bibliographystyle{mn2e}
\bibliography{Post-correlationFiltering}

\label{lastpage}

\end{document}